\documentstyle[12pt,htb]{article}

\newcommand{\beq}{\begin{equation}}
\newcommand{\eeq}{\end{equation}}
\newcommand{\beqn}{\begin{eqnarray}}
\newcommand{\eeqn}{\end{eqnarray}}
\newcommand{\stackM}{\stackrel{\scriptstyle >}{{ }_{\sim}}}
\newcommand{\stackm}{\stackrel{\scriptstyle <}{{ }_{\sim}}}
\newcommand{\ba}{\begin{array}}
\newcommand{\ea}{\end{array}}  

\begin{document}
\thispagestyle{empty}
\def\prepnum{UAB-FT-397}
\def\data{}
\def\hepnum{hep-ph/9607485}
\begin{flushright}
{\parbox{3.5cm}{
\prepnum

\data

\hepnum
}}
\end{flushright}


\vspace{3.5cm}
\begin{center}
\begin{large}
\begin{bf}
QUANTUM EFFECTS ON 
$t\rightarrow H^{+}\,b$ IN THE MSSM: A  WINDOW TO 
``VIRTUAL'' SUPERSYMMETRY?\\
\end{bf}
\end{large}
\vspace{1cm}
J.A. COARASA, David GARCIA, Jaume GUASCH,\\
Ricardo A. JIM\'ENEZ, Joan SOL\`A\\
\vspace{0.25cm} 
Grup de F\'{\i}sica Te\`orica\\ 
and\\ 
Institut de F\'\i sica d'Altes Energies\\ 
\vspace{0.25cm} 
Universitat Aut\`onoma de Barcelona\\
08193 Bellaterra (Barcelona), Catalonia, Spain\\
\end{center}
\vspace{1cm}
\hyphenation{super-symme-tric}
\hyphenation{com-pe-ti-ti-ve}
\hyphenation{e-mer-ging}
\begin{center}
{\bf ABSTRACT}
\end{center}
\begin{quotation}
\noindent
We analyze the one-loop effects (strong and electroweak)
on the unconventional top quark decay mode
$t\rightarrow H^{+}\, b$ within the MSSM. The results are presented
in the on-shell renormalization scheme with a physically well motivated
definition of $\tan\beta$.
The study of this process at the quantum level is useful to   
unravel the potential supersymmetric nature of the charged Higgs 
emerging from that decay. As compared with 
the  standard mode $t\rightarrow W^{+}\,b$,
the corrections to $t\rightarrow H^{+}\, b$ 
are large, slowly
decoupling and persist at a sizeable level even for all 
sparticle masses
well above the LEP 200 discovery range. 
As a matter of fact, the potential size of the SUSY effects,
which amount to corrections of several ten percent, could 
counterbalance the standard QCD corrections and even make them to appear
with the ``wrong'' sign. 
Therefore, if the charged Higgs decay of the top quark 
is kinematically allowed --a possibility which is not excluded
by the recent measurements of the branching ratio
$BR(t\rightarrow W^{+}\,b)$ at the Tevatron -- 
it could be an invaluable laboratory  
to search for ``virtual'' supersymmetry.
While a first significant test of these effects
could possibly be performed at the upgraded Tevatron, a
more precise verification would most likely
be carried out in future experiments 
at the LHC.  
\end{quotation}
 
\baselineskip=6.5mm  
 
\newpage
 
\begin{Large}
{\bf 1. Introduction}
\end{Large}
\vspace{0.5cm}

Recently, the Standard Model (SM) of the strong and electroweak interactions 
has been crowned with the discovery of the penultimate
building block of its theoretical structure: the top 
quark, $t$\,\cite{Tevatron}.
At present the best determination of the
top-quark mass at the Tevatron reads as follows\,\cite{Grannis}:
\beq
m_t=175\pm 6\,\,GeV\,.
\label{eq:CDFD0}
\eeq 
While the SM has been a most successful framework to describe the
phenomenology of the strong and electroweak interactions
for the last thirty years,
the top quark itself stood, at a purely theoretical level --namely, on
the grounds of requiring internal consistency, such as gauge invariance 
and renormalizability--as a firm prediction of the SM
since the very confirmation of the existence of the bottom quark and
the measurement of its weak isospin quantum numbers\,\cite{Zerwas}.
With the finding of the top quark, the matter content of the SM has been fully
accounted for by experiment. Still, the last building block of the SM,
viz. the fundamental Higgs scalar, has not been found yet, which means that
in spite of the great significance of the
top quark discovery the theoretical mechanism by which all particles 
acquire their masses in the SM remains experimentally unconfirmed. 
Thus, it is not clear at present whether the SM will remain as the last word
in the phenomenology of the strong and electroweak interactions around
the Fermi's scale or whether it will be eventually subsumed within a larger 
and more fundamental theory. The search for physics beyond the SM, therefore,
far from been accomplished, must continue with redoubled efforts. 
Fortunately, the peculiar nature of the top quark (in particular its large
mass--in fact, perhaps the heaviest particle 
in the SM!-- and its characteristic interactions with the scalar particles)  
may help decisively to unearth any vestige of physics beyond the SM. 

We envisage at least four wide avenues of interesting new 
physics potentially conveyed by top quark 
dynamics and which could offer us the clue to solving the nature of
the spontaneous symmetry breaking (SSB) mechanism, to wit:
i) The ``Top Mode'' realization(s) of the SSB mechanism, 
i.e. SSB without fundamental Higgs scalars, but rather
through the existence of $t\bar{t}$ condensates\,\cite{Bardeen};
ii) The extended Technicolour Models; also without Higgs particles, and
giving rise to residual non-oblique interactions of the top quark with
the weak gauge bosons\,\cite{Chikuvula};
iii) The non-linear (chiral Lagrangian)
realization of the $SU(2)_L\times U(1)_Y$
gauge symmetry\,\cite{Georgi}, which may either accommodate or dispense with
the Higgs scalars. It can also generate additional (i.e. non-standard)
non-oblique interactions of the top quark with the weak gauge
bosons\,\cite{Peccei};  
and iv) The supersymmetric (SUSY)
realization of the SM, such as the Minimal Supersymmetric Standard Model
(MSSM)\,\cite{MSSM}, where also a lot of potential new
phenomenology spurred by top and Higgs physics might
be creeping in here and there. 
Hints of this new phenomenology may show up either in the form
of direct or virtual effects from supersymmetric Higgs particles or from 
the ``sparticles'' themselves
(i.e. the $R$-odd\,\cite{MSSM}  partners of the SM particles),
in particular
from the top-squark (``stop'') which is the SUSY counterpart of the top quark. 
Due to the huge mass of the latter, one expects that the 
top-stop system is one of the most preferential chiral
supermultiplets to which the Higgs sector
should couple. Therefore, top quark dynamics is deemed to be an ideal
environment for Higgs phenomenology and a most suitable SUSY trigger, if
SUSY is there at all.

In this paper, we shall focus our attention on the fourth large avenue of
hypothetical physics beyond the SM, namely on the (minimal) SUSY extension 
of the SM, the MSSM, which is at present the most predictive framework
for physics beyond
the SM and, in contradistinction to all other approaches, it has the virtue of 
being a fully-fledged Quantum Field Theory. Most important, on the experimental
side the global fit analyses to all indirect precision
data within the MSSM are at least as good as in the SM;
in particular,
the MSSM analysis implies that
$m_t=172\pm 5$\,\cite{LangMaryland,WdeBoer},
a result which is
compatible with the above mentioned experimental determinations of $m_t$.

In the MSSM the spectrum of Higgs-like particles  and of
Yukawa couplings is far and away richer than in the SM. In this respect,
a crucial fact affecting the results of our work is that in such a framework the
bottom-quark Yukawa coupling may counterbalance the smallness of the bottom
mass, $m_b\simeq 5\,GeV$,
at the expense of a large value of $\tan\beta$ -- the ratio of the vacuum
expectation values (VEV's) of the two 
Higgs doublets (Cf. Section 2) -- the upshot being that
the top-quark and bottom-quark Yukawa couplings
(normalized with respect to the $SU(2)$ gauge coupling)
as they stand in the superpotential, take on the form
\begin{equation}
\lambda_t\equiv {h_t\over g}={m_t\over \sqrt{2}\,M_W\,\sin{\beta}}\;\;\;\;\;,
\;\;\;\;\; \lambda_b\equiv {h_b\over g}={m_b\over \sqrt{2}
\,M_W\,\cos{\beta}}\,.
\label{eq:Yukawas} 
\end{equation}
Thus, depending on the actual value of $\tan\beta$, $\lambda_b$ and $\lambda_t$
can be of the same order of magnitude, perhaps even showing up 
in ``inverse hierarchy'': $\lambda_b>\lambda_t$ for $\tan\beta> m_t/m_b$. 
Notice that due to the perturbative bound
$\tan\beta\stackrel{\scriptstyle <}{{ }_{\sim}}70$ one never reaches
a situation where $\lambda_t<<\lambda_b$.
In a sense, $\lambda_b\simeq\lambda_t$  could be judged as a natural 
relation in the MSSM; it can even be a necessary relation in
specific SUSY-GUT models, e.g. those
based on $t,b$ and $\tau$ Yukawa coupling unification\,\cite{SO10},
at least at the unification point. Furthermore, one
expects that if such a relation holds, then it is not just the
top-stop system, but also the bottom-sbottom chiral supermultiplet
that could play a momentous role in the quantum physics of the
top and bottom quarks.
Indeed, since the Higgs sector of the MSSM doubles
that of the SM, and it comes associated with the 
fermionic SUSY partners --the so-called higgsinos--, one expects that 
in the limit $\lambda_b\stackM \lambda_t$ there should
occur a very stimulating dynamics triggered by the presence of a rich
variety of potentially large Yukawa-like interactions formed out of
the top/stop-bottom/sbottom-Higgs/higgsino fields.

A particularly brilliant
form of this dynamics, on which we shall focus our attention,
is revealed through the study of the quantum effects on
the non-standard top quark decay into a charged 
Higgs: $t\rightarrow H^+\,b$.
This decay, which has already deserved some attention in the early
literature on the subject\,\cite{Bigi,Kunszt},
is not at all excluded by the recent measurements
(at the Tevatron) of the branching
ratio of the standard top quark decay, $t\rightarrow W^+\,b$, as will 
be discussed in more detail in Section 2.
The quantum effects on $t\rightarrow H^+\,b$, which we wish to compute in
the framework of the MSSM at one-loop, can be both strong and electroweak like. 
Of these the conventional strong  corrections (QCD) mediated 
by gluons have already been treated in detail\,\cite{CD}. Also 
the subset of strong supersymmetric corrections mediated by gluinos, stop and 
sbottom squarks, i.e. the SUSY-QCD corrections, has been
discussed in Ref.\cite{GJS}.
Here, therefore, we will come to grips with the
remaining part --as a mater of fact, the largest and most
difficult part -- of the MSSM corrections: namely, the multifarious   
electroweak supersymmetric corrections produced by squarks, sleptons,
charginos, neutralinos and supersymmetric Higgs bosons, which we shall
combine with the total strong (QCD +SUSY-QCD) corrections to obtain the 
MSSM correction. 

In the present study\footnote{A preliminary presentation of these results
was given in Ref.\cite{JSP}.}, we will closely follow
the systematic pathway
adopted in our previous treatment of the supersymmetric
quantum corrections to
the canonical decay $t\rightarrow W^+\,b$\,\cite{GJSH,DHJJS}. However, 
because of the Higgs particle in the final state,  
we have to incorporate the details of 
the renormalization of the Higgs sector of the MSSM, which substantially alter
the analytical
counterterm structure of the $t\,b\,H^{\pm}$-vertex as compared to the
conventional $t\,b\,W^{\pm}$-vertex. This also has dramatical consequences 
on the quantitative side, as we shall see.
Moreover, we have to include the quantum
corrections from the supersymmetric Higgs bosons
themselves. In spite of their
negligible effect on the width of the standard decay
$t\rightarrow W^+\,b$\,\cite{GH} -- a fact which was not obvious a priori
since there are potentially large non-oblique Higgs interactions originating
from the Yukawa couplings (\ref{eq:Yukawas}) -- 
their impact on the alternative 
decay $t\rightarrow H^+\,b$ must also be  
carefully evaluated. In certain circumstances, they
can be comparable to the virtual effects from the genuine
supersymmetric particles (sparticles). 
 
It is worth mentioning that the one-loop analysis of the unconventional top 
quark decay  $t\rightarrow H^+\,b$ is motivated not only by its obvious
interest on its
own as a laboratory to test the spontaneous symmetry breaking mechanism
beyond the SM, but also as a way to characterize the SUSY nature of the
charged Higgs emerging from that decay.
In fact, we shall see that important ($\sim 50\%$) SUSY radiative
corrections can be
obtained in certain regions of the MSSM parameter space. 
In one of these regions,
$\tan\beta \sim m_t/m_b$ is large enough such that $\lambda_b\simeq\lambda_t$.
Incidentally, we note that these are typical values of $\tan\beta$ 
characterizing one of the possible regions of 
the MSSM parameter space highlighted
in the literature\,\cite{GJS2}
(other regions have also been exploited \,\cite{Rb2}) 
in order to alleviate some formerly claimed
``anomalies'' observed in the hadronic decay ratios $R_b$ and $R_c$ 
of the $Z$-boson into $b\,\bar{b}$ and $c\,\bar{c}$,
and in general to try to improve the global
fit analyses of the MSSM to electroweak precision data.
However, most of the fuss about the $Z$-boson ``anomalies''
seems to have declined;                                                                                                                                                                                            
at present, $R_c$ is fully understood within the SM, and there only
remains a marginal $1.8\,\sigma$ discrepancy in $R_b$ with respect to the
SM prediction\,\cite{Warsaw}. It should, nevertheless, be clear
that the situation with the MSSM is not bad at all. As mentioned above, both
the SM and the MSSM can comfortably accommodate the present high precision
data with basically the same level of
statistical significance\,\cite{WdeBoer}.

Thus the MSSM is at present in very good shape and we are
perfectly entitled to
deal with the decay $t\rightarrow H^+\,b$ from the SUSY point of view.
In particular, we wish to quest for 
generic regions of the MSSM parameter space where
the decay rate of $t\rightarrow H^+\,b$ is competitive with the
SM decay  $t\rightarrow W^+\,b$ and 
at the same time to look for regions where it is
as much sensitive as possible to 
virtual supersymmetric effects.
For definiteness, in this paper we take the point of view that the
study of the decay $t\rightarrow H^+\,b$ is worthwhile provided that
its branching ratio is operative at a level
$BR(t\rightarrow H^+\,b)> 10\%$,
a condition which is not excluded by present Tevatron data (see Section 2).
From the theoretical point of view, this is fully guaranteed provided that
$\tan\beta$ is large enough ($\stackM 30$).
In these conditions, the SUSY-QCD one-loop corrections can typically
be in the $50\%$ range and the
electroweak effects  induced by the
supersymmetric Yukawa couplings
$\lambda_t$ and $\lambda_b$ can be rather large
(typically $20\%$) and so all of them are liable
to being measured in future experiments at the Tevatron and/or at the LHC.
We shall see that there are scenarios where these electroweak supersymmetric
corrections could basically constitute the total quantum effect from the MSSM,
i.e. the net effect after including the complete (standard plus supersymmetric)
QCD corrections.
  
The paper is organized as follows. In Section 2 the 
lowest order relations concerning the Higgs sector and the top quark decay 
are given. We also discuss the status of the charged Higgs decay of the
top quark in the light of the recent data from Tevatron, and the
prospects for its detection. Section 3 discusses the renormalization of 
the $t\,b\,H^{\pm}$-vertex in the on-shell scheme with a physically
well motivated definition of $\tan\beta$. 
In Section 4 we present the full analytical 
formulae for the one-loop corrected partial width
$\Gamma(t\rightarrow H^+\,b)$ in 
the MSSM. The numerical analysis and discussion, as well as the conclusions,
are delivered in Section 5, where we also comment on the feasibility
of measuring the computed quantum effects in hadron colliders.
Finally, we devote three appendices, respectively on
SUSY Lagrangians, renormalization details
and one-loop functions,
in order to make the paper as self-contained as possible.


\vspace{0.75cm} 
\begin{Large}
{\bf 2. Lowest order relations and
determination of\\
\indent
 $BR (t\rightarrow H^+\,b)$ from experiment}
\end{Large}
\vspace{0.5cm}

In this paper we wish to emphasize the possibility that a 
charged pseudoscalar, $H^{\pm}$, involved in a possible unconventional 
decay of the
top-quark, $t\rightarrow H^+\,b$, be the charged Higgs
of the MSSM\,\footnote{In the MSSM
there are several additional, more exotic, $2$-body decays
of the top quark and also a host of 
$3$-body final states worth studying\,\cite{GuaschS}.}.
A charged Higgs is necessary in the MSSM since
Supersymmetry requires the existence of at least two Higgs
$SU(2)_L$-doublets with opposite weak-hypercharges to give masses to 
matter and gauge fields (Cf. Appendix A): 
\beq
H_1=\left(\begin{array}{c}
H_1^0 \\ H_1^{-}
\end{array} \right)
\ \ \ (Y=-1)\,,\ \ \ \ \ 
H_2=\left(\begin{array}{c}
H_2^{+} \\ H_2^0 
\end{array} \right)\ \ \ (Y=+1)\,. 
\label{eq:H1H2} 
\eeq
Because of the SUSY constraints, the structure of the Higgs potential
of the MSSM constructed out of the two doublets (\ref{eq:H1H2}) takes
on the form\,\cite{Hunter}:
\beqn
V &=& m_1^2\,|H_1|^2+m_2^2\,|H_2|^2-m_{12}^2\,\left(
\epsilon_{ij}\,H_1^i\,H_2^j+{\rm h.c.}\nonumber\right)\\
&+&{1\over 8}(g^2+g'^2)\,\left(|H_1|^2-|H_2|^2\right)^2
+{1\over 2}\,g^2\,|H_1^{\dagger}\,H_2|^2\,,
\label{eq:potential}
\eeqn
where $m_1^2,m_2^2,m_{12}^2$ are soft SUSY-breaking masses and 
$g,g'$ are the $SU(2)_L\times U(1)_Y$ gauge coupling constants. 
After SSB, the physical content of the Higgs sector of the MSSM consists of
one CP-odd (``pseudoscalar'') neutral Higgs, $A^0$, two CP-even neutral
Higgs bosons, $h^0, H^0$, and a charged Higgs boson, $H^{\pm}$. 
Upon due account of the physical gauge sector, the masses of
the various spinless bosons are determined
in terms of just three parameters, which can be chosen to be
the two vacuum expectation values (VEV's) $<H_2^0>=v_2$, 
$<H_1^0>=v_1$, giving masses to the top and bottom quarks respectively,
and one physical
Higgs mass. However, due to the SSB constraint
\beq 
v^2\equiv v_1^2+v_2^2=2 M_W^2/g^2=2^{-3/2} G_F^{-1}\simeq (174\,GeV)^2\,,
\label{eq:vv1v2}
\eeq 
where $G_F$ is Fermi's constant in $\mu$-decay, in the end only two parameters 
suffice to completely specify the MSSM Higgs masses at the tree-level. 
Moreover, since we are interested in the 
decay process $t\rightarrow H^+\,b$, it is
natural to take $M_{H^{\pm}}$ as the physical input mass rather than 
$M_{A^0}$, as frequently done in other contexts. 
As the second independent parameter, one can take
the ratio of the two VEV's: $\tan\beta=v_2/v_1$. Then, in lowest order,
we have the relations\,\cite{Hunter}
\beqn
M_{A^0}^2 &=&M_{H^{\pm}}^2-M_W^2\,,\nonumber\\
M_{H^0,h^0 }^2&=&\frac{1}{2}\left(M_{A^0}^2+M_Z^2\pm
\sqrt{(M_{A^0}^2+M_Z^2)^2-4M_Z^2 M_{A^0}^2\cos^2{2\beta}}\right)\,,
\label{eq:Higgsmasses}
\eeqn 
where $M_{h^0}< M_{H^0}$.
It is well-known that these formulas become modified at 
one-loop\,\cite{Higgsloop}.  
In our case, once $M_{H^{\pm}}$ is fixed, 
the other Higgs masses enter the decay rate of $t\rightarrow H^+\,b$ only
through virtual corrections. Moreover, 
the renormalization of the masses also induces a renormalization of the
CP-even mixing angle\,\cite{Higgsloop}.
Although these one-loop effects are necessary
to guarantee a neutral Higgs spectrum fully compatible
with the phenomenological bounds
near the $\tan\beta\simeq 1$ region,
one expects that in practice a one-loop shift in the masses and couplings
just entails a small (two-loop order) 
correction to the decay rate.
We have explicitly checked this fact (Section 5).
As for the supersymmetric charged Higgs, the present
LEP 1.5 bound $M_{A^0}\stackM 60\,GeV$\,\cite{Moriond} on the CP-odd state
translates into
a lower limit $M_{H^{\pm}}\stackM 100\,GeV$, which is not significantly
modified by the radiative corrections (when $M_{A^0}$ is taken as
an input)\,\cite{Higgsloop}.

The charged Higgs can be, as noted
above, very sensitive to bottom-quark interactions. Specifically,
after expressing the two-doublet Higgs fields of the MSSM in terms
of the corresponding mass-eigenstates, 
the interaction Lagrangian describing the $t\,b\,H^{\pm}$-vertex  
reads as follows\,\cite{Hunter}: 
\beq
{\cal L}_{Hbt}={g\,V_{tb}\over\sqrt{2}M_W}\,H^-\,\bar{b}\,
[m_t\cot\beta\,P_R + m_b\tan\beta\,P_L]\,t+{\rm h.c.}\,,
\label{eq:LtbH}
\eeq 
where $V_{tb}$ is the corresponding
Cabibbo-Kobayashi-Maskawa matrix element, and
$P_{L,R}=(1/2)(1\mp\gamma_5)$ are the projection operators on LH and RH
fermions. 
On the phenomenological side, one should not dismiss the possibility that 
the bottom-quark Yukawa coupling could play a
momentous role in the physics of the top quark, to the extend of
drastically changing standard expectations on top-quark observables,
particularly on the top-quark width. Of course, this is possible because of the
potential $\tan\beta$-enhancement of that Yukawa coupling.

In the ``$\alpha$-parametrization'', where the input
parameters are $(\alpha, M_W, M_Z, M_H, m_f,...)$, the coupling
$g$  on eq.(\ref{eq:LtbH})
stands for $e/s_W$, where $\alpha\equiv \alpha_{\rm e.m.}(q^2=0)=e^2/4\pi$ and
 $s_W^2\equiv 1-c^2_W\equiv 1-M_W^2/M_Z^2$.
An alternative framework (``$G_F$-parametrization'') based on the set of inputs  
$(G_F, M_W, M_Z, M_H, m_f,...)$ is also useful, 
especially at higher orders in perturbation theory (Cf. Section 3).
At the tree-level, the relation between the two parametrizations is trivial:
\beq
{G_F\over\sqrt{2}}={\pi\alpha\over 2 M_W^2 s_W^2}\,.
\label{eq:schemes0}
\eeq
From the Lagrangian (\ref{eq:LtbH}), the tree-level width of
the unconventional top quark decay into a charged Higgs boson 
in the $G_F$-parametrization reads:  
\beqn
\Gamma^{(0)}(t\rightarrow H^+\,b)
 &=&\left({G_F\over 8\pi\sqrt{2}}\right){|V_{tb}|^2\over m_t}
\ \lambda^{1/2}
(1, {m_b^2\over m_t^2},{\mHps\over m_t^2})\nonumber\\
& & \times[(m_t^2+m_b^2-\mHps) (m_t^2\cot^2\beta+m_b^2\tan^2\beta)+4m_t^2m_b^2]\,,
\label{eq:treeH}
\eeqn    
where
\begin{equation}
\lambda^{1/2} (1, x^2, y^2)\equiv\sqrt{[1-(x+y)^2][1-(x-y)^2]}\,.
\end{equation}
It is useful to compare eq.(\ref{eq:treeH}) with the tree-level 
width of the canonical top quark decay in the SM:
\beqn
\Gamma^{(0)}(t\rightarrow W^+\,b) & = &
\left({G_F\over 8\pi\sqrt{2}}\right){|V_{tb}|^2\over m_t} 
\ \lambda^{1/2} (1, {m_b^2\over m_t^2}, {M_W^2\over m_t^2})
\nonumber\\
& & \times [M_W^2(m_t^2+m_b^2)+ (m_t^2-m_b^2)^2-2M_W^4]\,.
\label{eq:treeW}
\eeqn
The ratio between the two partial widths becomes more transparent upon
neglecting the kinematical bottom mass
contributions, while retaining all the Yukawa coupling effects: 
\beq
{\Gamma^{(0)}(t\rightarrow H^+\,b)\over \Gamma^{(0)}(t\rightarrow W^+\,b)}=
{\left(1-{M_{H^{+}}^2\over m_t^2}\right)^2\,
\left[{m_b^2\over m_t^2}\,\tan^2\beta+\cot^2\beta\right]
\over
\left(1-{M_W^2\over m_t^2}\right)^2\,\left(1+2{M_W^2\over m_t^2}\right)}\,.
\label{eq:ratioHW} 
\eeq     
We see from it that if $M_{H^{\pm}}$ is not much heavier than $M_W$, then
there are two regimes, namely a low and a high $\tan\beta$ regime,
where the decay rate of the
unconventional top quark decay becomes sizeable 
as compared to the conventional decay. They can be 
defined approximately as follows:
i) Low $\tan\beta$ regime: $\tan\beta<2$, and 
ii) High $\tan\beta$ regime: $\tan\beta\geq m_t/m_b\simeq 35$.
The critical regime of the decay
$t\rightarrow H^+\,b$ occurs at the intermediate value 
$ \tan\beta= \sqrt{m_t/m_b}\sim 6$, where the partial width has a pronounced
dip. Around this value, the canonical decay $t\rightarrow W^+\,b$ is dominant
over the charged Higgs decay; more specifically, for
$3\stackm\tan\beta\stackm 15$
the decay rate of the mode $t\rightarrow H^+\,b$ is basically irrelevant as
compared to the standard mode: $BR(t\rightarrow H^+\,b)<10\%$.
Therefore, a detailed study of the 
quantum effects within that interval is of no practical interest.

Even though the approximate perturbative regime
for $\tan\beta$ extends over the wide range
\beq
0.5\stackm\tan\beta\stackm 70\,,
\label{eq:tanbeta}
\eeq
we shall emphasize the results obtained in
the phenomenologically interesting high $\tan\beta$
region (typically $\tan\beta\stackM 30$). As for the low $\tan\beta$ range,
while $BR (t\rightarrow H^+\,b)$ can also be sizeable it turns out that
the corresponding quantum
effects are generally much smaller than in the high $\tan\beta$ case
(Cf. Section 5). Still, we find that in the very low
$\tan\beta$ segment $0.5\stackm\tan\beta\stackm 1$ these effects
can be of some phenomenological interest and we shall also report on them.  
As mentioned in Section 1, we do have some theoretical motivation
to contend that at least
one of the two regimes i) or ii) may apply.
Therefore, it is justified to focus our attention on $t\rightarrow H^+\,b$,
not only as a possibility on its own, but also because
there may be realistic situations where it could have a non-negligible
branching ratio.

As a matter of fact, and despite naive expectations, the non-SM branching ratio
$BR (t\rightarrow H^+\,b)$ is not as severely constrained
as apparently dictated by the recent measurements of the standard branching ratio
at the Tevatron, namely,
$BR (t\rightarrow W^+\,b)\stackM 70\%$\,\cite{Tartarelli}.
To assess this fact, notice that the former result
strictly applies only 
under the assumption that the sole source of top quarks in $p\bar{p}$
collisions is the  standard Drell-Yan pair
production mechanism
 $q\,\bar{q}\rightarrow t\,\bar{t}$\,\cite{CPYuan}. Now, the observed
cross-section is equal to the Drell-Yan 
production cross-section
convoluted over the parton distributions times the
squared branching ratio. Schematically,
\beq
\sigma_{\rm obs.}=\int dq\,d\bar{q}\,\,\sigma (q\,\bar{q}\rightarrow t\,\bar{t})\,
\times |BR (t\rightarrow W^+\,b)|^2\,.
\label{eqn:productionSM}
\eeq
However, in the framework of the MSSM, we rather expect a generalization of this
formula in the following way:
\beqn
\sigma_{\rm obs.}&=&\int dq\,d\bar{q}\,\,\sigma (q\,\bar{q}\rightarrow t\,\bar{t})\,
\times |BR (t\rightarrow W^+\,b)|^2\nonumber\\
&+& \int dq\,d\bar{q}\,\,\sigma (q\,\bar{q}\rightarrow \tilde{g}\,\bar{\tilde{g}})\,
\times |BR (\tilde{g}\rightarrow t\,\bar{\tilde{t}}_1)|^2\times
|BR (t\rightarrow W^+\,b)|^2\nonumber\\
&+& \int dq\,d\bar{q}\,\,\sigma (q\,\bar{q}\rightarrow 
\tilde{b}_a\,\bar{\tilde{b}}_a)\,
\times |BR (\tilde{b}_a\rightarrow t\,\chi^-_1)|^2\times
|BR (t\rightarrow W^+\,b)|^2+...\,,
\label{eq:productionMSSM}
\eeqn
where $\tilde{g}$ stand for the gluinos,
$\tilde{t}_1$ for the lightest stop and $\tilde{b}_a (a=1,2)$ for the sbottom quarks.
One should also include electroweak and QCD radiative corrections to all these
production cross-sections within the MSSM. For some of these processes
calculations already exist in the literature showing that one-loop effects
can be important\,\cite{Kim,Unpub}.

It should be clear that the observed cross-section on eq.(\ref{eq:productionMSSM})
refers not only to the standard
$bW\,bW$ events, but to all kind of final states that can simulate them. 
Thus, effectively, we should substitute $BR (t\rightarrow W^+\,b)$ in that
formula by $BR (t\rightarrow X\,b)$, and then sum the cross-section
over $X$, where $X$ is any state that leads to an observed 
pattern of leptons and jets similar to those resulting from $W$-decay.
In particular, 
$X=H^{\pm}$ would contribute (see below) to the $\tau$-lepton signature,
if $\tan\beta$ is large enough. Similarly, there can be
direct top quark decays into SUSY
particles that could mimic the SM decay of the top quark\,\cite{GuaschS}.    
Notwithstanding, even in the absence of 
the $X$ contributions, eq.(\ref{eq:productionMSSM}) shows
that if there are alternative (non-SM) sources of top quarks subsequently decaying
into the SM final state, $W^+\,b$, one cannot rigorously
place any stringent upper bound on
$BR (t\rightarrow W^+\,b)$ from the present data.
The only restriction being an approximate lower
bound $BR (t\rightarrow W^+\,b)\stackM 40-50\%$ in order to guarantee the
purported standard top quark events at the Tevatron\,\cite{Tevatron}. 
Thus, from these considerations it is not excluded
that the non-SM branching ratio of the top quark, 
$BR (t\rightarrow $``new''$)$, could be
as big as the SM one, i.e. $\sim 50\%$.

Notice that at present one cannot exclude eq.(\ref{eq:productionMSSM})
since the observed form of the conventional $t\rightarrow W^+\,b$ final state
involves missing energy, as it is also the case for the decays
comprising supersymmetric particles.
A first step to improve
this situation would be to compute some of the 
additional top quark production cross-sections in the MSSM under given hypotheses
on the SUSY spectrum. Recently, the inclusion of the 
$q\,\bar{q}\rightarrow \tilde{g}\,\bar{\tilde{g}}$ mechanism followed by
the $\tilde{g}\rightarrow t\,\bar{\tilde{t}}_1$ decay has been considered
in Ref.\cite{KaneM}, where it is claimed that
$BR (t\rightarrow \tilde{t}_1\,\chi_1^0)\simeq 50\%$.
 By the same token, one cannot place any compelling
restriction on $BR (t\rightarrow H^+\,b)$ from the present FNAL data.
In particular, if $\tan\beta$ is large and there exists a relatively light
chargino with a non-negligible higgsino component, the third mechanism suggested on
eq.(\ref{eq:productionMSSM}), namely 
$q\,\bar{q}\rightarrow \tilde{b}_a\,\bar{\tilde{b}}_a$ followed by
$\tilde{b}_a\rightarrow t\,\chi^-_1$, could also be a rather efficient
non-SM source of top quarks. Moreover,
if $100\,GeV\stackm M_{H^{\pm}}\stackm 150\,GeV$, 
then a sizeable portion of the top quarks will decay into a charged Higgs.
Thus, if either
$m_t+m_{\tilde{t}_1}\stackm m_{\tilde{g}}\stackm 300\,GeV$
and/or $m_t+m_{\chi_i^-}\stackm m_{\tilde{b}_a}\stackm 300\,GeV$ , so that at
least one of the alternative SUSY sources of top quark final states 
contributing to eq.(\ref{eq:productionMSSM}) is available
(and $m_{\tilde{g}}, m_{\tilde{b}_a}$ are not too heavy so that the production
cross-section is not too phase-space suppressed), then 
one may equally argue that
a large branching ratio $BR (t\rightarrow H^+\,b)\simeq 50\%$ is not incompatible
with the present measurement of the top quark cross-sections\,\cite{Tevatron}. 
This could be most likely the case if the frequently 
advocated SUSY decay 
$t\rightarrow \chi_1^0\,\tilde{t}_1$ is kinematically forbidden. Nonetheless,
even if it is allowed, it is non-enhanced 
in our preferential large $\tan\beta$ region, in contrast
to $t\rightarrow H^+\,b$.

Furthermore, it is worth mentioning that the decay mode $t\rightarrow H^+\,b$
has a distinctive signature 
which could greatly help in its detection, viz. the fact that at large
$\tan\beta$ the emergent charged Higgs 
would seldom decay into a pair of quark jets, but rather into
a $\tau$-lepton and associated neutrino. This follows from inspecting the
ratio
\beqn  
{\Gamma(H^{+}\rightarrow\tau^{+}\nu_{\tau})\over
\Gamma(H^{+}\rightarrow c\bar{s})}&=&\frac{1}{3}
\left(\frac{m_{\tau}}{m_c}\right)^2
{\tan^2\beta\over
(m^2_s/ m^2_c)\tan^2\beta+\cot^2\beta}\nonumber\\
&\rightarrow & \frac{1}{3}\left(\frac{m_{\tau}}{m_s}\right)^2
>10\ \ \ \ \ \ ({\rm for}\ \tan\beta>\sqrt{m_c/m_s}\stackM 2)\,,
\label{ratiotaucs}
\eeqn
where we see that the identification of the charged Higgs decay of the top quark
could be a matter of measuring a departure from the universality prediction
for all lepton channels. In practice, $\tau$-identification is possible at the
Tevatron\,\cite{tauTev,Matorras}; and the feasibility of tagging the excess of events
with one isolated $\tau$-lepton as compared to events with an additional lepton
has also been substantiated by studies of the LHC collaborations \,\cite{Atlas}.
The experimental signature for $t\bar{t}\rightarrow H^+\,H^-\,b\,\bar{b}$
would differ from $t\bar{t}\rightarrow W^+\,W^-\,b\,\bar{b}$ by
an excess of final states with two $\tau$-leptons and two b-quarks and large missing
transverse energy.

A preliminary study in this direction by the CDF 
collaboration at the Tevatron\,\cite{Conway}
has been able to exclude a large portion of the
$(\tan\beta, M_{H^{\pm}})$-plane characterized by $\tan\beta\stackM 60$
and $M_{H^{\pm}}$ below a given value which varies with $\tan\beta$.
For extremely high 
$\tan\beta\stackM {\cal O}(100)$, the uppermost excluded mass region is
$M_{H^{\pm}}\stackm 140\,GeV$.  
However, within the interval $\tan\beta=60-80$, the allowed upper
limit on $M_{H^{\pm}}$ varies
very fast with $\tan\beta$.
In particular, the MSSM permissible values $M_{H^{\pm}}\stackM 100\,GeV$ 
(compatible with $M_{A^0}\stackM 60\,GeV$) are not manifestly excluded for
$\tan\beta$ equal or below the perturbative bound $\tan\beta=70$,
eq.(\ref{eq:tanbeta}).
We shall nevertheless err on the conservative side and assume 
that $\tan\beta\leq 60$ throughout our analysis. Thus,
as far as the high $\tan\beta$ regime is concerned, we will for definiteness 
optimize our results in the safe, and phenomenologically interesting, 
high $\tan\beta$ segment
\beq
30\leq\tan\beta\leq 60\,.
\label{eq:tanbeta2}
\label{eq:htanr}
\eeq
To round off the $\tau$-lepton business,
it has recently been shown that it should be fairly easy to  
discriminate between the $W$-daughter $\tau$'s and the 
$H^{\pm}$-daughter $\tau$'s by just taking advantage 
of the opposite states of $\tau$ polarization resulting from the $W^{\pm}$
and $H^{\pm}$ decays; the two polarization states can be 
distinguished by
measuring the charged and neutral contributions to the $1$-prong $\tau$-jet
energy (even without identifying the individual meson states)
\,\cite{GuraRoy,Raychau}. 

In short, there are good prospects for
detecting the decay $t\rightarrow H^+\,b$, if it is kinematically accessible.
Unfortunately, on the sole basis of 
computing tree-level effects we cannot find out whether the charged Higgs
emerging from that decay is supersymmetric or not. Quantum effects, however, can.



\vspace{1.25cm} 
\begin{Large}
{\bf 3. Renormalization of the $t\,b\,H^{+}$-vertex}
\end{Large}
\vspace{0.5cm}

Proceeding closely in parallel with our supersymmetric approach 
to the conventional decay $t\rightarrow W^+\,b$\,\cite{GJSH,DHJJS}, 
we shall address
the calculation of the one-loop corrections to the partial width
of $t\rightarrow H^+\,b$ in the MSSM within the context of the on-shell
renormalization framework\,\cite{RossTaylor}\footnote{For a 
comprehensive exposition, see e.g. Refs.\cite{Yellow}-\cite{Jegerlehner}.}. 
Again we may use both the $\alpha$ or the $G_F$ parametrizations.
At one-loop order, we shall call
the former the ``$\alpha$-scheme'' and the latter the ``$G_F$-scheme''.
In the ``$\alpha$-scheme'', the
structure constant $\alpha\equiv \alpha_{\rm em}(q^2=0)$ 
and the masses of the gauge bosons, fermions and scalars are
the renormalized parameters: $(\alpha, M_W, M_Z, M_H, m_f, M_{SUSY},...)$ 
--$M_{SUSY}$ standing for the collection of renormalized sparticle masses. 
Similarly, the ``$G_F$-scheme'' is characterized by the set of inputs
$(G_F, M_W, M_Z, M_H, m_f, M_{SUSY},...)$. Beyond lowest order,
the relation between the two on-shell schemes is no longer given by 
eq.(\ref{eq:schemes0}) but by
\beq
{G_F\over\sqrt{2}}={\pi\alpha\over 2 M_W^2 s_W^2}
(1+\Delta r^{MSSM})\,,
\label{eq:DeltaMW}
\eeq 
where $\Delta r^{MSSM}$ is the prediction of the parameter
$\Delta r$\,\cite{Yellow} in the MSSM\footnote{A dedicated study of
$\Delta r^{MSSM}$ has been presented in Ref\,\cite{GS}.}.
 
Let us sketch the renormalization procedure affecting the parameters and
fields related to the $t\,b\,H^{\pm}$-vertex, whose interaction Lagrangian
was given on eq.(\ref{eq:LtbH}). 
In general, the renormalized  MSSM Lagrangian
${\cal L}\rightarrow {\cal L}+\delta{\cal L}$
is obtained following a similar pattern as in the SM, i.e.
by attaching multiplicative renormalization constants to each free
parameter and field: $g_i\rightarrow (1+\delta g_i/g_i)g_i$,
$\Phi_i\rightarrow Z^{1/2}_{\Phi_i}\Phi_i$. As a matter of fact, field 
renormalization (and so Green's functions renormalization) 
is unessential and can be either omitted or be carried out
in many different ways without altering physical ($S$-matrix) amplitudes.
In our case, in the line of Refs.\cite{GJSH,DHJJS} we shall
use minimal field renormalization, i.e. one renormalization constant per
gauge symmetry multiplet\,\cite{BSH}. In this way  the counterterm 
Lagrangian, $\delta{\cal L}$, as well as the various Green's functions
are automatically gauge-invariant.
Specifically, for the quark
fields under consideration, we have
\beqn
& &\left(\begin{array}{c}
t_L \\ b_L
\end{array} \right)  
\rightarrow  Z_L^{1/2}\,
\left(\begin{array}{c}
t_L \\ b_L
\end{array} \right)
\rightarrow \left(\begin{array}{c}
{(Z_L^t)}^{1/2}t_L \\ {(Z_L^b)}^{1/2}b_L
\end{array} \right)\,,   
\nonumber\\
& & b_R \rightarrow {(Z_R^b)}^{1/2}b_R\,, \ \ \
t_R \rightarrow{(Z_R^t)}^{1/2}t_R\,.
\label{eq:FR}
\eeqn
Here $Z_i=1+\delta Z_i$ are the doublet ($Z_L$) and singlet ($Z_R^{t,b}$)
field renormalization constants for the top and bottom quarks.
Although in the minimal field renormalization scheme there is only one
fundamental constant, $Z_L$, per matter doublet, it is useful
to work with $Z_L^b=Z_L$ and $Z_L^t$, where the latter differs from the
former by a {\it finite}
renormalization effect\,\cite{BSH}.
To fix all these constants one starts from the usual on-shell 
mass renormalization condition for fermions, $f$, together with the
``${\rm residue}=1$'' condition on the renormalized propagator. These are
completely standard procedures, and 
in this way one obtains \footnote{We follow the
notation of Ref.\cite{GJSH}, which is close enough to that of Ref.\cite{BSH}, 
but differs from it in several respects, in particular
in the sign conventions for the self-energy functions. Moreover, we understand
that in all formulas defining counterterms we are taking the real part of the
corresponding functions.} 
\beq
{\delta m_f\over m_f}=-\left[{\Sigma^f_L(m_f^2)+\Sigma^f_R(m_f^2)\over 2}
+\Sigma^f_S(m_f^2)\right]\,,
\label{eq:deltamf}
\eeq 
and
\beqn
\delta Z_{L,R}^f &=& \Sigma^f_{L,R}(m_f^2)+m_f^2[\Sigma^{f\,\prime}_L
(m_f^2)+\Sigma^{f\,\prime}_R(m_f^2)
+2\Sigma^{f\,\prime}_S(m_f^2)]\,,
\label{eq:DSRC}
\eeqn
where we have 
decomposed the fermion self-energy according to
\begin{equation}
\Sigma^f(p)=\Sigma^f_L(p^2)\not{p}\,P_L+\Sigma^f_R(p^2)\not{p}\,P_R
+m_f\,\Sigma^f_S(p^2)\,,
\label{eq:Sigma}
\end{equation}
and used the notation $\Sigma'(p)\equiv \partial\Sigma(p)/\partial p^2$.
 
One also assigns doublet renormalization constants to the two Higgs doublets
(\ref{eq:H1H2}) of the MSSM:
\beq
\left(\begin{array}{c}
H_1^0 \\ H_1^{-}
\end{array} \right)\rightarrow Z_{H_1}^{1/2}
\left(\begin{array}{c}
H_1^0 \\ H_1^{-}
\end{array} \right)\,,\ \ \ \ \ 
\left(\begin{array}{c}
H_2^{+} \\ H_2^0 
\end{array} \right)\rightarrow Z_{H_2}^{1/2}
\left(\begin{array}{c}
H_2^{+} \\ H_2^0 
\end{array} \right)\,.
\label{eq:ZH1H2}  
\eeq
The renormalization of the gauge sector is related to that
of the Higgs sector. In particular, 
we point out the presence in our decay process
$t\rightarrow H^+\,b$ of the (one-loop induced) mixing term 
$H^{\pm}-W^{\pm}$ for the bare fields (Cf. Appendix B), which
must be renormalized away for the physical fields $H^{\pm}$ and $W^{\pm}$.
In order to generate the corresponding Lagrangian counterterm  
we write
\beq
W^{\pm}_{\mu}\rightarrow {(Z_2^W)}^{1/2} W^{\pm}_{\mu}\pm
i\,{\delta Z_{HW}\over M_W}\,\partial_{\mu} H^{\pm}\,.
\label{eq:ZHW}
\eeq
Therefore, from
\beq 
{\cal L}_{Wbt} =  
{g\over \sqrt{2}}\,W^{-}_{\mu}\, \bar{b}\,\gamma^{\mu}\,P_L\,t
 +{\rm h.c.}
\eeq
we obtain
\beqn 
\delta{\cal L}_{HW} &= & 
 -i\,\delta Z_{HW}{g\over \sqrt{2} M_W}\,\partial_{\mu} H^{-}\,
 \bar{b}\,\gamma^{\mu}\,P_L\,t +{\rm h.c.}\nonumber\\
& \rightarrow & \,\delta Z_{HW}{g\over \sqrt{2} M_W}\,H^{-}\,
\left[m_t\,\bar{b}\,\,P_R\,t
-m_b\,\bar{b}\,\,P_L\,t\right]+{\rm h.c.}\,,
\label{eq:ZHW2}
\eeqn
and in this way it adopts the form of the original vertex (\ref{eq:LtbH}).
In the above expression (\ref{eq:ZHW}), $Z_2^W=1+\delta Z_2^W$ is the usual
$SU(2)_L$ gauge triplet renormalization
constant given by the formula
\beq
\delta Z_2^W  = 
\left.  {\Sigma_{\gamma}(k^2)\over k^2}\right|_{k^2=0}-2{c_W
\over s_W}\,{\Sigma_{\gamma Z}(0)\over M_Z^2}
+{c_W^2\over s_W^2}
\left({\delta M_Z^2\over M_Z^2}-{\delta M_W^2\over M_W^2}\right)\,,
\label{eq:gtriplet}
\eeq
and
\beq
\delta M_W^2=-\Sigma_W(k^2=M_W^2)\,\,,\ \ \ \ \
\delta M_Z^2=-\Sigma_Z(k^2=M_Z^2)\,,
\label{eq:MCT}
\eeq
are the gauge boson mass counterterms
enforced by the usual on-shell mass renormalization conditions.
Furthermore, $\delta Z_{HW}$ on eqs.(\ref{eq:ZHW})-(\ref{eq:ZHW2})
is a dimensionless constant associated to the
wave-function renormalization mixing among the bare $H^{\pm}$
and $W^{\pm}$ fields.
Its relation with the doublet renormalization constants, $Z_{H_i}=1+\delta Z_{H_i}$,
is the following (see Appendix B):
\beq
\delta Z_{HW}=\sin\beta\cos\beta\,
\left[{1\over2}\,(\delta Z_{H_2}-\delta Z_{H_1})+
{\delta\tan\beta\over\tan\beta}\right]\,,
\eeq
where $\delta\tan\beta$ is a counterterm associated to the renormalization of 
$\tan\beta$ (see below).

In practice, the most straightforward way to compute
$\delta Z_{HW}$ is from
the unrenormalized mixed self-energy $\Sigma_{HW}(k^2)$
in the unitary gauge, where it takes on the simplest form:
\beq
\delta Z_{HW}={\Sigma_{HW}(M_{H^{\pm}}^2)\over M_W^2}\,.
\eeq
However, since we shall perform the rest of the calculation in the
Feynman gauge\,\cite{BSH},
it is worth considering the computation of $\delta Z_{HW}$ in
that gauge (see Appendix B), where the discussion
is slightly more complicated due to the presence of Goldstone bosons
($G^{\pm}$) leading to additional ($H^{\pm}-G^{\pm}$) mixing terms
among the bare fields. 
The corresponding expression for $\delta Z_{HW}$ is, however,
formally identical in both gauges. 

For the $SU(2)_L$ gauge coupling constant, we have 
\beq
g \rightarrow  (1+\frac{\delta g}{g}) g={(Z_1^W)}\,{(Z_2^W)}^{-3/2}\, g\,, 
\eeq
where $Z_1^W$ refers to the renormalization constant associated to the
triple vector boson vertex. Therefore, from charge renormalization,
\beq
\frac{\delta\alpha}{\alpha} =    
-\left.  {\Sigma_{\gamma}(k^2)\over k^2}\right|_{k^2=0}-2{s_W
\over c_W}\,{\Sigma_{\gamma Z}(0)\over M_Z^2}\,,
\eeq
and the bare relation $\alpha=g^2\,s^2_W/4\pi\rightarrow
\alpha+\delta\alpha=(g^2+\delta g^2)\,(s^2_W+\delta s^2_W)/4\pi$,  
one gets for the counterterm to $g$:  
\beq
\frac{\delta g^2}{g^2} =\frac{\delta\alpha}{\alpha}-
{c_W^2\over s_W^2}
\left({\delta M_Z^2\over M_Z^2}-{\delta M_W^2\over M_W^2}\right)\,,
\label{eq:dg}
\eeq 
and as a by-product
\beq
\delta Z_1^W =  \frac{1}{2}\frac{\delta g^2}{g^2}
+\frac{3}{2} \delta Z_2^W\,.
\eeq
Let us now outline the renormalization of the Higgs sector of 
the MSSM\cite{Higgsloop,CPRD}.
Depending on the particular problem at hand, the renormalization procedure 
may adopt the  CP-odd state $A^0$ as the basic field
on which to set the mass and wave-function renormalization conditions. 
In the present work, however, since the external
Higgs particle is charged, we rather take $H^{\pm}$ as the basic field.
Its mass and field renormalization constants are defined by
\beq
M_{H^{\pm}}^2\rightarrow M_{H^{\pm}}^2+\delta M_{H^{\pm}}^2\,,\ \ \ \ \ 
 H^{\pm}\rightarrow Z_{H^{\pm}}^{1/2} H^{\pm}\,.
\eeq 
The charged Higgs field renormalization constant,   
$Z_{H^{\pm}}=1+\delta Z_{H^{\pm}}$, is of course related to the
fundamental doublet renormalization
constants introduced on eq.(\ref{eq:ZH1H2}),
as follows (Cf. Appendix B):
\beq
\delta Z_{H^{\pm}}=\sin^2\beta\, \delta Z_{H_1}+\cos^2\beta\, \delta Z_{H_2}\,.
\eeq 
The structure of the renormalized self-energy is
\beq
\hat{\Sigma}_{H^{\pm}}(k^2)= \Sigma_{H^{\pm}}(k^2)+\delta M_{H^{\pm}}^2
-(k^2-M_{H^{\pm}}^2)\,\delta Z_{H^{\pm}}\,,
\eeq
where $\Sigma_{H^{\pm}}(k^2)$ is the corresponding unrenormalized self-energy. 

In order to determine the counterterms,
we impose the following renormalization conditions: 

i) On-shell mass renormalization condition:
\beq
\hat{\Sigma}_{H^{\pm}}(M_{H^{\pm}}^2)=0\,,
\eeq                                                                                                                                            

ii) ``Residue $=1$'' condition for the renormalized propagator
at the pole mass:
\beq
\left. {\partial\hat{\Sigma}_{H^{\pm}}( k^2)
\over \partial k^2}\right|_{k^2=M_{H^{\pm}}^2}
\equiv\, \hat{\Sigma}_{H^{\pm}}^{\prime}(M_{H^{\pm}}^2) =0\,.
\eeq                                                                                                                                            
From these conditions one derives
\beqn  
\delta M_{H^{\pm}}^2 &=& -\Sigma_{H^{\pm}}(M_{H^{\pm}}^2)\,,\nonumber\\
\delta Z_{H^{\pm}} &=& +\Sigma_{H^{\pm}}^{\prime}(M_{H^{\pm}}^2)\,.
\eeqn
Although not needed in our calculation, it is clear that with these
settings the neutral Higgs fields will undergo an additional finite wave
function renormalization.       
                                                                                                                                        
Consider next the renormalization of the Higgs potential in 
the MSSM, eq.(\ref{eq:potential})\,\cite{Higgsloop}.
After expanding the neutral components $H_1^0$ and $H_2^0$ 
around their VEV's $v_1$ and
$v_2$, the one-point functions of the resulting CP-even fields 
are required to vanish, i.e. the tadpole counterterms are constrained to exactly 
cancel the tadpole diagrams, so that the renormalized tadpoles are zero and the
quantities $v_{1,2}$ remain as the VEV's of the renormalized Higgs potential.
Notwithstanding, at this stage a prescription to renormalize
$\tan\beta=v_2/v_1$, 
\beq
\tan\beta\rightarrow\tan\beta+\delta\tan\beta\,,
\eeq
is still called for. There are many possible strategies. The ambiguity is 
related to the fact that this parameter is just a Lagrangian parameter 
and as such it is not a physical observable.
Its value beyond the tree-level is
renormalization scheme dependent. (The situation is similar to the definition
of the weak mixing angle $\theta_W$, or equivalently of
$\sin^2\theta_W$.)
However, even within a given scheme, e.g. the
on-shell renormalization scheme, there are some ambiguities that must be
fixed. 
For example, we may wish to define $\tan\beta$ in a 
process-independent (``universal'') way
as the ratio $v_2/v_1$ between the true VEV's after renormalization of the
Higgs potential\,\cite{Higgsloop,CPRD}. 
In this case a consistent choice (i.e. a choice capable of renormalizing
away the tadpole contributions) is to simultaneously 
shift the VEV's and the mass parameters
of the Higgs potential, eq.(\ref{eq:potential}), 
\beqn 
v_i &\rightarrow & Z_{H_i}^{1/2} (v_i+\delta v_i)\,,\nonumber\\
m_i^2 &\rightarrow & Z_{H_i}^{1\over 2}\,(m_i^2+\delta m_i^2)\,,\nonumber\\
m_{12}^2 &\rightarrow & Z_{H_1}^{1\over 2}\,Z_{H_2}^{1\over 2}
\,(m_{12}^2+\delta m_{12}^2)\,,
\eeqn
($i=1,2$) in such a way that
$\delta v_1/v_1=\delta v_2/v_2$.  This choice generates the following
counterterm for $\tan\beta$ in that scheme (Cf. Appendix B):
\beq
{\delta\tan\beta\over\tan\beta}=
{1\over 2}\,\left(\delta Z_{H_2}-\delta Z_{H_1}\right)\,.
\label{eq:ddtan}
\eeq 
Nevertheless, this procedure looks very formal and
one may eventually like to 
fix the on-shell renormalization condition on $\tan\beta$ in a more physical
way, i.e. by relating it to some concrete physical observable, so that it is
the measured value of this observable that is taken as an input rather 
than the VEV's of the Higgs potential. 
Following this practical attitude, we choose as a physical observable 
the decay width of the charged Higgs boson into $\tau$-lepton and associated
neutrino: $H^{+}\rightarrow\tau^{+}\nu_{\tau}$. 
As it has been argued in Section 2, this should be a good choice,
because: i) When $t\rightarrow H^+\,b$ is allowed, the decay
$H^{+}\rightarrow\tau^{+}\nu_{\tau}$
is the dominant decay of $H^{\pm}$ already for $\tan\beta\stackM 2$;
ii) From the experimental point of view
there is a well-defined method to separate the
final state $\tau$'s originating from $H^+$-decay from those coming out of
the conventional decay $W^{+}\rightarrow\tau^{+}\nu_{\tau}$, so that 
$H^{+}\rightarrow\tau^{+}\nu_{\tau}$ should be physically accessible;
and iii) At high $\tan\beta$, the charged Higgs decay of the top quark  
can have a sizeable branching ratio.  

The interaction Lagrangian describing the decay 
$H^{+}\rightarrow\tau^{+}\nu_{\tau}$ is 
directly proportional to $\tan\beta$,  
\beq
{\cal L}_{H\tau\nu}={g\, m_{\tau}\tan\beta\over\sqrt{2}M_W}
\,H^-\,\bar{\tau}\,P_L\,\nu_{\tau}+{\rm h.c.}\,,
\label{eq:LtaunuH}
\eeq
and the relevant decay width is proportional to
$\tan^2\beta$. Whether in the $\alpha$-scheme or in the $G_F$-scheme, it
reads:
\beq
\Gamma(H^{+}\rightarrow\tau^{+}\nu_{\tau})=
{\alpha m_{\tau^{+}}^2\,M_{H^{+}}\over 8 M_W^2 s_W^2}\,\tan^2\beta= 
{G_F m_{\tau^{+}}^2\,M_{H^{+}}\over 4\pi\sqrt{2}}\,\tan^2\beta\, 
(1-\Delta r^{MSSM})\,,
\label{eq:tbetainput}
\eeq
where we have used the relation (\ref{eq:DeltaMW}). 
By measuring this decay width 
one obtains a physical definition of $\tan\beta$ 
which can be used beyond the tree-level.
A combined measurement
of $M_{H^{\pm}}$ and $\tan\beta$ from charged Higgs decaying
into $\tau$-lepton
in a hadron collider has been described
in the literature\,\cite{Godbole,Atlas}
by comparing the size of the various signals for charged
Higgs boson production, such as the multijet channels accompanied by a
$\tau$-jet or a large missing $p_T$, and the two-$\tau$-jet
channel.
At the upgraded Tevatron, the
conventional mechanisms $gg(q\bar{q})\rightarrow t\bar{t}$ followed by
$t\rightarrow H^+\,b$ have been studied and
compared with the usual $t\rightarrow W^+\,b$,
and the result is that for $M_{H^{\pm}}\simeq 100\,GeV$ the charged Higgs
production is at least as large as the $W^{\pm}$ production, apart from a gap
around 
$\tan\beta\simeq 6$\,\cite{Godbole}.

Insofar as the determination of the counterterm $\delta\tan\beta$
in our scheme, it can be
fixed unambiguously from 
our Lagrangian definition of $\tan\beta$ on eq.(\ref{eq:LtaunuH}) and 
the renormalization procedure described above. 
It is straightforward to find:
\beq
{\delta\tan\beta\over \tan\beta}
={\delta v\over v}-\frac{1}{2}\delta Z_{H^\pm}
+\cot\beta\, \delta Z_{HW}+ 
\Delta_{\tau}\,.
\label{eq:deltabeta}
\eeq    
Notice the appearance of the vacuum counterterm
\beq
{\delta v\over v}=\frac{1}{2}{\delta v^2\over v^2}=\frac{1}{2}
\frac{\delta M_W^2}{M_W^2}-\frac{1}{2}\frac{\delta g^2}{g^2}\,,
\label{eq:dv2}
\eeq
which is associated to $v^2=v^2_1+v^2_2$, and whose structure is fixed from
eq.(\ref{eq:vv1v2}). 
The last term on eq.(\ref{eq:deltabeta}), 
\beq
\Delta_{\tau}=-{\delta m_{\tau}\over m_{\tau}}
-\frac{1}{2}\delta Z_L^{\nu_{\tau}}-\frac{1}{2}\delta Z_R^{\tau}-F_{\tau}\,,
\label{eq:deltatau}
\eeq    
is the (finite) process-dependent part of the counterterm. Here
$\delta m_{\tau}/ m_{\tau}$, $\delta Z_L^{\nu_{\tau}}$ and
$\delta Z_R^{\tau}$ are
obtained from eqs.(\ref{eq:deltamf}) and (\ref{eq:DSRC})
(with $m_{\nu_{\tau}}=0$ ); 
they represent the contribution from the
mass and wave-function renormalization of the $(\nu_{\tau},\tau)$-doublet,
including the finite renormalization of the neutrino 
leg. Finally, $F_{\tau}$ on eq.(\ref{eq:deltatau}) is the form
factor describing the vertex 
corrections to the amplitude of $H^{+}\rightarrow\tau^{+}\nu_{\tau}$.

On comparing eqs.(\ref{eq:ddtan}) and (\ref{eq:deltabeta}) we see that
the first definition of $\tan\beta$ appears as though it is free from
process-dependent contributions.
In practice, however,  
process-dependent terms are inevitable, irrespective of the definition
of $\tan\beta$. In fact, the 
definition of $\tan\beta$ where
 $\delta v_1/v_1=\delta v_2/v_2$\,\cite{Hollikprivate} will
also develop process-dependent contributions,
as can be seen by trying to relate the ``universal'' value
of $\tan\beta$ in that scheme with a physical quantity directly read off 
some physical observable. For instance, if $M_{A^0}$ 
is heavy enough, one may define $\tan\beta$ as follows:
\beqn
{\Gamma (A^0\rightarrow b\,\bar{b})\over
\Gamma (A^0\rightarrow t\,\bar{t})}&=&
\tan^4\beta\,{m_b^2\over m_t^2}\,\left(1-{4\,m_t^2\over
 M_{A^0}^2}\right)^{-1/2}\,\left[1+4\,\left({\delta v_2\over v_2}-
{\delta v_1\over v_1}\right)
\right.\nonumber\\
& &\left.+2\,\left({\delta m_b\over m_b} 
+\frac{1}{2}\delta Z_L^b +\frac{1}{2}\delta Z_R^b
-{\delta m_t\over m_t} 
-\frac{1}{2}\delta Z_L^t-\frac{1}{2}\delta Z_R^t\right)
 +\delta V\right]\,,
\label{eq:tanbeta3}
\eeqn
where we have neglected $m_b^2\ll M_{A^0}^2$, and $\delta V$ stands for the vertex
corrections to the decay processes $A^0\rightarrow b\,\bar{b}$ and 
$A^0\rightarrow t\,\bar{t}$.
Since the sum of the mass and wave-function renormalization
terms along with the vertex corrections is UV-finite, one can consistently choose
$\delta v_1/v_1=\delta v_2/v_2$ leading to eq.(\ref{eq:ddtan}).
Hence, deriving $\tan\beta$ from eq.(\ref{eq:tanbeta3}) 
unavoidably incorporates also some
process-dependent contributions.

Any definition of $\tan\beta$ is
in principle as good
as any other; and in spite of the fact that
the corrections themselves may show some dependence
on the choice of the particular definition,
the physical observables should not depend at all on that choice.
However, it can be a practical matter what definition to use
in a given situation.   
For example, our definition of $\tan\beta$ given on eq.(\ref{eq:tbetainput})
should be most adequate for $M_{H^{\pm}}<m_t-m_b$ and large $\tan\beta$, since then
$H^+\rightarrow\tau^+\,\nu_{\tau}$ is the dominant decay of $H^+$,
whereas the definition based on
eq.(\ref{eq:tanbeta3}) requires also a large value of $\tan\beta$ (to avoid
an impractical suppression of the $b\,\bar{b}$ mode); moreover, in order to 
be operative, it also requires
a much heavier charged Higgs boson, since 
$M_{H^{\pm}} \simeq M_{A^0}>2\,m_t$ when
the decay $A\rightarrow t\bar{t}$ is kinematically open in the MSSM.
(Use of light quark final states would, of course, be extremely
difficult from the practical
point of view.)

Within our context, we use 
eq.(\ref{eq:deltabeta}) for $\delta\tan\beta/\tan\beta$ in order 
to compute the one-loop corrections to our decay
$t\rightarrow H^+\,b$.                                                                                                                                          
Putting all the pieces together, the counterterm Lagrangian for the vertex
$t\,b\, H^{+}$ follows right away from the bare Lagrangian 
(\ref{eq:LtbH}) after
re-expressing everything in terms of renormalized parameters and fields in the 
on-shell scheme. It takes on the form :
\beq
\delta{\cal L}_{Hbt}={g\over\sqrt{2}\,M_W}\,H^-\,\bar{b}\left[
\delta C_R\ m_t\,\cot\beta\,\,P_R+
\delta C_L\ m_b\,\tan\beta\,P_L\right]\,t
+{\rm h.c.}\,,
\label{eq:LtbH2}
\eeq
with
\beqn
\delta C_R &=& {\delta m_t\over m_t}-{\delta v\over v}
+\frac{1}{2}\,\delta Z_{H^+}+\frac{1}{2}\,\delta Z_L^b+\frac{1}{2}
\,\delta Z_R^t
-{\delta\tan\beta\over\tan\beta}+\delta Z_{HW}\,\tan\beta\,,\nonumber\\
\delta C_L &=& {\delta m_b\over m_b}-{\delta v\over v}
+\frac{1}{2}\,\delta Z_{H^+}+\frac{1}{2}\,\delta Z_L^t+\frac{1}{2}
\,\delta Z_R^b
+{\delta\tan\beta\over\tan\beta}-\delta Z_{HW}\,\cot\beta\,,
\eeqn
and where we have set $V_{tb}=1$  ($V_{tb}=0.999$ within
$\pm 0.1\%$, from unitarity of the CKM-matrix under the assumption of
three generations).


\vspace{0.75cm} 
\begin{Large}
{\bf 4. One-loop Corrected $\Gamma(t\rightarrow H^+\,b)$ in the MSSM}
\end{Large}
\vspace{0.5cm}

As stated in Section 2, the study of the decay $t\rightarrow H^+\,b$
is worthwhile in the small ($\tan\beta<2$), and most conspicuously in
the high ($\tan\beta\geq 30$) $\tan\beta$ region, where the branching 
ratio can be comparable to the one of the standard decay $t\rightarrow W^+\,b$.
These are, therefore, the regions on which we will focus our search for
potentially significant (strong and electroweak like) SUSY quantum
corrections to $t\rightarrow H^+\,b$. 
As for the strong effects, they can be rather large and
have been evaluated in Ref.\,\cite{GJS}; here we shall not dwell any longer
on their detailed structure apart from including them in our numerical analysis
and adding some useful remarks in Section 5.

On the electroweak side, one may also expect
sizeable quantum corrections from  enhanced Yukawa couplings
of the type (\ref{eq:Yukawas}).
In the relevant $\tan\beta$ regions mentioned above, 
the latters yield the leading electroweak contributions  and
in these conditions we will neglect the
pure gauge corrections from transversal gauge bosons in the Feynman gauge. 
Moreover, as already stressed in Section 2, the branching ratio of the 
charged Higgs mode in the intermediate $\tan\beta$ region is too small to speak of, 
so that the detailed structure of the radiative corrections in this range
is irrelevant.

In the following we will describe
the relevant electroweak one-loop supersymmetric diagrams entering
the amplitude of  
$t\rightarrow H^+\,b$ in the MSSM.  At the tree-level, the
only  Feynman diagram is the one in Fig.1. 
At the one-loop, we have the diagrams exhibited in Figs.2-6. 
The computation of the one-loop diagrams requires to use the full structure of
the MSSM Lagrangian. The explicit form of the most relevant pieces of this
Lagrangian, together with the necessary SUSY notation, is provided in 
Appendix A.

Specifically, Fig.2 shows the electroweak
SUSY vertices involving squarks, charginos and neutralinos. 
In all these diagrams a sum over all indices is taken for
granted.
The supersymmetric Higgs particles of the MSSM and
Goldstone bosons (in the Feynman gauge) contribute a host of one-loop vertices
as well (see Fig.3). 
As for the various self-energies, they will be treated as counterterms
to the vertices. Their structure is dictated by the
Lagrangian (\ref{eq:LtbH2}). 
Thus, Fig.4 displays the counterterms $C_{b1}-C_{t4}$ generated from the
external bottom and top quark lines; they include contributions from
supersymmetric particles, Higgs bosons and Goldstone bosons.
Similarly, Fig.5 contains the counterterms $C_{H1}-C_{H4}$ associated to the
self-energy of the
external charged Higgs boson. A variant of the latter contribution is the
mixed $W^+-H^+$ self-energy counterterms $C_{M1}-C_{M3}$ shown in Fig.6.

Although we have displayed only the process dependent diagrams,
the full analysis should also include the SUSY and Higgs/Goldstone
boson contributions to the various
universal vacuum  polarization effects comprised in our counterterms.
However, the
calculation of all these pieces has already been discussed in detail 
long ago in the literature\,\cite{GrifSol,Bertolini} and thus the
lengthy formulae accounting for these results
will not be explicitly quoted here. Their contribution
is not $\tan\beta$-enhanced, but since we wish to
compute the full supersymmetric contribution in the relevant regions of
the MSSM parameter space, those 
effects will be included in our numerical code.
Finally,
the smaller --though numerically overwhelming -- subset of strong supersymmetric 
one-loop graphs are displayed in Fig.2 of Ref.\cite{GJS}. We will use
the formulae from the latter reference in the present analysis
to produce the total (electroweak+strong) SUSY correction
to our process. 

Next let us report on the contributions from
the various vertex diagrams and counterterms
in the on-shell renormalization scheme.
The generic structure of any
renormalized vertex function, $\Lambda$, in Figs.2-3 is composed of two form 
factors $F_L$, $F_R$ plus the counterterms.
Therefore, on making use of the formulae of Section 3, one immediately finds:
\beq
\Lambda = {i\,g\over\sqrt{2}\,M_W}
\,\left[m_t\,\cot\beta\,(1+\Lambda_R)\,P_R
 + m_b\,\tan\beta\,(1+\Lambda_L)\,P_L\right]\,,
\label{eq:AtbH}
\eeq
where
\beqn
\Lambda_R & = & F_R+{\delta m_t\over m_t}
+\frac{1}{2}\,\delta Z_L^b+\frac{1}{2}\,\delta Z_R^t-\Delta_{\tau}\nonumber\\
& - & {\delta v^2\over v^2}+\delta Z_{H^+}+(\tan\beta-\cot\beta)\,\delta Z_{HW}
 \,,\nonumber\\
\Lambda_L &=& F_L+{\delta m_b\over m_b}
+\frac{1}{2}\,\delta Z_L^t+\frac{1}{2}\,\delta Z_R^b
+\Delta_{\tau}\,.
\label{eq:lambdaLR}
\eeqn
In the following the analytical contributions to the vertex 
form factors and counterterms will be specified diagram by
diagram.

\vspace{0.5cm} 
\begin{large}
{\bf 4.1 SUSY vertex diagrams}
\end{large}
\vspace{0.25cm}

In this section we will make intensive use of the definitions and
formulae of Appendix~A. We refer the reader there for questions
about notation and conventions. 
Following the labelling of Feynman graphs in Fig.2 we write
down the terms coming from virtual SUSY particles.
\begin{itemize}
\item {Diagram $(V_{S1})$:}
Making use of the coupling matrices of eqs.~(\ref{V1Apm}) and (\ref{eq:QLQR}) we 
introduce the shorthands\footnote{Lower indices are summed over, whereas
upper indices (some of them within parenthesis) are just for notational 
convenience.}
\begin{equation}
  \label{V1Apmdef}
  \Apm\equiv\Apmit\ \ \mbox{and}\ \ \Apmz\equiv\Apmat\,,
\end{equation}
and define the combinations (omitting indices also for $\QaiL,\QaiR$)
\begin{eqnarray}
  \label{V1matrices}
    \RRLo =\cbt\Apc\HRc\Amz\,,&\ \ \ 
   &\RRRo =\cbt\Amc\HRc\Amz\,,\nonumber\\
    \LRLo =\cbt\Apc\HRc\Apz\,,&\ \ \ 
   &\LRRo =\cbt\Amc\HRc\Apz\,,\nonumber\\
    \RLLo =\sbt\Apc\HLc\Amz\,,&\ \ \ 
   &\RLRo =\sbt\Amc\HLc\Amz\,,\nonumber\\
    \LLLo =\sbt\Apc\HLc\Apz\,,&\ \ \ 
   &\LLRo =\sbt\Amc\HLc\Apz\,.           
\end{eqnarray}

The contribution from diagram $(V_{S1})$ to the form factors $F_L$ and $F_R$
is then
\begin{eqnarray}
  \label{V1FF}
  F_L&=&M_L\left[\LLRo\Czt+\right.\nonumber\\
  &+&\mb\,\left(\mt\RRLo+\ma\LRLo+\mb\LLRo+\mi\LLLo\right)\Cot\nonumber\\
  &+&\mt\,\left(\mt\LLRo+\ma\RLRo+\mb\RRLo+\mi\RRRo\right)
     \left(\Coo-\Cot\right)\nonumber\\
  &+&\left.\left(\mt\mb\RRLo+\mt\mi\RRRo+\ma\mb\LRLo+\mi\ma\LRRo\right)
     \Cz\right]\,,\nonumber\\
  F_R&=&M_R\left[\RRLo\Czt+\right.\nonumber\\
  &+&\mb\,\left(\mt\LLRo+\ma\RLRo+\mb\RRLo+\mi\RRRo\right)\Cot\nonumber\\
  &+&\mt\,\left(\mt\RRLo+\ma\LRLo+\mb\LLRo+\mi\LLLo\right)
     \left(\Coo-\Cot\right)\nonumber\\
  &+&\left.\left(\mt\mb\LLRo+\mt\mi\LLLo+\ma\mb\RLRo+\mi\ma\RLLo\right)
     \Cz\right]\,,
\end{eqnarray}
where the overall coefficients $M_L$ and $M_R$ are the following:
\begin{equation}
  \label{MLMR}
  M_L=-\frac{ig^2\mw}{\mb\tb}\ \ \ \ M_R=-\frac{ig^2\mw}{\mt\ctb}\,.
\end{equation}
The notation for the various $3$-point functions is summarized in 
Appendix~C. On eq.~(\ref{V1FF}) they must be evaluated with arguments:
\begin{equation}
  \label{V1Cs}
  C_{*}=C_{*}\left(p,p',\msta,\ma,\mi\right)\,.
\end{equation}

\item {Diagram $(V_{S2})$:}
For this diagram --which in contrast to the
others is finite-- we also use the matrices on
eqs.~(\ref{V1Apm}) and (\ref{gdef}), and introduce the shorthands
\begin{equation}
  \label{V2Apmdef}
   \Apmb\equiv A_{\pm b\alpha}^{(b)}\ \ \mbox{and}\ \ \Apmt\equiv\Apmat\,,
\end{equation}
to define the products of coupling matrices
\begin{eqnarray}
  \label{V2matrices}
  \RLt =\Gba\Apbc\Amt\,,&\ \ \ &\RRt =\Gba\Ambc\Amt\,,\nonumber\\
  \LLt =\Gba\Apbc\Apt\,,&\ \ \ &\LRt =\Gba\Ambc\Apt\,.           
\end{eqnarray}
The contribution to the form factors $F_L$ and $F_R$ from this diagram is
\begin{eqnarray}
  \label{V2FF}
  F_L&=&\frac{M_L}{2\mw}\left[
  \mb\LLt\Cot+\mt\RRt\left(\Coo-\Cot\right)-\ma\LRt\Cz\right]\,,\nonumber\\
  F_R&=&\frac{M_R}{2\mw}\left[
  \mb\RRt\Cot+\mt\LLt\left(\Coo-\Cot\right)-\ma\RLt\Cz\right]\,,
\end{eqnarray}
the coefficients $M_L$, $M_R$ being those of eq.~(\ref{MLMR}) and the 
scalar $3$-point functions now evaluated with arguments
\begin{equation}
  \label{V2Cs}
  C_{*}=C_{*}\left(p,p',\ma,\msta,\msbb\right)\,.
\end{equation}

\item {Diagram $(V_{S3})$:}
For this diagram  we will need
\begin{equation}
  \label{V3Apmdef}
  \Apm\equiv\Apmib\ \ \mbox{and}\ \ \Apmz\equiv\Apmab\,,
  \end{equation}
and again omitting indices we shall use
\begin{eqnarray}
  \label{V3matrices}
    \RRLth =\cbt\Apzc\HRc\Am\,,&\ \ \ 
   &\RRRth =\cbt\Amzc\HRc\Am\,,\nonumber\\
    \LRLth =\cbt\Apzc\HRc\Ap\,,&\ \ \ 
   &\LRRth =\cbt\Amzc\HRc\Ap\,,\nonumber\\
    \RLLth =\sbt\Apzc\HLc\Am\,,&\ \ \ 
   &\RLRth =\sbt\Amzc\HLc\Am\,,\nonumber\\
    \LLLth =\sbt\Apzc\HLc\Ap\,,&\ \ \ 
   &\LLRth =\sbt\Amzc\HLc\Ap\,.           
\end{eqnarray}
{}From these definitions the contribution of diagram $(V_{S3})$ to the form
factors can be obtained by performing the following changes in that of
diagram $(V_{S1})$, eq.~(\ref{V1FF}): 
\begin{itemize}
\item Everywhere on eqs.~(\ref{V1FF}) and (\ref{V1Cs}) replace 
$\mi\leftrightarrow\ma$ and $\msta\leftrightarrow\msba$.
\item Replace on eq.~(\ref{V1FF}) couplings from (\ref{V1matrices}) 
with those of (\ref{V3matrices}).
\item Include a global minus sign.
\end{itemize}

\end{itemize}


\vspace{0.5cm} 
\begin{large}
{\bf 4.2 Higgs vertex diagrams}
\end{large}
\vspace{0.25cm}

Now we consider contributions arising from the exchange of virtual
Higgs particles
and Goldstone bosons in the Feynman gauge,
as shown in Fig.3. We follow the vertex
formula for the form factors by the value of the
overall coefficient $N$ 
and by the arguments of the corresponding $3$-point functions.
\begin{itemize}
\item {Diagram $(V_{H1})$:}
\begin{eqnarray*}
  F_L&=&N\,[\mbs (\Cot-\Cz)+\mts \ctbs (\Coo-\Cot) ]\,,\\
  F_R&=&N\mbs [\Cot-\Cz +\tbs (\Coo-\Cot) ] \,,\\
  N&=&\mp\frac{ig^2}{2}\left(1-\frac{\{\mHs,\mhs\}}{2\mws}\right)
        \frac{\{\ca , \sa \}}{\cbt}\{\cbma , \sbma\}\,, \\
  C_{*}&=&C_{*}\left(p,p',\mb,\mHp,\{\mH,\mh\}\right)\,.
\end{eqnarray*}

\item {Diagram $(V_{H2})$:}
\begin{eqnarray*}
  F_L&=&N\ctb[\mts (\Coo -\Cot )+\mbs(\Cz -\Cot )]\,,\\
  F_R&=&N\mbs\tb (2\Cot -\Coo -\Cz ) \,,\\
  N&=&\frac{ig^2}{4} \frac{\{\ca , \sa \}}{\cbt}\{\sbma  , \cbma \}
        \left(\frac{\mHps}{\mws}-\frac{\{\mHs  , \mhs \}}
{\mws}\right)\,, \\
  C_{*}&=&C_{*}\left(p,p',\mb,\mw,\{\mH,\mh\}\right)\,.
\end{eqnarray*}

\item {Diagram $(V_{H3})$:}
\begin{eqnarray*}
  F_L&=&N\mts [\ctbs\Cot +\Coo -\Cot -\Cz ] \,,\\
  F_R&=&N\,[\mbs\tbs\Cot  +\mts (\Coo -\Cot-\Cz )]\,, \\
  N&=&-\frac{ig^2}{2}\frac{\{\sa , \ca \}}{\sbt}\{\cbma  , \sbma \}
        \left(1-\frac{\{\mHs  ,\mhs \}}{2\mws} \right) \,,\\
  C_{*}&=&C_{*}\left(p,p',\mt,\{\mH,\mh\},\mHp\right)\,.
\end{eqnarray*}

\item {Diagram $(V_{H4})$:}
\begin{eqnarray*}
  F_L&=&N\mts (2\Cot-\Coo+\Cz)\ctb \,,\\
  F_R&=&N\,[-\mbs\Cot +\mts (\Coo-\Cot-\Cz)]\tb\,, \\
  N&=&\mp\frac{ig^2}{4}\frac{\{\sa , \ca \}}{\sbt}\{\sbma  , \cbma \}
        \left(\frac{\mHps}{\mws}-\frac{\{\mHs  ,\mhs \}}
{\mws}\right) \,,\\
  C_{*}&=&C_{*}\left(p,p',\mt,\{\mH,\mh\},\mw\right)\,.
\end{eqnarray*}

\item {Diagram $(V_{H5})$:}
\begin{eqnarray*}
  F_L&=&N\,[\mbs(\Cot+\Cz)+\mts(\Coo-\Cot)]\,, \\
  F_R&=&N \mbs \tbs (\Coo+\Cz)\,, \\
  N&=&-\frac{ig^2}{4}\left(\frac{\mHps}{\mws}-\frac{\mAs}
{\mws}\right)\,, \\
  C_{*}&=&C_{*}\left(p,p',\mb,\mw,\mA\right)\,.
\end{eqnarray*}

\item {Diagram $(V_{H6})$:}
\begin{eqnarray*}
  F_L&=& N\mts\ctbs (\Coo+\Cz)\,, \\
  F_R&=& N\,[\mbs\Cot +\mts (\Coo-\Cot+\Cz)]\,, \\
  N&=&-\frac{ig^2}{4}\left(\frac{\mHps}{\mws}-\frac{\mAs}
{\mws}\right) \,,\\
  C_{*}&=&C_{*}\left(p,p',\mt,\mA,\mw\right)\,.
\end{eqnarray*}

\item {Diagram $(V_{H7})$:}
\begin{eqnarray*}
  F_L&=&N\,[(2\mbs\Coo+\Czt+2(\mts-\mbs)(\Coo-\Cot ))\ctbs
        +2\mbs(\Coo+2\Cz)]\mts\,, \\
  F_R&=&N\,[(2\mbs\Coo+\Czt+2(\mts-\mbs)(\Coo-\Cot ))\tbs
        +2\mts(\Coo+2\Cz)]\mbs\,, \\
  N&=&\pm\frac{ig^2}{4\mws}\frac{\sa \ca}{\sbt \cbt}\,, \\
  C_{*}&=&C_{*}\left(p,p',\{\mH,\mh\},\mt,\mb\right)\,.
\end{eqnarray*}

\item {Diagram $(V_{H8})$:}
\begin{eqnarray*}
  F_L&=& N\mts\ctbs\,\Czt\,,  \\
  F_R&=& N\mbs\tbs\,\Czt\,,   \\
  N&=&\mp \frac{ig^2}{4\mws}\,, \\
  C_{*}&=&C_{*}\left(p,p',\{\mA,\mz\},\mt,\mb\right)\,.
\end{eqnarray*}

\end{itemize}

In the equations above, 
it is understood that the CP-even mixing angle, $\alpha$, is renormalized 
into $\alpha_{\rm eff}$ by the one-loop Higgs mass relations\,\cite{Higgsloop}.

As for the SUSY and Higgs contributions to the counterterms, they are much
simpler since they just involve $2$-point functions. Thus we shall present
the full electroweak results by adding up the various sparticle and Higgs
effects. In the following formulae, we append labels referring to the
specific diagrams on Figs.4-6. 

\newpage
\vspace{0.5cm} 
\begin{large}
{\bf 4.3 Counterterms}
\end{large}
\vspace{0.25cm}

\begin{itemize}
\item {Counterterms $\delta m_f\,,\delta Z_L^f\,,\delta Z_R^f$:}
For a given down-like fermion $b$, and corresponding
isospin partner $t$, 
the fermionic self-energies receive contributions

\begin{eqnarray}
  \label{SUSYSelfb}
  \Sigma_{\{L,R\}}^b(p^2)&=&\left.\Sigma_{\{L,R\}}^b(p^2)
   \right|_{(C_{b1})+(C_{b2})}\nonumber\\
 &=&-ig^2\left[\abs{\Apmit}^2\Bo\left(p,\mi,\msta\right)
  +\frac{1}{2}\abs{\Apmab}^2\Bo\left(p,\ma,\msba\right)\right]\,,\nonumber\\
  \mb\Sigma_S^b(p^2)&=&\left.\mb\Sigma_S^b(p^2)
   \right|_{(C_{b1})+(C_{b2})}\nonumber\\
 &=&ig^2\left[\mi {\rm Re}\left(\Apitc\Amit\right)
    \Bz\left(p,\mi,\msta\right)\right.\nonumber\\
 &&+\left.\frac{1}{2}\ma {\rm Re}\left(\Amabc\Apab\right)
    \Bz\left(p,\ma,\msba\right)\right]\,,
\end{eqnarray}
from SUSY particles, and
\begin{eqnarray}
  \label{HiggsSelfb}
  \Sigma_{\{L,R\}}^b(p^2)&=&\left.\Sigma_{\{L,R\}}^b(p^2)
   \right|_{(C_{b3})+(C_{b4})}\nonumber\\
  &=&\frac{g^2}{2i\mws}\left\{m_{\{t,b\}}^2\left[
   \{\ctbs,\tbs\} \Bo(p,\mt,\mHp)+\Bo(p,\mt,\mw)
   \right]\right. \nonumber\\
  &+&\frac{\mbs}{2\cbts}\left[\cas\,\Bo(p,\mb,\mH)
   +\sas\,\Bo(p,\mb,\mh)\right. \nonumber\\
  &&\ \ \ +\left.\left. \sbts\,\Bo(p,\mb,\mA)
   +\cbts\,\Bo(p,\mb,\mz)\right]\right\}\,,\nonumber\\
  \Sigma_S^b(p^2)&=&\left.\Sigma_S^b(p^2)
   \right|_{(C_{b3})+(C_{b4})}\nonumber\\
 &=&-\frac{g^2}{2i\mws}\left\{\mts\left[
   \Bz(p,\mt,\mHp)-\Bz(p,\mt,\mw)\right]\right.\nonumber\\
  &+&\frac{\mbs}{2\cbts}\left[\cas\,\Bz(p,\mb,\mH)+
   \sas\,\Bz(p,\mb,\mh)\right.\nonumber\\
  &&\ \ \ -\left.\left. \sbts\,\Bz(p,\mb,\mA)
   -\cbts\,\Bz(p,\mb,\mz)\right]\right\}\,,
\end{eqnarray}
from Higgs and Goldstone bosons in the Feynman gauge.
To obtain the corresponding expressions for an up-like fermion, $t$, just
perform the label substitutions $b\leftrightarrow t$
on eqs. (\ref{SUSYSelfb})-(\ref{HiggsSelfb}); and 
on eq.~(\ref{HiggsSelfb}) 
replace $\sa\leftrightarrow \ca$ and
$\sbt\leftrightarrow \cbt$ (which also implies replacing
$\tb\leftrightarrow \ctb$).

Introducing the above expressions into eqs.~(\ref{eq:deltamf})-(\ref{eq:DSRC})
one immediately obtains the SUSY contribution 
to the counterterms $\delta m_f\,,\delta Z_{L,R}^f$.

\item {Counterterm $\delta Z_{H^\pm}$}:
\begin{eqnarray}
  \label{dZHSusy}
  \delta Z_{H^\pm}&=&\left.\delta Z_{H^\pm}
                     \right|_{(C_{H1})+(C_{H2})+(C_{H3})+(C_{H4})}
    =\,\Sigma_{H^\pm}'(\mHps)\nonumber\\
  &=&-\frac{ig^2N_C}{\mws}\left[(\mbs\tbs+\mts\ctbs)
   (\Bo+\mHps\Bo'+\mbs\Bz')\right. \nonumber\\
  &&\left.\mbox{\hspace{1.5cm}}+2\mbs\mts\Bz'\right](\mHp,\mb,\mt)\nonumber\\
  &&+\frac{ig^2}{2\mws}N_C\sum_{ab}\abs{\Gba}^2
   \Bz'(\mHp,\msbb,\msta)  \nonumber\\
  &&-2ig^2\sum_{i\alpha}\left[
\left(\abs{Q^L_{\alpha i}}^2\cbts+\abs{Q^R_{\alpha i}}^2\sbts\right)
  (\Bo+\mHps \Bo'+\mas\Bz')\right.                              \nonumber\\
  &&\left.\phantom{\abs{Q^L_i}^2}+2\mi\ma Re
   \left(Q^L_{\alpha i}Q^{R*}_{\alpha i}\right)
   \sbt\cbt\Bz'\right](\mHp,\ma,\mi)\,.
\end{eqnarray}
Notice that diagram $(C_{H3})$ gives a vanishing contribution to
$\delta Z_{H^\pm}$.

\item {Counterterm $\delta Z_{HW}$}:
\begin{eqnarray}
  \label{dZHWSusy}
  \delta Z_{HW}&=&\left.\delta Z_{HW}\right|_{(C_{M1})+(C_{M2})+(C_{M3})}
 =\frac{\Sigma_{HW}(\mHps)}{\mws}\nonumber\\
 &=&-\frac{ig^2N_C}{\mws}\left[\mbs\tb(\Bz+\Bo)
  +\mts\ctb\Bo\right](\mHp,\mb,\mt)  \nonumber\\
 &&-\frac{ig^2N_C}{2\mws}\sum_{ab}\Gba\Rot 
  R_{1b}^{(b)*}\left[2\Bo+\Bz\right](\mHp,\msbb,\msta) \nonumber\\
 &&+\frac{2ig^2}{\mw}\sum_{i\alpha}
  \left[\ma\left(\HiaR\CL+\HiaL\CR\right)(\Bz+\Bo)\right.  \nonumber\\
 &&+\left.\mi\left(\HiaL\CL+\HiaR\CR\right)\Bo\right]
  (\mHp,\ma,\mi)\,. 
\end{eqnarray}
A sum is understood over all generations.
\end{itemize}

Finally, the evaluation of $\Delta_{\tau}$ on eq.(\ref{eq:deltatau}) 
yields similar bulky analytical formulae, which follow
after computing diagrams akin to those in Figs.2-6
for the MSSM corrections to $H^+\rightarrow \tau^+\,{\nu}_{\tau}$.
We refrain from quoting them explicitly here. The numerical
effect, though, will 
be explicitly given in Section 5. 

We are now ready to furnish the corrected width of $t\rightarrow H^+\,b$
in the MSSM. It just follows after computing the interference between
the tree-level amplitude and the one-loop amplitude. It is convenient to
express the result  as a relative correction with respect to
the tree-level width  both
in the $\alpha$-scheme and in the $G_F$-scheme.  In the former we obtain
the relative MSSM correction
\beqn
\delta_{\alpha}^{MSSM} & = & {\Gamma -\Gamma^{(0)}_{\alpha}
\over \Gamma^{(0)}_{\alpha}}\nonumber\\ 
&=&\frac{N_L}{D}\,[2\,Re(\Lambda_L)]+
\frac{N_R}{D}\,[2\,Re(\Lambda_R)]+\frac{N_{LR}}{D}\,
[2\,Re(\Lambda_L+\Lambda_R)]\,,
\label{eq:deltaalpha}
\eeqn
where the corresponding lowest-order width is
\beq
\Gamma^{(0)}_{\alpha}=\left({\alpha\over s^2_W}\right)
\,{D\over 16\, M_W^2\,m_t}\,
\lambda^{1/2} (1, {m_b^2\over m_t^2},{\mHps\over m_t^2})\,,
\label{treeHalpha}
\eeq
with
\beqn
D &=& (m_t^2+m_b^2-\mHps)\,(m_t^2\cot^2\beta+m_b^2\tan^2\beta)
+4m_t^2m_b^2\,,\nonumber\\
N_L & = & (m_t^2+m_b^2-\mHps)\,m_b^2\tan^2\beta\,,\nonumber\\
N_R & = & (m_t^2+m_b^2-\mHps)\,m_t^2\cot^2\beta\,,\nonumber\\
N_{LR} & = & 2m_t^2m_b^2\,.
\eeqn 
From these equations it is obvious that at low $\tan\beta$ the relevant 
quantum effects basically come from the contributions to the form factor $\Lambda_R$
whereas at high $\tan\beta$ they come from $\Lambda_L$.   

Using  eq.(\ref{eq:DeltaMW}) we find that the relative MSSM
correction in the $G_F$-parametrization reads
\beq
\delta_{G_F}^{MSSM}= {\Gamma-\Gamma^{(0)}_{G_F}
\over \Gamma^{(0)}_{G_F}}=\delta_{\alpha}^{MSSM}-\Delta r^{MSSM}\,,
\label{eq:alphagf1}
\eeq
where the tree-level width in the $G_F$-scheme,
 $\Gamma^{(0)}_{G_F}$, is given by eq.(\ref{eq:treeH}) and is related to
eq.(\ref{treeHalpha}) through
\beq
\Gamma^{(0)}_{\alpha}=\Gamma^{(0)}_{G_F}\,(1-\Delta r^{MSSM})\,.
\label{eq:alphagf}
\eeq 


\vspace{0.75cm} 
\begin{Large}
{\bf 5. Numerical Analysis and Discussion }
\end{Large}
\vspace{0.5cm}

Quantum effects should be able to
discriminate whether the charged
Higgs emerging from the decay $t\rightarrow H^+\,b$ is supersymmetric or not,
for the MSSM provides a well defined prediction of the typical size of
these effects using the present bounds on sparticle masses. 
Some work on radiative corrections to the decay width of $t\rightarrow H^+\,b$ has
already appeared in the literature. In particular, the conventional 
QCD corrections have been evaluated\,\cite{CD} and found to significantly
reduce the partial width. The SUSY-QCD corrections are also substantial and have
been analyzed, only in part in Refs.\cite{Koenig,LiYangHu}, and in more detail
in Ref.\cite{GJS}. Nevertheless, the electroweak corrections produced by
the roster of genuine (R-odd) sparticles have not been considered at all yet.
As for the virtual effects mediated by the Higgs bosons,
a first treatment is given in Refs.\cite{Mendez} and \cite{LiHuYang}.
However, these references disagree in several parts of the calculation,
and moreover they are both incomplete
calculations on their own, for they fully ignore the Higgs effects  
associated to the bottom quark
Yukawa coupling, which could in principle be significant in 
the large $\tan\beta$ region. On the other hand, even though the latter kind of 
Higgs effects have been discussed
in the literature in other renormalization schemes based on alternative definitions
of $\tan\beta$\,\cite{CPRD,Yamada2,Yamada3},
a detailed analysis including the genuine SUSY effects themselves has never
been attempted.  
Thus, if only for completeness, we are providing here
not only a dedicated treatment of the $R$-odd 
contributions mediated by the sparticles of the MSSM, but also
the fully-fledged pay-off of the supersymmetric Higgs effects.

Before presenting the results of the complete numerical
analysis, it should be clear that the bulk of the high $\tan\beta$ corrections
to the decay rate of $t\rightarrow H^+\,b$ in the MSSM is expected
to come from SUSY-QCD. This could already be foreseen from what is known in
SUSY GUT models\,\cite{SO10}; in fact, in this context 
a non-vanishing sbottom mixing  (which we also assume
in our analysis) may lead to important SUSY-QCD quantum effects 
on the bottom mass, $m_b= m_b^{GUT}+\Delta m_b$, where $\Delta m_b$ is 
proportional to 
$M_{LR}^b\rightarrow -\mu\tan\beta $ at sufficiently high
$\tan\beta$.
These are finite threshold effects that one has to include
when matching the SM and MSSM renormalization group equations (RGE) at the
effective supersymmetric threshold
scale, $T_{SUSY}$, above which the RGE evolve according to the MSSM
$\beta$-functions in the $\overline{MS}$ scheme\,\cite{CPW,MC}.
In our case, since the bottom mass is an input parameter for the on-shell scheme,
these effects obviously have a different physical meaning, 
but are formally the same; they are just fed into
the mass counterterm $\delta m_b/m_b$
on eq.(\ref{eq:lambdaLR}) and contribute to it with opposite sign
($\delta m_b/m_b=-\Delta m_b +...$)
\footnote{In the alternative framework of Ref.\cite{GJS},
the SUSY-QCD corrections have been computed assuming no mixing in the sbottom mass
matrix. Nonetheless, the typical size of the SUSY-QCD corrections does not change
as compared to the present approach (in which we do assume a non-diagonal
sbottom matrix) the
reason being that in the absence of sbottom mixing, i.e. $M_{LR}^b=0$,
the contribution $\delta m_b/m_b\propto -\mu\tan\beta$ at large $\tan\beta$
is no longer possible but, in contrast, the vertex correction
does precisely inherits this dependence and
compensates for it (Cf. Appendix A).
The drawback of an scenario based on
$M_{LR}^b=0$, however, is that when it is combined with a large value
of $\tan\beta$ it may lead to a value of
$A_b$ which overshoots the natural range expected for this parameter. }. 

Explicitly, when viewed in terms of diagrams of the 
electroweak-eigenstate basis, the relevant finite corrections from
the bottom mass counterterm 
are generated by mixed LR-sbottoms and gluino loops (Cf. Fig.7a):
\beqn
\left({\delta m_b\over m_b}\right)_{\rm SUSY-QCD} &=&
{2\alpha_s(m_t)\over 3\pi}\,m_{\tilde{g}}\,M_{LR}^b\,
I(m_{\tilde{b}_1},m_{\tilde{b}_2},m_{\tilde{g}})\nonumber\\
&\rightarrow & -{2\alpha_s(m_t)\over 3\pi}\,m_{\tilde{g}}\,\mu\tan\beta\,
I(m_{\tilde{b}_1},m_{\tilde{b}_2},m_{\tilde{g}}) \,,
\label{eq:dmbQCD}
\eeqn
where the last result holds for sufficiently large $\tan\beta$ and for
$\mu$ not too small as compared to $A_b$. We have introduced 
the positive-definite function (Cf. Appendix C)
\beq
I(m_1,m_2,m_3)\equiv 16\,\pi^2 i\,C_0(0,0,m_1,m_2,m_3)=
{m_1^2\,m_2^2\ln{m_1^2\over m_2^2}
+m_2^2\,m_3^2\ln{m_2^2\over m_3^2}+m_1^2\,m_3^2\ln{m_3^2\over m_1^2}\over
 (m_1^2-m_2^2)\,(m_2^2-m_3^2)\,(m_1^2-m_3^2)}\,.
\label{eq:I123}
\eeq
In addition, we could also foresee potentially large (finite)
SUSY electroweak effects
from $\delta m_b/m_b$. They are induced by
$\tan\beta$-enhanced Yukawa couplings of the type (\ref{eq:Yukawas}). 
Of course, these effects
have already been fully included in the calculation presented in
Section 3 that we have performed in the mass-eigenstate basis, but it is illustrative
of the origin of the leading contributions to pick them 
up again directly from the diagrams in the
electroweak-eigenstate basis. In this case, from loops involving mixed LR-stops 
and mixed charged higgsinos (Cf. Fig.7b), one finds:
\beqn
\left({\delta m_b\over m_b}\right)_{\rm SUSY-Yukawa} &=&
-{h_t\,h_b\over 16\pi^2}\,\, {\mu\over m_b}\,m_t\,M_{LR}^t
I(m_{\tilde{t}_1},m_{\tilde{t}_2},\mu)\nonumber\\
&\rightarrow &
-{h_t^2\over 16\pi^2}\,\mu\tan\beta\,A_t\,
I(m_{\tilde{t}_1},m_{\tilde{t}_2},\mu)\,,
\label{eq:dmbEW}
\eeqn
where again the last expression holds for large enough $\tan\beta$.

Notice that, at variance with eq.(\ref{eq:dmbQCD}), the Yukawa coupling
correction (\ref{eq:dmbEW}) dies away with increasing $\mu$. 
Setting $h_t\simeq 1$ at high $\tan\beta$, and assuming that
there is no large hierarchy between the sparticle masses, the ratio
between (\ref{eq:dmbQCD}) and (\ref{eq:dmbEW}) is given, in good approximation,
by $4\,{m_{\tilde{g}}/ A_t}$
times a slowly varying function of the masses of
order $1$, where the (approximate) proportionality
to the gluino mass reflects the very slow decoupling rate of
the latter\,\cite{GJS}.

In view of the present bounds on the gluino mass, and
since $A_t$ (as well as $A_b$) cannot increase
arbitrarily (Cf. Appendix A), we expect that the SUSY-QCD effects
can be dominant and even overwhelming for sufficiently heavy gluinos.
Unfortunately, in contradistinction to the SUSY-QCD case, there are also plenty of 
additional vertex contributions  both from the Higgs sector 
and from the stop-sbottom/gaugino-higgsino sector
where those Yukawa couplings enter once again the game.
So if one wishes to trace the origin of the leading
contributions in the electroweak-eigenstate
basis, a similar though somewhat more involved 
exercise has to be carried out also for vertex functions. Of course,
all of these effects are automatically included in our
calculation of Section 3 within the framework of
the mass-eigenstate basis\footnote{The mass-eigenstate basis
is extremely convenient to carry out the numerical analysis, but it does not
immediately provide a ``physical interpretation'' of the results.
The electroweak-eigenstate basis, in contrast, is a better bookkeeping device
to trace the origin of the most relevant effects, but as a drawback the
intricacies of the full analytical calculation
can be (in general) abhorrent.}. 

We may now pass on to the numerical analysis of the over-all quantum effects.
After explicit computation of the various loop diagrams, the results are 
conveniently cast in terms of the relative
correction with respect to the tree-level width:
\beq
\delta={\Gamma_H-\Gamma^{(0)}_H\over \Gamma^{(0)}_H}
\equiv{\Gamma (t\rightarrow H^+\,b)-\Gamma^{(0)}(t\rightarrow H^+\,b)
\over \Gamma^{(0)}(t\rightarrow H^+\,b)}\,.
\label{eq:pito}
\eeq
In what follows we understand that $\delta$ defined by eq.(\ref{eq:pito})
is  $\delta_{\alpha}$ -- Cf. eq.(\ref{eq:deltaalpha}) -- i.e.
we shall always give our corrections with respect to the tree-level width
$\Gamma_{\alpha}^0$ in the $\alpha$-scheme. The corresponding correction with
respect to the tree-level width in the $G_F$-scheme is simply given by 
eq.(\ref{eq:alphagf1}), where $\Delta r^{MSSM}$ was object of a particular
study \,\cite{GS} and therefore it can be easily incorporated, if necessary.
Notice, however, that $\Delta r^{MSSM}$ is already tightly
bound by the experimental data on $M_Z=91.1863\pm 0.0020\,GeV$ 
at LEP\,\cite{LEPEWWG} and the ratio
$M_{W}/M_Z$ in
$p\bar{p}$, which lead
 to $M_W=80.356\pm 0.125\,GeV$. 
Therefore, even without doing the exact theoretical calculation
of $\Delta r$ within the MSSM, we
already know from
\beq
\Delta r=1-{\pi\alpha\over \sqrt{2}\,G_F}\,{1\over M_W^2\,(1-M_W^2/M_Z^2)}\,,
\label{eq:deltar}
\eeq
that $\Delta r^{MSSM}$ must lie in the experimental interval 
$\Delta r^{\rm exp}\simeq 0.040\pm 0.018$. 

Now, since the corrections computed
in Section 3 can typically be about one order of magnitude larger than 
$\Delta r^{\rm MSSM}$, the bulk of the quantum effects 
on $t\rightarrow H^+\,b$ is already comprised in the relative
correction (\ref{eq:pito}) in the $\alpha$-scheme
\footnote{For the standard decay $t\rightarrow W^+\,b$, the
situation is quite different since the SM electroweak
corrections\,\cite{TopSM} and the maximal SUSY electroweak 
corrections\,\cite{GJSH} in the $\alpha$-scheme are much smaller than for the
decay $t\rightarrow H^+\,b$ , namely they are of the order
of $\Delta r$. Therefore, for the standard decay
$t\rightarrow W^+\,b$ there is a significant cancellation between the corrections 
in the $\alpha$-scheme and $\Delta r$ in most of the $\tan\beta$ range 
resulting in a substantially diminished
correction in the $G_F$-scheme.}.
Furthermore, in the conditions under study, 
only a small fraction of $\Delta r^{MSSM}$ is supersymmetric\,\cite{GS}, and
we should not be dependent on isolating this universal, relatively small,
part of the total SUSY correction to
$\delta$. To put in a nutshell:
if there is to be any hope to measure supersymmetric quantum effects on the
charged Higgs decay of the top quark, they should better come from the potentially
large, non-oblique, corrections computed in Section 3. The SUSY effects contained
in $\Delta r^{MSSM}$\,\cite{GS}, instead,  will be measured in a much more efficient 
way from a high precision  ($\delta M_W^{\rm exp}=\pm 40\,MeV$) determination of 
$M_W$ at LEP $200$.   
 
Another useful quantity is the branching ratio
\beq
B_H\equiv BR (t\rightarrow H^+\,b) ={\Gamma_H\over
\Gamma_W +\Gamma_H+\Gamma_{SUSY}}\,,
\label{eq:BR}
\eeq
where $\Gamma_W\equiv\Gamma (t\rightarrow W^+\,b)$ and 
$\Gamma_{SUSY}$ stands for decays of the top quark into SUSY particles.
In particular, the potentially important SUSY-QCD mode
$t\rightarrow \tilde{t}_1\,\tilde{g}$
is kinematically forbidden in most part of our analysis where 
we usually assume $m_{\tilde{g}}={\cal O} (300)\,GeV$.
There may also be the competing electroweak SUSY decays
$t\rightarrow \tilde{t}_1\,\chi^0_{\alpha}$ and
$t\rightarrow \tilde{b}_1\,\chi^+_i$ for some $\alpha=1,...,4$ and
some $i=1,2$.  The latter, however, 
is also phase space obstructed in most of
our explored parameter space, since we
typically assume $m_{\tilde{b}_1}=150\,GeV$.
The decay $t\rightarrow \tilde{t}_1\,\chi^0_{\alpha}$,
instead, is almost always open, but it is not $\tan\beta$-enhanced in our
favourite segment (\ref{eq:tanbeta2}).  
However, when studying the branching ratio (\ref{eq:BR})
as a function of the squark and gluino masses, we do include the effects from all
these supersymmetric channels whenever they are kinematically open. Thus in
general $\Gamma_{SUSY}$ on eq.(\ref{eq:BR}) is given by
\beq
\Gamma_{SUSY}=\Gamma (t\rightarrow \tilde{t}_1\,\tilde{g})+
\sum_{\alpha}\Gamma (t\rightarrow \tilde{t}_1\,\chi^0_{\alpha}) +
\sum_i\Gamma (t\rightarrow \tilde{b}_1\,\chi^+_i)\,.
\label{eq:gsusy}
\eeq  
The various terms contributing to this equation are computed at the
tree-level. Recently, the
SUSY-QCD corrections to some of these supersymmetric modes have been
evaluated and in some cases may be important\,\cite{Djuadi}.
Similarly, we treat the computation of the partial width of the standard mode 
$t\rightarrow W^+\,b$ at the tree-level. 
This is justified since, as shown in 
Refs.\cite{GJSH,DHJJS,GH,YangLi}, this decay 
cannot in general 
develop large supersymmetric radiative corrections, or at least as large as to be 
comparable to those affecting the charged Higgs mode (for the same value of the
input parameters). 
The reason for it stems from the very different
structure of the counterterms for both decays; in particular, the
standard decay mode of the top quark does not involve the mass renormalization
counterterms for the external fermion lines, and as a consequence the
aforementioned large quantum effects associated
to the bottom quark self-energy at high $\tan\beta$ are not possible.

Figures 8-20 and 22-23 display a clear-cut r\'esum\'e of our numerical results. 
We wish to point out that 
they have been thoroughly checked. Scale independence of
$\delta$, eq.(\ref{eq:pito}),
and cancellation of UV-divergences have been explicitly verified. 
Most of the
analytical and numerical calculations have been doubled. In particular,
we have constructed two independent numerical
codes and checked that the two
approaches perfectly agree at different stages.    
In all our numerical evaluations we have imposed the
following restriction on the non-SM contributions to the
$\rho$-parameter \,\cite{LangMaryland}:
\beq
\delta\rho_{\rm new}<0.003\,.
\eeq
To start with, we concentrate on the case $\mu<0$, which we study in
Figs.8-20. (The case $\mu>0$ is studied apart in Figs.22-23 and will be
commented later on.)    
We observe that, for negative $\mu$, the leading SUSY-QCD effects on
$\delta$ are positive. This means that
in these circumstances the potentially large strong supersymmetric effects
are in frank competition with the conventional 
QCD corrections, which are also very large and stay always negative
as will be discussed later on.

Needless to say, a crucial parameter to be investigated is $\tan\beta$. 
In Fig.8 we plot the tree-level width, $\Gamma_{0} (t\rightarrow H^+\,b)$, and 
the total partial width, $\Gamma_{MSSM} (t\rightarrow H^+\,b)$,
comprising all the MSSM effects, as a function of $\tan\beta$.
A typical set of parameters is chosen
well within canonical expectations (see below); the individual influence of each one
of them is tested in Figs.10 to 20.
Also shown in Fig.8 is the (tree-level) partial width of the standard
top quark decay $t\rightarrow W^+\,b$, which is (as noted above) 
far less sensitive to quantum corrections.
For convenience, we have included in Fig.8 a plot of
$\Gamma_{\rm QCD} (t\rightarrow H^+\,b)$, i.e. the partial width that would be
obtained in the presence of only the standard QCD corrections.
In practice, this is
tantamount to saying that $\Gamma_{\rm QCD} (t\rightarrow H^+\,b)$ is 
the partial width that would be expected in the absence of SUSY effects, for 
the electroweak non-supersymmetric
corrections turn out to be negligible versus the ordinary QCD effects. 

From eq.(\ref{eq:deltaalpha}) it is clear that,
for large (resp. small) $\tan\beta$, the renormalized
form factor yielding the bulk of the SUSY contribution is 
$\Lambda_L$ (resp. $\Lambda_R$).  
To appraise the relative importance of the 
various types of MSSM effects on $\Gamma (t\rightarrow H^+\,b)$, in Figs.9a-9b we
provide plots for the correction 
to the partial width, eq.(\ref{eq:pito}), and to
the branching ratio, eq.(\ref{eq:BR}), as a function of $\tan\beta$, reflecting the
various individual contributions.
Specifically, we show in Fig.9a:
\begin{itemize}
\item{(i)} The supersymmetric electroweak
contribution from genuine ($R$-odd) sparticles (denoted $\delta_{\rm SUSY-EW}$), i.e. 
from sfermions (squarks and sleptons), charginos and neutralinos;
\item{(ii)} The electroweak contribution from non-supersymmetric ($R$-even) 
particles ($\delta_{EW}$). It is composed of two distinct types
of effects, namely, those from Higgs 
and Goldstone bosons (collectively called
``Higgs'' contribution, and denoted $\delta_{\rm Higgs}$) plus
the leading  SM effects\,\cite{BSH}
from conventional fermions ($\delta_{\rm SM}$):
\beq
\delta_{EW}=\delta_{\rm Higgs}+\delta_{\rm SM}\,;
\eeq
The remaining non-supersymmetric electroweak effects
are subleading and are neglected.
\item{(iii)} The strong supersymmetric contribution
(denoted by $\delta_{\rm SUSY-QCD}$) from squarks and gluinos; 
\item{(iv)}
The strong contribution from conventional quarks
and gluons (labelled $\delta_{\rm QCD}$);
 and
\item{(v)}
The total MSSM contribution, $\delta_{\rm MSSM}$,
namely, the net sum of all the previous contributions:
\beq
\delta_{\rm MSSM}=\delta_{\rm SUSY-EW}+\delta_{EW}+
\delta_{\rm SUSY-QCD}+\delta_{\rm QCD}.
\label{eq:individ}
\eeq
\end{itemize}
In Fig.9b we reflect the impact of the MSSM on the branching ratio,  
as a function of $\tan\beta$; also shown are the tree-level value of
the branching ratio and the latter quantity after including the
(non-supersymmetric) $QCD$ corrections. 
A typical common set of inputs has been chosen
in Figs.9a-9b such that the supersymmetric electroweak
corrections reinforce the
strong supersymmetric effects (SUSY-QCD). For this set of inputs, the total MSSM
correction to the partial width of $t\rightarrow H^+\,b$ 
is positive for $\tan\beta> 20$ (approx.).
Remarkably enough, this is so in spite of
the huge negative effects induced by QCD. In fact, we see that 
the gluon effects are overridden by the gluino effects provided $\tan\beta$
is sufficiently large,
to be concrete for $\tan\beta\geq 30$. Beyond this value, the strenght of the
supersymmetric loops
becomes rapidly overwhelming; e.g. at the representative value $\tan\beta=m_t/m_b=35$ 
we find $\delta_{\rm MSSM}\simeq +27\%$; 
and at $\tan\beta\simeq 50$, which
is the preferred value claimed by $SO(10)$ Yukawa coupling unification
models\,\cite{SO10},
the correction is already $\delta_{\rm MSSM}\simeq +55\%$.
Quite in contrast, at that $\tan\beta$ one would expect, in the
absence of SUSY effects, a (QCD) correction
of about $-57\%$, i.e. virtually of the same size but opposite in sign!. 

Coming back to Fig.8, we see that,
after including the SUSY effects, the partial width of $t\rightarrow H^+\,b$
equals the partial width of the standard decay $t\rightarrow W^+\,b$
near the ``$SO(10)$'' point
$\tan\beta=50$. (The meeting point is actually a bit earlier in $\tan\beta$, 
after taking into account the known\,\cite{GJSH,DHJJS}, 
negative, SUSY corrections to
$t\rightarrow W^+\,b$, but this effect is
not shown in the figure since it is relatively small.)
Now, for the typical set of parameter values introduced
in Fig.8, the top quark decay width into SUSY particles,
eq.(\ref{eq:gsusy}), is rather tiny. Thus it is not surprising
that in these conditions the branching ratio
of the charged Higgs mode can be remarkably high:
 $BR (t\rightarrow H^+\,b)\simeq 50\%$,
i.e. basically $50\%-50\%$ versus the standard decay mode.
In contrast, the branching ratio without SUSY effects (i.e. essentially the
QCD-corrected branching ratio) is much
smaller: at the characteristic $SO(10)$
value, $\tan\beta=50$, it barely reaches $20\%$.  
Clearly, if the SUSY quantum effects are there,
they could hardly be missed!.
 
As noted before, even though the dominant MSSM effects are, by far, 
the QCD and SUSY-QCD ones, they have opposite signs.
Therefore, there is a crossover point of the
two strongly interacting dynamics, where the conventional QCD 
loops are fully counterbalanced by the SUSY-QCD loops.
This leads to a funny situation, namely,
that at the vicinity of that point the total MSSM correction is given by just the 
subleading, albeit non-negligible, electroweak supersymmetric 
contribution: $\delta_{\rm MSSM}\simeq \delta_{\rm SUSY-EW}$.
The crossover point occurs at $\tan\beta\stackM 32\simeq m_t/m_b$,
where $\delta_{\rm SUSY-EW}\stackM 20$.
For larger and larger $\tan\beta$ beyond $m_t/m_b$,
the total (and positive) MSSM correction grows very fast,
as we have said, since the
SUSY-QCD loops largely overcompensate the standard QCD corrections.
As a result, the net effect on the partial width
appears to be opposite in sign to what might
naively be ``expected'' (i.e. the QCD sign).   
Of course, this is not a general
result since it depends on the actual values of the MSSM parameters. In the
following we wish to explore the various parameter dependences and
in particular we want to assess whether a favourable
situation as the one just described is likely to happen in an ample portion of
the MSSM parameter space. 
In particular, the value $\tan\beta=m_t/m_b=35$ will be chosen
in all our plots where
that parameter must be fixed. We consider it as representative of
the low end of the  high $\tan\beta$ segment, eq.(\ref{eq:htanr}).
Thus $\tan\beta=m_t/m_b=35$ behaves as
a sort of threshold point beyond which
the MSSM quantum effects on $t\rightarrow H^+\,b$ take off so fast that
they should have indelible experimental consequences on top quark physics.

As regards to the non-supersymmetric electroweak
corrections, $\delta_{\rm EW}$, it is apparent from Fig.9a  that 
they are very small, especially 
in the high $\tan\beta$ segment. Also in the very low $\tan\beta$ 
segment, $0.5\stackm\tan\beta\stackm 1$,
$\delta_{EW}$ is relatively small; and this is so not only because both
$\delta_{\rm Higgs}$ and $\delta_{\rm SM}$ become never too large 
in absolute value, but also because in that region
there is a cancellation between $\delta_{\rm Higgs}<0$  and
$\delta_{\rm SM}>0$. 
As it happens, we end up with the fact
that the complicated Higgs effects result in a 
very tiny contribution, except in the 
very low $\tan\beta$ end, where e.g. they
can reach $-15\%$ at $\tan\beta\simeq 0.5$. In this corner
of the parameter space, $\delta_{\rm Higgs}$
becomes the dominant part of $\delta_{\rm MSSM}$,
being even larger than the QCD effects, which stay at the level of $-8\%$, and
also larger than the SUSY-QCD and SUSY-EW corrections, which remain below $+4\%$ 
and $-1\%$, respectively.

We have treated in detail the very low $\tan\beta$ segment by
including the one-loop
renormalization of the Higgs masses\,\cite{Higgsloop}. 
This is necessary in order to avoid that
the lightest CP-even Higgs mass either vanishes at $\tan\beta=1$ or becomes lighter
than the phenomenological bounds near that value. In passing, we have checked 
that the one-loop shift of the masses, as well as of the CP-even
mixing angle, $\alpha$,
has little impact on the partial width of $t\rightarrow H^+\,b$ in the
entire range of $\tan\beta$, eq.(\ref{eq:tanbeta}).
They entail 
at most an additional $5\%$ negative shift of $\delta_{\rm MSSM}$ in the very 
low $\tan\beta$ region\footnote{To perform that check, we have
included both the stop and sbottom contributions to the one-loop Higgs mass
relations. A set of $7$ independent parameters has been used
to fully characterize these effects, viz.
$(M_{H^\pm}, \mu, \tan\beta, m_{\tilde{b}_1}, m_{\tilde{t}_1}, A_b, A_t)$.
We refrain from writing out the cumbersome formulae\,\cite{Higgsloop}.}. 
It is precisely in this region where the Higgs effects 
could be expected of some relevance, and thus
where the renormalization of the  CP-even mixing angle
could have introduced some noticeable change in the neutral Higgs couplings.
Quite on the contrary, at high $\tan\beta$ the corresponding effect
is found to be of order one per mil and is thus negligibly small. 
On the other hand, a simple inspection of Figs.8 and 9b shows that
even in the very low $\tan\beta$ ballpark, where there may be some ten
percent effect from
the Higgs sector, the rising of the tree-level width is so fast that
it becomes very hard to isolate these corrections.
We conclude that, despite the rather large number of
diagrams involved, the over-all yield 
from the Higgs sector of the MSSM on $t\rightarrow H^+\,b$
is rather meagre in the whole $\tan\beta$ range (\ref{eq:tanbeta}). 
This fact is somewhat surprising and was not obvious
a priori, due to the presence of enhanced Yukawa couplings
(\ref{eq:Yukawas}) in the whole plethora of Higgs diagrams.
The cancellations involved are reminiscent of the scanty SUSY Higgs effects
obtained for the standard top quark decay $t\rightarrow W^+\,b$\,\cite{GH}.

We come now to briefly discuss the standard QCD effects up to ${\cal O}(\alpha_s)$,
which involve one-loop gluon corrections and gluon bremsstrahlung\,\cite{CD}.
As it is plain from Fig.9a, $\delta_{\rm QCD}$
is negative-definite and very important in the high $\tan\beta$
segment. It quickly saturates for $\tan\beta\stackM 10$ at a large value 
of order $-60\%$.
Therefore, the QCD effects need to be considered in order to isolate the
virtual SUSY signature\,\cite{CD}.
The leading behaviour of the standard QCD component in
the relative correction (\ref{eq:pito})  
can be easily assessed by considering the
following asymptotic formula
\beq
\delta_{QCD}=-\frac{2\, \alpha_s}{3\pi}\;\frac{\frac{8\pi^2-15}{12}
             (m^2_b\tan^2\beta+m^2_t\cot^2\beta)+
             3(4+\tan^2\beta-2\frac{M^2_{H^+}}{m^2_t}\cot^2\beta)
             m^2_b\ln\left(\frac{m^2_t}{m^2_b}\right)}
             {m^2_b\tan^2\beta+m^2_t\cot^2\beta}\,,
\label{eq:QCD1}
\eeq  
which we have obtained by expanding the exact one-loop formula up to
${\cal O}(m_b^2/m_t^2, M_{H^+}^2/m_t^2)$.
Here $\alpha_s\equiv \alpha_s(m_t^2)$, normalized as $\alpha_s(M_Z^2)\simeq 0.12$.
The big log factor 
$\ln(m^2_t/m^2_b)$ originates from the running b-quark mass evaluated 
at the top quark scale. The correction is seen to be always negative. 
We point out that while we have used the exact 
${\cal O}(\alpha_s)$ formula for the
numerical evaluation, the approximate expression given above is
sufficiently accurate to convey the general features
to be expected
both at low and at high $\tan\beta$.
In particular, for $m_b\neq 0$ and 
$\tan\beta$ in the relevant high segment (\ref{eq:tanbeta2}), 
the QCD correction becomes very large and saturates at the value
\beq
\delta_{QCD}=-\frac{2\, \alpha_s}{\pi}\;\left(\frac{8\pi^2-15}{36}
            + \ln\frac{m^2_t}{m^2_b}\right)\simeq -62\%\,\
 \ \ (\tan\beta>>\sqrt{m_t/m_b}\simeq 6)\,.
\label{eq:QCD2}
\eeq 
(The exact ${\cal O}(\alpha_s)$ formula gives slightly below $-60\%$.)
At low values of $\tan\beta$, the corrections are much smaller, as it follows from
the approximate expression 
$\delta_{\rm QCD}\simeq (-\alpha_s/\pi)(8\pi^2-15)/18\simeq -12\%$. 
We remark that for $m_b=0$ the dependence on $\tan\beta$
totally disappears from eq.(\ref{eq:QCD1}), so that 
one would never be able to suspect the large contribution (\ref{eq:QCD2}) in
the high $\tan\beta$ regime. The limit $m_b=0$, nevertheless, has been
considered for the standard QCD corrections
in some places of the literature\,\cite{LiYuan,Mehen} but,
as we have seen, it is untenable unless one concentrates on values 
of $\tan\beta$ of order $1$, in which case the 
relevance of our decay for SUSY is doomed to oblivion.
This situation is similar to the one
mentioned above concerning the SUSY-QCD corrections in the
limit $m_b=0$, 
which leads to an scenario totally blind to
the outstanding supersymmetric quantum effects obtained for 
$m_b\neq 0$ at high $\tan\beta$\,\cite{GJS}.
We stress that in spite of the respectable size of the standard QCD effects,
they become fast stuck at the saturation value (\ref{eq:QCD2}),
which is independent of $\tan\beta$. On the contrary, the
SUSY-QCD effects grow endlessly  with $\tan\beta$ and thus rapidly overtake 
the standard QCD prediction.

Worth noticing is the evolution of the quantities (\ref{eq:pito}) and (\ref{eq:BR})
as a function of the gluino mass (Cf. Figs.10a-10b). Of course, only
the SUSY-QCD component is sensitive to $m_{\tilde{g}}$.
Although the SUSY-QCD effects have been object of
a particular study in Ref.\cite{GJS},
we find it convenient, to ease comparison, to display the corresponding 
results in the very same
conditions in which the electroweak supersymmetric corrections are presented.
The steep falls in Fig.10a are associated to the presence of
threshold effects occurring at points satisfying
$m_{\tilde{g}}+m_{\tilde{t}_1}\simeq m_t$. 
An analogous situation was observed in Ref.\cite{GJSH,DHJJS} for the SUSY
corrections to
the standard top-quark decay. Away from the threshold points, the behaviour of
$\delta_{SUSY-QCD}$ is smooth and perfectly consistent with perturbation theory.
In Fig.10b, where the branching ratio (\ref{eq:BR}) is plotted, the steep falls
at the threshold points are no longer present since they are compensated for by the
simultaneous opening of the two-body supersymmetric mode
 $t\rightarrow \tilde{t}_1\,\tilde{g}$, for $m_{\tilde{g}}<m_t-m_{\tilde{t}_1}$.

We emphasize that the relevant gluino mass region for the decay $t\rightarrow H^+\,b$ is 
not the light gluino region, but the heavy one, the reason being that the
important self-energy correction mentioned above, eq.(\ref{eq:dmbQCD}), 
involves a  gluino mass insertion. 
As a consequence, virtually for any set of MSSM parameters, there is a well sustained
SUSY correction for any gluino mass above a certain value, in our case
$m_{\tilde{g}}\stackM 250-300\,GeV$.  
The correction raises with the gluino mass up to a long flat maximum  before
bending --very gently -- into the decoupling regime (far beyond $1\,TeV$). 
The fact that the decoupling rate of the gluinos appears to be 
so slow has an obvious phenomenological interest.

Next we consider in detail the sensitivity of our decay
on the higgsino-gaugino parameters $(\mu,M)$ characterizing 
the chargino-neutralino
mass matrices (Cf. Appendix A).
We start with the supersymmetric Higgs mixing 
mass, $\mu$. As already stated above, we will largely concentrate on
the $\mu<0$ case. 
Together with $A_t>0$ this yields $A_t\,\mu<0$,
which is a sufficient condition\,\cite{Ng} for the MSSM prediction on
$BR(b\rightarrow s\,\gamma)$ to be compatible with
experiment in the presence of a relatively light
charged Higgs boson  (as the one participating in the
top decay under study). In fact, it is known that charged Higgs
bosons of ${\cal O}(100)\,GeV$ interfere constructively with the
SM amplitude and would render
a final value of $BR(b\rightarrow s\,\gamma)$ exceedingly high. 
Fortunately, this situation
can be remedied in the MSSM since the alternative contribution from charginos
and stops tends to cancel the Higgs contribution provided that
$A_t\,\mu<0$. Furthermore, one must also
require relatively light values for the masses of the
lightest representatives of these sparticles, as well
as high values of $\tan\beta$\,\cite{Ng}; hence one is led to
a set of conditions 
which fit in with nicely to build up a favourable scenario
for the decay $t\rightarrow H^+\,b$. 

The evolution of the individual
contributions (\ref{eq:individ}), together with the total MSSM yield,
as a function of $\mu<0$, is shown
in Figs.11a-11b for given values of the other parameters.
We immediately gather from these figures
that the SUSY-QCD correction is extremely
sensitive to $\mu$. In fact, $\delta_{\rm SUSY-QCD}$ grows rather fast with $|\mu|$.
This is already patent at the level of the leading $\delta m_b/m_b$ 
effect given by eq.(\ref{eq:dmbQCD}).
In all figures where a definite $\mu<0$ is to be
chosen, we have taken the moderate value $\mu=-150\,GeV$. In this way, 
for $M\simeq |\mu|=150\,GeV$, we guarantee that the lightest chargino mass is above
the LEP $1.5$ phenomenological bound: $m_{\chi^{\pm}_1}\stackM 65\,GeV$\,\cite{Moriond}.
Concerning the electroweak contribution,
we noted above that the component $\delta m_b/m_b$, eq.(\ref{eq:dmbEW}),
actually decreases with $\mu$. However,
the $\mu$ dependences in the full $\delta_{\rm SUSY-EW}$ are more complicated than
in $\delta_{\rm SUSY-QCD}$ and cannot be read off
 eq.(\ref{eq:dmbEW}).
This is evident from  Fig.11a where the total  $\delta_{\rm SUSY-EW}$
is fairly insensitive to $\mu$; $\delta_{\rm MSSM}$, therefore, inherits
its marked $\mu$-dependence basically from the SUSY-QCD component.
As for the sensitivity of the corrections on the $SU(2)_L$-gaugino soft
SUSY-breaking parameter, $M$, Fig.12 conveys
immediately that it is virtually non-existent.  
We point out that the choice of $M$ and $\mu$
is always made in a range where the chargino 
masses are above $65\,GeV$.

There is some slight evolution of the corrections with $A_b$ (Fig.13a),
mainly on the SUSY-QCD side.
We realize that $\delta_{\rm SUSY-QCD}$ is not perfectly
symmetric with respect to the sign of $A_b$. 
Once the sign $\mu<0$ is chosen, the correction is 
larger for negative 
values of $A_b$ than for positive values. We have erred on the 
conservative side by choosing $A_b=+300\,GeV$ wherever this parameter
is fixed.
As far as $A_t$ is concerned, $\delta_{\rm SUSY-QCD}$ can only evolve as
a function of that parameter through vertex corrections,
which are proportional to $A_t\,\cot\beta$ (Cf. Appendix A); however, at large
$\tan\beta$ these are very depressed. The electroweak correction
$\delta_{\rm SUSY-EW}$, instead, is very much dependent on $A_t$; indeed,
a typical component exhibiting this behaviour 
is given by eq.(\ref{eq:dmbEW}),
which is linear in $A_t$. The full dependence, however, is not linear and is
recorded in Fig.13b. We realize that $\delta_{\rm SUSY-EW}$ and $\delta_{\rm MSSM}$
change sign with $A_t$. The shaded vertical band in Fig.13b is excluded by our
choice of parameters in Fig.8.

Another very crucial parameter to be investigated is the value of
$m_{\tilde{b}_1}$. 
This is because the SUSY-QCD correction hinges a great deal on the value of the
sbottom masses, as it is plain from eq.(\ref{eq:dmbQCD}).
As  a matter of fact, a too large a value of $m_{\tilde{b}_1}$  may upside down
the leadership of the SUSY-QCD effects.
As a phenomenological lower bound for all the squark masses we take
the absolute LEP $1.5$ bound
$m_{\tilde{q}}\geq 65\,GeV$\,\cite{Moriond}. 
However, as a typical mass value for all squarks other than the stop we use
$m_{\tilde{q}}\geq 150-200\,GeV (\tilde{q}\neq \tilde{t})$.
From Fig.14a we see that provided
$m_{\tilde{b}_1}\stackm 300\,GeV$ the SUSY-QCD effects remain dominant, but
they steadily go down the larger is $m_{\tilde{b}_1}$. The
electroweak correction $\delta_{SUSY-EW}$, on the other hand, is quite sustained 
with increasing $m_{\tilde{b}_1}$ and there are parameter configurations where for
sufficiently heavy sbottoms the supersymmetric electroweak effects are larger
than the SUSY-QCD effects. However, this is not the most likely situation.
The behaviour of the branching ratio is plotted in Fig.14b.

Obviously, the evolution of the SUSY-QCD
corrections with the stop masses is basically flat (Fig.15) since the leading
contribution is independent of $m_{\tilde{t}_1}$. Therefore, it is of little help
to use the strict lower mass bound $m_{\tilde{t}_1}\simeq 65\,GeV$ in 
our calculation instead of, say, the 
more conservative $m_{\tilde{t}_1}\simeq 100\,GeV$.
Nonetheless, if we wish to keep $\delta_{\rm MSSM}>0$,
we cannot go too far with  $m_{\tilde{t}_1}$,
for the electroweak correction is also seen to 
decrease with $m_{\tilde{t}_1}$.
Indeed, whereas for $m_{\tilde{t}_1}=65-100\,GeV$ one has
$\delta_{\rm SUSY-EW}\stackM 20\%$, for
 $m_{\tilde{t}_1}\stackM 250\,GeV$ one finds 
$\delta_{\rm SUSY-EW}\stackm 10\%$. For heavier stop masses, $\delta_{\rm MSSM}$
becomes zero or slightly negative. In this situation, the imprint of SUSY
lies in the fact that the total quantum effect is not as negative as predicted
by standard QCD, eq.(\ref{eq:QCD2}).
 
The influence from the sleptons and the other squarks is practically irrelevant
as it is borne out by Figs.16a-16b. 
They enter the correction through oblique (universal) quantum
corrections. The only exception are the $\tau$-sleptons $\tilde{\tau}_a$ 
(``staus''), since they are involved in the process-dependent (non-oblique) 
contribution eq.(\ref{eq:deltatau}), where the $\tau$-lepton Yukawa coupling becomes
enhanced at large $\tan\beta$.
For this reason, $\delta_{\rm SUSY-EW}$ in Fig.16b is
somewhat larger the smaller is the
$\tau$-sneutrino mass (assumed to be degenerate with the other sneutrinos).
In all our calculation we have fixed the common
sneutrino mass at $200\,GeV$.

We have also tested the variation of our results as a function of the
external particle masses, namely the masses of the
top quark, bottom quark and charged Higgs. 
As for the external fermion masses,
the corrections themselves are not very sensitive (see Figs.17a and 18a). 
Among the SUSY corrections, the most sensitive
one on $m_t$ (respectively on $m_b$) is $\delta_{\rm SUSY-EW}$
(resp. $\delta_{\rm SUSY-QCD}$).
The branching ratios also show 
some dependence (Figs.17b and 18b), especially on $m_b$.
This effect is mainly due to the variation of 
the tree-level partial widths
as a function of $m_t$ and $m_b$.
As for the charged Higgs mass, $M_{H^+}$, up to now it has been
fixed at $M_{H^+}=120\,GeV$. 
We confirm from  Fig.19a that
there is nothing special in the chosen value for that parameter since the
sensitivity of the correction is generally low, except near the
uninteresting boundary of the phase space where the 
branching ratio (Fig.19b) boils down to zero. 

We close our study of the corrections in the $\mu<0$ case
by plotting $\delta_{\tau}$ as a function of $\tan\beta$
(see Fig.20). By definition,  $\delta_{\tau}$ is that part of
$\delta_{\rm MSSM}$ originating from
the full process-dependent term $\Delta_{\tau}$, eq.(\ref{eq:deltatau}),
which stems from our  definition of $\tan\beta$ on eq.(\ref{eq:tbetainput}).
This piece of information is relevant enough.
In fact, it should be recalled that the quantum corrections described in the previous
figures are scheme dependent. In particular, they rely on our definition of
$\tan\beta$ given on eq.(\ref{eq:tbetainput}). What is {\it not} scheme
dependent, of course, is the predicted
value of the width and branching ratio (Figs.8 and 9b)
after including all the radiative corrections.
Now, from Fig.20 it is clear that 
the  $\Delta_{\tau}$-term is not negligible, 
and so there is a process-dependence in our
definition of $\tan\beta$, as it was announced in Section 3. 
At first sight, the $\delta_{\tau}$-effects are not
dramatic since they are small as compared to
$\delta_{\rm SUSY-QCD}$, but since the latter is cancelled out by
standard QCD we end up with 
$\delta_{\tau}$ being of the order (roughly half the size) of the 
electroweak correction $\delta_{\rm SUSY-EW}$.

The main source of process-dependent $\delta_{\tau}$-effects 
lies in the corrections generated by the $\tau$-mass counterterm,
$\delta m_{\tau}/ m_{\tau}$, and can be easily picked out in the 
electroweak-eigenstate basis (see Fig.21) much in the same way as we did for the
$b$-mass counterterm. There are, however, some differences, as can be
appraised by comparing the diagrams in Figs.7 and 21, where we see that
in the latter case the effect derives from diagrams
involving $\tau$-sleptons with gauginos or mixed gaugino-higgsinos.  
An explicit computation
of the diagrams (a) $+$ (b) in Fig.21 yields
\beqn
{\delta m_{\tau}\over m_{\tau}} &=&
{g'^2\over 16\pi^2}\,\mu\,M'\,\tan\beta\,
I(m_{\tilde{\tau}_1},m_{\tilde{\tau}_2},M')\nonumber\\
& + & {g^2\over 16\pi^2}\,\mu\,M\,\tan\beta\,
I(\mu,m_{\tilde{\nu}_{\tau}},M)\,,
\label{eq:dmtau12}
\eeqn
where $g'=g\,s_W/c_W$ and $M', M$ (Cf. Appendix A) are the soft SUSY-breaking
Majorana masses
associated to the bino $\tilde B$ and winos $\tilde{W}^{\pm}$, respectively,
and the function $I(m_1,m_2,m_3)$ is again given by eq.(\ref{eq:I123}).
In the formula above
we have projected, from the bino diagram in Fig.21a, only the leading piece which is
proportional to $\tan\beta$.
Even so, the contribution from the wino-higgsino diagram in Fig.21b is much larger. 
Numerical evaluation of the sum of the two contributions on
eq.(\ref{eq:dmtau12}) indeed shows that it reproduces to within
few percent the full numerical result (Cf. Fig.20) previously obtained in
the mass-eigenstate basis, thus 
confirming that eq.(\ref{eq:dmtau12}) gives the leading contribution. 
In practice, for a typical choice of parameters as in Fig.8,
this contribution is approximately cancelled out by part of the electroweak
supersymmetric corrections associated to the original process $t\rightarrow H^+\,b$,
and one is effectively left with eq.(\ref{eq:dmbEW}) as being the main source   
of electroweak supersymmetric quantum effects at high $\tan\beta$. 

Finally, the corrections corresponding to the case where $\mu>0$ are 
studied in Figs.22a and 22b.  
The problem with $\mu>0$ is that, then, the large SUSY-QCD corrections
have the same (negative) sign as the conventional QCD effects, and as
a consequence the total MSSM correction can easily blow up above $100\%$,
the branching ratio becoming negative!. 
To avoid this disaster (from the point of view of perturbation
theory), we enforce the SUSY-QCD correction to be smaller
than in the $\mu<0$ case
by assuming an ``obesse SUSY scenario''
characterized by  very large values for the
sbottom mass ($m_{\tilde{b}_1}=600\,GeV$) and the
gluino mass ($m_{\tilde{g}}=1000\,GeV$).
We also choose $A_t<0$ so that the electroweak SUSY 
correction becomes opposite
in sign to the SUSY-QCD correction (a feature that also applies in the
$\mu<0$ case, see Fig.13b) and in this way the total SUSY correction is
further lessened in absolute value. Incidentally, we
remark that the simultaneous sign change of both
$\mu$ and $A_t$ is also necessary in order to
keep  $A_t\,\mu<0$; as noted above, this is required in order
that the MSSM can be
compatible with $BR(b\rightarrow s\,\gamma)$ in the presence of
a relatively light charged Higgs.
In Fig.23 we bring forward the
effect of the new situation on the total partial width. 
In the present instance, the physical signature would
be to measure a  partial width significantly
smaller than the one predicted by QCD.
Clearly, the $\mu>0$ ($A_t\,\mu<0$) scenario is not
as appealing as the $\mu<0$ ($A_t\,\mu<0$) one.

In the end, from the explicit numerical analysis, we have confirmed
our expectations that the SUSY-QCD contribution
to $\Gamma (t\rightarrow H^+\,b)$ is generally dominant.
This conclusion would not hold only in some (unlikely) cases, e.g.
if the gluino is very light
and/or the lightest bottom squark is ``obesse'' as compared to
lightest top squark, i.e. if the former is
unusually much heavier than expected. 
Furthermore, by restricting ourselves to the case $\mu<0$ ($A_t\,\mu<0$)
we confirm that at large $\tan\beta$ and for typical values of the parameters
the total (standard plus supersymmetric) QCD correction largely 
cancels out, leaving a remainder
on the SUSY-QCD side (Figs.8-9).
In all circumstances the virtual Higgs effects remain comparatively very small.   
Around $\tan\beta=m_t/m_b\simeq 35$,
one is left with basically
the electroweak supersymmetric correction, $\delta_{SUSY-EW}$,
which can be sizeable enough to be pinned down by experiment.
However, as stated above, there is in general
a strong remainder, $\delta_{SUSY-QCD}+\delta_{QCD}>0$,
which grows very fast with $\tan\beta$ and it has the same sign as 
$\delta_{SUSY-EW}$. 
In this favourable scenario, 
the virtual SUSY effects could be spectacular.
This is true not only because in the
relevant window of parameter space the 
SUSY quantum corrections are by themselves rather
large, but also because they push into the opposite direction than the 
``expected'' standard QCD corrections. 
As a result, the relative deviation
between the MSSM prediction and the QCD prediction effectively
``doubles'' the size of the observable effect, a fact which is
definitely welcome from the experimental point of view. 

From all the previous discussion there is one fact
standing out which can be 
hardly overemphasized: If the charged Higgs decay mode of the top quark,
$t\rightarrow H^+\,b$, does show up  
with a branching ratio of order $10\%$ or above (perhaps even as
big as $50\%$), a fairly rich event 
statistics will be collected at the Tevatron
and especially at the LHC e.g. by making use
of the identification methods described in Section 2.  
If, in addition, it comes out that the dynamics underlying that decay is truly
supersymmetric, then the valuable quantum signatures that our calculation has
unveiled over an ample portion of the MSSM parameter
space should eventually become manifest and, for sure, we could not miss them. 

At present all the collected event statistics basically relies on
our experimental ability to recognize the top quark decays originating from
standard patterns (angular distribution, energy spectrum, 
jet topology etc.) associated to
the usual Drell-Yan production mechanism. 
Notwithstanding, we wish to point out that it should in principle be possible to  
clutch at the supersymmetric virtual corrections associated to the
vertex $t\,b\,H^{\pm}$ also
through an accurate measurement of the various 
inclusive top quark and Higgs boson production cross
sections in hadron colliders. As an example, in Fig.24 we sketch a few
alternative mechanisms which
would generate top quark production patterns  
heavily hinging
on the properties of the interaction $t\,b\,H^{\pm}$-vertex\,\cite{RicardToni}.
Thus, while this vertex could be 
responsible in part for the decay
of the top quark once it is produced,   
it might as well be at the root of the production process itself
at LHC energies, where it could take over from 
Drell-Yan production\,\cite{CPYuan}. 

We observe that
in some of these mechanisms a Higgs boson is produced in
association, but in some others (fusion processes)
the Higgs boson enters as a virtual particle.
Now, however different these production processes might be,
all of them are sensitive to the effective structure
of the $t\,b\,H^{\pm}$-vertex. Similar mechanisms can of course be depicted
involving the neutral Higgs bosons of the MSSM interacting with $t\,\bar{t}$ and
$b\,\bar{b}$  via enhanced Yukawa couplings\,\cite{RicardToni}. 
While it goes beyond the scope of this paper
to compute the SUSY corrections to the production processes
themselves, we have at least faced the detailed analysis of a partial
decay width which involves one of the
relevant production vertices. 
In this way, a definite prediction is made on the
properties of a physical observable and, moreover,
this should suffice both
to exhibit the relevance of the SUSY quantum effects
and to  demonstrate the necessity to incorporate these corrections 
in a future, truly 
comprehensive, analysis of the cross-sections, namely, an analysis where
one would include the quantum effects on all the relevant
production mechanisms within the framework 
of the MSSM.  
For this reason we think that   
in the future a precise measurement of the various (single and double) top
quark production cross-sections\,\cite{Willenbrock,Heinson} 
should be able to detect or to exclude
the $t\,b\,H^{\pm}$-vertex as well as the vertices $q\,\bar{q}\, A^0 (h^0,H^0)$
involving the neutral Higgs particles
of the MSSM and the third generation quarks $q=t,b$.

We conclude our discussion with the following remark.
Whereas, on the one hand, one expects that some top quark partial widths
will be determined with an accuracy
of $10\%$ at the upgraded Tevatron and perhaps better than $10\%$ 
at LHC\,\cite{CPYuan}, on the other hand we believe that
from the point of view of an {\sl inclusive} model-independent
measurement of  
the {\sl total} top-quark width, $\Gamma_t$, the future $e^+\,e^-$ supercollider
should be a better suited machine\,\cite{Fujii,Frey}.
For, in an inclusive measurement,
all possible non-SM effects will appear on top
of the corresponding SM effects already computed in the
literature\,\cite{TopSM}. Moreover, 
as shown in Ref.\cite{Fujii}, one hopes to be able to measure the total top-quark
width in $e^+\,e^-$ supercolliders at an unmatched precision
of $\sim 4\%$ on the basis of a
detailed analysis of the threshold effects in the cross-section, in particular
of the top momentum distribution and the
resonance contributions to the forward-backward
asymmetry in the $t\bar{t}$ threshold region. Under the assumption that
$\Gamma_H\simeq \Gamma_W$,  and that the
SUSY effects on $\Gamma_t$ are purely virtual effects, 
it follows that a large SUSY correction of, say $50\%$, to $t\rightarrow H^+\,b$ 
translates into a $20\%$ correction to $\Gamma_t$. 
This effect could not scape detection.
Thus, the combined information
from a future $e^+\,e^-$ supercollider
and from present and medium term hadron machines
can be extremely useful to pin down the nature of the observed effects.
Our general conclusion is, therefore, extremely encouraging:
In view of the potentially large size and large variety of manifestations,
quantum effects on top quark and Higgs boson
physics could be the clue to the discovery of ``virtual'' Supersymmetry.

\vspace{0.5cm}

{\bf Acknowledgements}:

One of us (JS) is thankful to W. Hollik for useful conversations on the
renormalization framework used in our work and its relation to other schemes
based on alternative definitions of $\tan\beta$. 
He is also very grateful to M. Carena and C. Wagner for pointing out interesting
connections of our work with SUSY GUT's and for enjoyable discussions.
Thanks are extended to G.L. Kane and H. Baer for remarks on SUSY
phenomenology at the Tevatron, and to F. Matorras and D.P. Roy for 
helpful information on $\tau$ physics at the LHC.
Private talks on the experimental aspects of present and future top quark
physics at the Tevatron and at the LHC with
A. Heinson, S. Willenbrock and with the experimental colleagues
in our Institute, especially with M. Bosman, M. Cavalli-Sforza
and M. Mart\'\i nez, are gratefully acknowledged. 
Finally, JS wishes to thank the hospitality
and financial support provided by the CERN Theory Division where this work was
finished. 
This work has also  been partially supported by CICYT 
under project No. AEN93-0474. The work of DG and JG has been financed by 
grants of the Comissionat per a Universitats i Recerca, Generalitat de 
Catalunya.


\newcommand{\titulo}[1]{{
\bf #1}}
\appendix
\setcounter{equation}{0}
\renewcommand{\theequation}{A.\arabic{equation}}

\vspace{0.75cm} 
\begin{Large}
{\bf Appendix A.}
\end{Large}
\begin{large}
{\bf  Interaction Lagrangian of the MSSM in the 
mass-eigenstate basis}
\end{large}
\vspace{0.5cm}

In this appendix we single out specific pieces of the MSSM Interaction
Lagrangian involved in the decay rate of the process $t\rightarrow H^+\,b$.
Although the Lagrangian of the MSSM is well-known\,\cite{MSSM}, it is
always useful to project explicitly the relevant pieces and to cast them in
a most suitable form for specific purposes.  As a matter of fact, we have
produced a complete set of Feynman rules for the MSSM using an algebraic
computer code based on MATHEMATICA\,\cite{Mathematica}\footnote{We have
  corrected several mistakes in the subset of rules presented in
  Ref.\cite{Hunter} involving sparticles and Higgses.}.  Here we limit
ourselves to quote the Lagrangian interactions affecting the
process-dependent parts of our decay and omit the interaction pieces needed
to compute the universal counterterm structures, whose SM parts are
well-known\,\cite{BSH} and the corresponding SUSY contributions are also
available in the literature since long time ago\,\cite{GrifSol}.  All our
interactions are expressed in the mass-eigenstate basis.

Within the context of the MSSM we need two Higgs superfield doublets
\begin{equation}
 \hat{H}_1= \left(\begin{array}{c}
\hat{H}_1^0 \\ \hat{H}_1^{-}
\end{array} \right)
 \;\; \;\;\;,\;\;\;\;\;
 \hat{H}_2= \left(\begin{array}{c}
\hat{H}_2^{+} \\ \hat{H}_2^0 
\end{array} \right)\,,  
\end{equation}
with weak hypercharges $Y_{1,2}=\mp 1$. The (neutral components of the)
corresponding scalar Higgs doublets give mass to the down (up) -like
quarks through the VEV $<H^0_1>=v_1$ ($<H^0_2>=v_2$). 
This is seen from the structure of the MSSM superpotential\,\cite{MSSM}
\begin{equation}
\hat{W}=\epsilon_{ij}[h_b\,\hat{H}_1^i\hat{Q}^j\hat{D}+
h_t\hat{H}_2^j\hat{Q}^i\hat{U}-\mu\hat{H}_1^i\hat{H}_2^j]\,,
\label{eq:W}
\end{equation}
where we have singled out only the Yukawa couplings of
the third quark-squark generation, $(t,b)-(\tilde{t},\tilde{b})$, as a 
generical fermion-sfermion generation of chiral matter superfields
$\hat{Q}$, $\hat{U}$ and $\hat{D}$. 
Their respective scalar (squark) components are: 
\begin{equation}
\tilde{Q}=\left(\begin{array}{c}
\tilde{t'}_L \\ \tilde{b'}_L
\end{array} \right) \; \;\; ,\;\;\;\tilde{U}=\tilde{t'}_R^{*}
\; \;\; ,\;\;\;\tilde{D}=\tilde{b'}_R^{*}\,,
\label{eq:QUD} 
\end{equation}
with weak hypercharges $Y_Q=+1/3$, $Y_U=-4/3$ and $Y_D=+2/3$. 
The primes in (\ref{eq:QUD}) denote the fact that
$\tilde{q'}_a=\{\tilde{q'}_L, \tilde{q'}_R\}$ are weak-eigenstates, not 
mass-eigenstates.
The ratio
\begin{equation}
  \tan\beta={v_2\over v_1}\,,
  \label{eq:v2v1}
\end{equation}
is a most relevant parameter throughout our analysis.

We briefly describe the necessary SUSY formalism entering 
our computations:
\begin{itemize}
\item The fermionic partners of the weak-eigenstate gauge bosons and
  Higgs bosons, called gauginos, $\tilde{B}$, $\tilde{W}$, and
  higgsinos, $\tilde{H}$, respectively. From them we construct the
  fermionic mass-eigenstates, the so-called charginos and neutralinos,
  by, first, forming the following three sets of two-component Weyl
  spinors:
  \begin{equation}
    \Gamma_i^{+} = \{-i\tilde{W}^{+}, \tilde{H_2^{+}}\}\ , \;\;\;
    \Gamma_i^{-} = \{-i\tilde{W}^{-}, \tilde{H_1^{-}}\}\ , \nonumber
  \end{equation}
  \begin{equation}
    \Gamma_{\alpha}^{0} =
    \{-i\tilde{B^{0}}, -i\tilde{W_3^{0}}, \tilde{H_2^0}, \tilde{H_1^0}\}\ , 
    \nonumber  
  \end{equation}
  which get mixed up when the neutral Higgs fields acquire nonvanishing
  VEV's and then diagonalizing the resulting ``ino'' mass Lagrangian
 \begin{eqnarray}
  {\cal L}_M&=&-<\Gamma^{+}|
              \pmatrix{
                  M                       &  \sqrt{2} M_W \cbtt \cr
                  \sqrt{2} M_W \sbtt &  \mu 
              }
              |\Gamma^{-}>\nonumber\\
              &-&{1\over2}
              <\Gamma^0|
              \pmatrix{
M'               & 0                 & M_Z\sbtt s_W & -M_Z\cbtt s_W\cr
0                & M                 &-M_Z\sbtt c_W & M_Z\cbtt c_W \cr
M_Z\sbtt s_W & -M_Z\sbtt c_W & 0                & -\mu\cr
-M_Z\cbtt s_W& M_Z\cbtt c_W  & -\mu             &0
              }
              |\Gamma^0>\nonumber\\
              &+&h.c.\ ,
              \label{eq:CNMM}
  \end{eqnarray}
  where we remark the presence of the parameter $\mu$ introduced above
  and of the soft SUSY-breaking Majorana masses $M$ and $M'$, usually
  related as $M'/ M=(5/3)\,\tan^2{\theta_{W}}$, and where $\cbtt=\cbt$ 
  and $\sbtt=\sbt$.  The corresponding
  mass-eigenstates\footnote{We use the following notation:
    first Latin indices a,b,...=1,2 are reserved for sfermions,
    middle Latin indices i,j,...=1,2 for charginos, and first Greek
    indices $\alpha, \beta,...=1,\ldots ,4$ for neutralinos. } 
  (charginos and neutralinos) are the following:
  \begin{equation}
    \Psi_i^{+}= \pmatrix{
                U_{ij}\Gamma_j^{+} \cr V_{ij}^{*}\bar{\Gamma}_j^{-}
                }
              \; \;\; \;\;,\;\;\;\;\;
              \Psi_i^{-}= C\bar{\Psi_i}^{-T} =\pmatrix{
                  V_{ij}\Gamma^{-}_j \cr U_{ij}^{*}\bar{\Gamma}_j^{+} 
                }\ ,  
              \label{eq:cinos}
              \nonumber
  \end{equation}
  and
  \begin{equation}
    \Psi_{\alpha}^0= \pmatrix{
                N_{\alpha\beta}\Gamma_{\beta}^0 \cr 
                N_{\alpha\beta}^{*}\bar{\Gamma}_{\beta}^0
                } =  
              C\bar{\Psi}_{\alpha}^{0T}\ ,
              \label{eq:ninos} 
              \nonumber
  \end{equation}
  where the matrices $U,V,N$ are defined through
  \begin{equation}
    U^{*}{\cal M}V^{\dagger}=diag\{M_1,M_2\}\;\;\;\;\;,\;\;\;\;\;
    N^{*}{\cal M}^0N^{\dagger}=diag\{M_1^0,\ldots, M_4^0\}\ .  \nonumber
\label{eq:UVN}
  \end{equation}    
In practice, we have performed the calculation with real matrices
 $U$, $V$ and $N$, so we have been using unphysical mass-eigenstates
(associated to non-positively definite chargino-neutralino masses).
The transition from our unphysical mass-eigenstate basis
$\{\Psi\}\equiv\{\Psi_i^{\pm}, \Psi_\alpha^0\}$ into the physical
mass-eigenstate basis 
$\{\chi\}\equiv\{\chi_i^{\pm}, \chi_\alpha^0\}$ can be done by
 introducing a set of
$\epsilon$ parameters as follows: for every chargino-neutralino $\Psi$ 
whose mass matrix eigenvalue is $M_i, M_{\alpha}$
the proper physical state, $\chi$, is given by
\beq
\chi=\left\{\ba{ll}
\Psi&\,\ {\rm if}\ \ \epsilon=1\\
\pm\gamma_5\,\Psi &\,\ {\rm if}\ \ \epsilon=-1\,,
\ea\right.
\eeq
and the physical masses for charginos and neutralinos are 
$m_{\chi^{\pm}_i}=\epsilon\,M_i$ and $m_{\chi_{\alpha}^0}=\epsilon\,M_{\alpha}^0$,
respectively. Needless to say, 
in this real formalism one is supposed to
propagate the $\epsilon$ parameters accordingly 
in all the relevant couplings, as shown 
in detail in Ref.\,\cite{GuaschS}. 
This procedure is
entirely equivalent\,\cite{Gunion}
to use complex diagonalization matrices insuring that physical
states are characterized by a set of positive-definite mass eigenvalues; 
and for this reason we have
maintained the complex notation in all our formulae in Section 4.
Whereas for computations with real sparticles the
distinction matters\,\cite{GuaschS}, for virtual sparticles 
the $\epsilon$ parameters cancel out, and so one could use either basis
$\{\Psi\}$ or $\{\chi\}$ without the inclusion of the
$\epsilon$ coefficients.
We have stressed here the differences between the two bases just to make clear
what are the physical 
chargino-neutralino states, when they are referred to in the text.   

  Among the gauginos we also have the strongly interacting gluinos, 
  $\sg^r$ $(r=1,\ldots,8)$,
  which are the fermionic partners of the gluons.
\item As for the scalar partners of quarks and leptons,
they are called squarks,
  $\tilde{q}$, and sleptons, $\tilde{l}$, respectively.
Again we will use the third quark-squark 
generation $(t,b)-(\tilde{t},\tilde{b})$ as a
   generical fermion-sfermion generation.  The squark
  mass-eigenstates,
  $\tilde{q}_a=\{\tilde{q}_1, \tilde{q}_2\}$, if we neglect
  intergenerational mixing, are obtained from the weak-eigenstate ones
  $\tilde{q'}_a=\{\tilde{q'}_1\equiv \tilde{q}_L,\,\, \tilde{q'}_2\equiv
  \tilde{q}_R\}$, through 
\beqn
 \tilde{q'}_a&=&\sum_{b}
     R_{ab}^{(q)}\tilde{q}_b,\nonumber\\ R^{(q)}&
     =&\pmatrix{ \cos{\theta_q} & -\sin{\theta_q} \cr 
         \sin{\theta_q} & \cos{\theta_q}}
\;\;\;\;\;\;
       (q=t, b)\,.
\label{eq:rotation}
\eeqn
 The rotation matrices in
    (\ref{eq:rotation}) diagonalize the corresponding stop and sbottom
    mass matrices:
  \begin{equation}
    {\cal M}_{\tilde{q}}^2 =\pmatrix{
      M_{\tilde{q}_L}^2+m_q^2+\cos{2\beta}(\TqL-\Qq s_W^2) M_Z^2 & 
      m_q M_{LR}^q\cr 
      m_q M_{LR}^q &
      M_{\tilde{q}_R}^2+m_q^2+ \cos{2\beta}\,\Qq\,s_W^2\, M_Z^2
      }\,,
    \label{eq:stopmatrix}
    \label{eq:sbottommatrix}
    \end{equation}
\beq
R^{(q)\dagger} {\cal M}_{\tilde{q}}^2 R^{(q)}=
      diag\{m_{\tilde{q}_2}^2,m_{\tilde{q}_1}^2\}
\ \ \ \ \ (m_{\tilde{q}_2}\geq m_{\tilde{q}_1})\,, 
\eeq    
    with $\TqL$ the third component of weak isospin, $Q$ the electric
    charge, and
 $M_{{\tilde{q}}_{L,R}}$ the soft SUSY-breaking squark masses\,\cite{MSSM}. (By
    $SU(2)_L$-gauge invariance, we must have
    $M_{\tilde{t}_L}=M_{\tilde{b}_L}$, whereas $M_{{\tilde{t}}_R}$,
    $M_{{\tilde{b}}_R}$ are in general independent parameters.)
The mixing angle on eq.(\ref{eq:rotation}) is given by
\beq
 \tan2{\theta_q} = {2\,m_q\,M_{LR}^q\over
M_{\tilde{q}_L}^2-M_{\tilde{q}_R}^2+\cos{2\beta}(\TqL-2\Qq s_W^2) M_Z^2}\,,
       \label{eq:thetarot}
\eeq
where
\beq
 M_{LR}^t=A_t-\mu\cot\beta\,,\ \ \ \ \  M_{LR}^b=A_b-\mu\tan\beta\,,
\eeq
are, respectively, the stop and sbottom off-diagonal mixing terms
on eq.(\ref{eq:sbottommatrix}). Furthermore,
$\mu$ is the SUSY Higgs mass parameter in the superpotential, and $A_{t,b}$
are the trilinear soft SUSY-breaking parameters. We shall assume that
$|A_{t,b}|<3 M_{\tilde{Q}}$, where $M_{\tilde{Q}}$
is the average soft SUSY-breaking mass
appearing in the mass matrix (\ref{eq:stopmatrix}); this relation
roughly corresponds to the necessary, though not sufficient, condition for
the absence of colour-breaking minima\,\cite{Frere}.

The charged slepton mass-eigenstates can be obtained in a similar way
after straightforward substitutions in the mass matrices, with the only 
proviso that there is no ${\tilde\nu}_R$, so that  ${\tilde\nu}_L$ is 
itself the sneutrino
mass-eigenstate, hence $R^{(\tilde\nu)}_{ab}=0$ unless $a=b=1$
where $R^{(\tilde\nu)}_{11}=1$.
  \end{itemize}

Next let us describe the relevant pieces of the MSSM interaction Lagrangian
involving the fields defined above.

\begin{itemize}
\item{\bf fermion--sfermion--(chargino or neutralino)}

After translating the 
allowed quark-squark-higgsino/gaugino interactions into the
mass-eigenstate basis, one finds 
\begin{eqnarray}
  \label{LcqsqLR}
  {\cal L}_{\Psi q\tilde{q}}&=&
  g\,\sum_{a=1,2}\sum_{i=1,2}\left( 
    -\sta^*\cbim\left(\Apit\PL+\Amit\PR\right)\,b
    -\sba^*\cbip\left(\Apib\PL+\Amib\PR\right)\,t 
  \right) \nonumber\\
   +&\displaystyle{\frac{g}{\sqrt{2}}}&
    \sum_{a=1,2}\sum_{\alpha=1,\ldots ,4}\left(
    -\sta^*\nba\left(\Apat\PL+\Amat\PR\right)\,t
    +\sba^*\nba\left(\Apab\PL+\Amab\PR\right)\,b
  \right) \nonumber\\
   +&\mbox{\rm h.c.}&\ 
\end{eqnarray}
where $\Apmit,\ \Apmib,\ \Apmat,\ \Apmab$ are
\begin{equation}
\begin{array}{lll}
  \label{V1Apm}
  \Apit &=& \Rotc\Uo^*-\lt\Rttc\Ut^*\, ,\\
  \Amit &=& -\lb\Rotc\Vt\, ,\\
  \Apat &=& \Rotc\left(\Nt^*+\YL\tth\No^*\right)
                       +\sqrt{2}\lt\Rttc\Nth^*\, ,\\
  \Amat &=& \sqrt{2}\lt\Rotc\Nth
                       -\YRt\tth\Rttc\No\, ,\\
  \Apib &=& \Robc\Vo^*-\lb\Rtbc\Vt^*\, ,\\
  \Amib &=& -\lt\Robc\Ut\, ,\\
  \Apab &=& \Robc\left(\Nt^*-\YL\tth\No^*\right)
                       -\sqrt{2}\lb\Rtbc\Nf^*\, ,\\
  \Amab &=& -\sqrt{2}\lb\Robc\Nf
                       +\YRb\tth\Rtbc\No\, .
\end{array}
\end{equation}
with $\YL$ and $Y_R^{t,b}$ the weak hypercharges of the left-handed $SU(2)_L$
doublet and right-handed singlet fermion, and $\lt$ and 
$\lb$ are -- Cf. eq.(\ref{eq:Yukawas}) -- the potentially significant Yukawa
couplings normalized to the $SU(2)_L$ gauge coupling constant $g$.

\item{\bf quark--squark--gluino}
\begin{equation}
{\cal L}_{\sg q\tilde{q}}=- \frac{g_s}{\sqrt{2}}
\tilde q_{a,k}^{ *}\, \bar\sg_r 
\left(\lambda^r\right)_{kl} \left(R_{1a}^{(q)*} \PL - R_{2a}^{(q)*} \PR \right) q_l
+\mbox{ h.c.}
\label{eq:Lqsqglui}\end{equation}
where $\lambda^r$ are the Gell-Mann matrices. This is just the SUSY-QCD Lagrangian
written in the squark mass-eigenstate basis.

\item{\bf squark--squark--Higgs}
\begin{equation}
  \label{LHsqsq}
  {\cal L}_{H^\pm\tilde{q}\tilde{q}}=\frac{g}{\sqrt{2}\mw}
  H^-\sba^*G_{ab}\,\stb\ +\ 
  \mbox{\rm h.c.} 
\end{equation}
where we have introduced the coupling matrix
\begin{eqnarray}
  G_{ab}&=&R_{cb}^{(t)}R_{da}^{*(b)}\gccdd\label{Gdef}\nonumber\\
  \gccdd&=&\pmatrix{
    \mbs\tb+\mts\ctb-\mws\sin 2\beta & 
    \mb\left(\mu+A_b\tb\right) \cr
    \mt\left(\mu+A_t\ctb\right) & \mt\mb\left(\tb+\ctb\right)
    }\,.  \label{gdef}
\end{eqnarray}
\item{\bf chargino--neutralino--charged Higgs}
\begin{equation}
  \label{LHcn}
  {\cal L}_{H^\pm\Psi^\mp\Psi^0}=
  -g\,H^-\nba\left(\HiaRc\PL+\HiaLc\PR\right)\cip\ +\
  \mbox{\rm h.c.} 
\end{equation}
with
\begin{equation}
\left\{
\begin{array}{lcl}
\QaiL&=&U_{i1}^*N_{\alpha 3}^*+
\frac{1}{\sqrt{2}}\left(N_{\alpha 2}^*+\tth N_{\alpha 1}^*\right)U_{i2}^*\\
\QaiR&=&V_{i1}N_{\alpha 4}-
\frac{1}{\sqrt{2}}\left(N_{\alpha 2}+\tth N_{\alpha 1}\right)V_{i2}\,.
\end{array}
\right.
\label{eq:QLQR}
\end{equation}

\item{\bf Gauge interactions}

In our calculation we only need the sparticle interactions with the $W^\pm$:

{\bf -quarks}
  \begin{equation}
    {\cal L}_{W^\pm q q}=\frac{g}{\sqrt{2}}\, \bar t \gamma^\mu \PL b\,
    W_\mu^++\mbox{ h.c.} \label{eq:Lqqw}
    \label{eq:Wq}
  \end{equation}
{\bf -squarks}
  \begin{equation}
    {\cal L}_{W^\pm\tilde{q}\tilde{q}}=i \frac{g}{\sqrt{2}} R_{1a}^{(t)*}
    R_{1b}^{(b)} W^+_\mu \tilde{t}_a^{ *}\stackrel{\leftrightarrow}
    {\partial^\mu}\tilde{b}_b+\mbox{ h.c.}
    \label{eq:LsqsqW}
  \end{equation}
{\bf -charginos and neutralinos}
  \begin{equation}
    {\cal L}_{W^\pm\Psi^\mp\Psi^0}=g\, \bar\Psi_\alpha^0 \gamma^\mu
    \left(C_{\alpha i}^L \epsilon_\alpha \epsilon_i\PL + C_{\alpha i}^R \PR
    \right)\Psi_i^+\, W^-_\mu+\mbox{ h.c.}
    \label{eq:WCN}
  \end{equation}
  \begin{equation}
    \left\{
      \begin{array}{lcl}
        C_{\alpha i}^L&=&\frac{1}{\sqrt{2}}N_{\alpha 3}U_{i 2}^*-N_{\alpha
          2}U_{i 1}^*\\ C_{\alpha i}^R&=&-\frac{1}{\sqrt{2}}N_{\alpha 4}^*
        V_{i 2}-N_{\alpha 2}^* V_{i 1}\,.  \end{array}\right.
  \end{equation}
\end{itemize} 


\vspace{0.75cm} 
\begin{Large}
{\bf Appendix B.}
\end{Large}
\begin{large}
{\bf The Counterterm $\delta Z_{HW}$ }
\end{large}
\vspace{0.5cm}
\setcounter{equation}{0}
\renewcommand{\theequation}{B.\arabic{equation}}

We discuss the structure of $\delta Z_{HW}$  both in
the unitary gauge and in the Feynman gauge.
In order to renormalize the mixed self-energy $\Sigma ^{HW}(k^2)$, 
a mixing counterterm  must be
introduced  in the wave function of $W^\pm_\mu$:
\begin{equation}
W^\pm_\mu \rightarrow {\left(Z_2^W\right)}^{1/2}
 W^\pm_\mu \pm i \frac{\delta Z_{H
W}}{M_W}\partial_\mu H^\pm\,.
\label{eq:contrados}\end{equation}

{\bf B.1 Unitary gauge}

If one chooses to work in the unitary gauge, then the
renormalization is straightforward:
\begin{equation}
L_{UG}=-\frac{1}{4}F_{\mu\nu}F^{\mu\nu}+M_W^2 W^+_\mu W{^-}^\mu \rightarrow
L_{ct}=i M_W \delta Z_{H W} (W^-_\mu
 \partial^\mu H^+ - W^+_\mu \partial^\mu H^-)\,.
\label{eq:lct}
\end{equation}
In this gauge the corresponding renormalized 2-point Green function reads
(Fig.25a):
\begin{equation}
\frac{i}{k^2-M_{H^\pm}^2} \left[k^\nu\frac{-i\Sigma_{HW} (k^2)}{M_W}
+ ik^\nu~M_W^2~\frac{\delta Z_{HW}}{M_W}\right]
\frac{-i\left(g_{\mu\nu}-\frac{k_\mu k_\nu}{M_W^2}\right) }{k^2-M_W^2}\,. 
\label{eq:green}\end{equation}
Thus, a renormalized self-energy can be defined as follows:
\begin{equation}
\hat\Sigma_{HW}(k^2)=\Sigma_{HW}(k^2)- M_W^2 \delta Z_{HW}
\label{eq:sigmahat}\end{equation}
Now we must impose a renormalization condition on $\hat\Sigma ^{HW}(k^2)$; 
and we
choose it in a way that the physical Higgs does not mix with the physical
$W^\pm$:
\begin{equation}
\hat \Sigma_{HW} (M_{H^\pm}^2) =0 \Longrightarrow
\delta Z_{HW}=\frac{\Sigma_{HW}(M_{H^\pm}^2)}{M_W^2}\,.
\label{eq:renorm}
\end{equation}

{\bf B.2 Feynman gauge}

As we carry out our calculation in the Feynman gauge, we would also
like to perform
the renormalization of the Higgs sector in that gauge. The Lagrangian
is sketched as follows\,\cite{BSH}:
\begin{equation}
L=L_{C}+L_{GF}+L_{FP}\,.
\end{equation}
where $L_{GF}$ stands for the gauge-fixing term in that gauge,
\beq
L_{GF}=-F^{+}\,F^{-}+...\,\ \ \ \ \ (F^{\pm}\equiv \partial^{\mu}W^+_{\mu}\mp iM_W\,G^+)
\label{eq:GF}
\eeq
and $L_{FP}=\bar{\eta}^a\,\left(\partial F^a/\partial\theta^b\right)\,\eta^b$ 
is the Faddeev-Popov ghost Lagrangian constructed from FP and 
anti-FP Grassmann scalar fields
$\eta, \bar{\eta}$.  Since we are interested in the
charged gauge-Higgs ($W^{\pm}-H^{\pm}$)
and charged Goldstone-Higgs ($G^{\pm}-H^{\pm}$) mixing terms in that gauge, 
we have singled out just the relevant term on eq.(\ref{eq:GF}). 

The relationship with the original weak-eigenstate
 components on eq.(A1) is the following: 
 \beq  
  \pmatrix{-H_1^+\cr H_2^+}
   =\pmatrix{ c_{\beta} & -s_{\beta} \cr 
         s_{\beta} & c_{\beta}
  } \pmatrix{G^+\cr H^+}\,.   
  \label{eq:mixGH}
 \eeq    
As is well-known, although the classical Lagrangian, $L_{C}$, also contains
a nonvanishing mixing among the weak gauge boson
fields, $W^{\pm}$, and the Goldstone boson fields, $G^{\pm}$, namely
\beq
{\cal L}_{GW}=i M_W\,W_\mu^- \partial^\mu G^+ +{\rm h.c.}\,,
\eeq
the latter is cancelled (in the action) by a piece 
contained in $L_{GF}$.
Now, after substituting the renormalization transformation 
for the Higgs doublets,
eq.(\ref{eq:ZH1H2}), on the Higgs boson kinetic term with
$SU(2)_L\times U(1)_Y$ gauge covariant
derivative, one easily projects out the following relevant counterterms
\beqn
\delta {\cal L}&= & \delta Z_{H^\pm}\,\partial_\mu H^+ \partial^\mu H^{-}
+ \delta Z_{G^\pm}\,\partial_\mu G^+ \partial^\mu G^-\nonumber\\
&+& \delta Z_{HG}\,\left(\partial_\mu H^- \partial^\mu G^+
+ {\rm h.c.}\right) 
+\delta Z_{HW}\,\left(i M_W\,W_\mu^-\partial^\mu H^+
+ {\rm h.c.} \right)+... 
\label{eq:lclas}
\eeqn
where
\beq
\pmatrix{\delta Z_{H^{\pm}}\cr \delta Z_{G^{\pm}} }
   =\pmatrix{ c_{\beta}^2 & s_{\beta}^2 \cr 
         s_{\beta}^2 & c_{\beta}^2}
   \pmatrix{\delta Z_{H_1}\cr \delta Z_{H_2} }\,,
\label{eq:deltaZs1}
\eeq
and
\beqn   
\delta Z_{HG} &=& s_{\beta}\,c_{\beta}\,\left(\delta Z_{H_2}
-\delta Z_{H_1}\right)\,,\nonumber\\
\delta Z_{HW} &=& s_{\beta}\,c_{\beta}\,\left[{1\over 2}\left(\delta Z_{H_2}
-\delta Z_{H_1}\right)+{\delta\tan\beta\over\tan\beta}\right]\,.
\label{eq:deltaZs2}
\eeqn
The renormalization transformation for the VEV's of the Higgs potential
(\ref{eq:potential}),
\beqn
v_i\rightarrow Z_{H_i}^{1/2} (v_i+\delta v_i)
=\left(1+{\delta v_i\over v_i}+{1\over 2}\,\delta Z_{H_i}\right)\,v_i\,,
\label{eq:renvevs}
\eeqn
implies that the counterterm to $\tan\beta$
is related to the fundamental
counterterms in the Higgs potential by
\beq
{\delta\tan\beta\over\tan\beta}={\delta v_2\over v_2}-{\delta v_1\over v_1}
+{1\over 2}\,\left(\delta Z_{H_2}-\delta Z_{H_1}\right)\,.
\label{eq:cdeltatan}
\eeq
If one imposes the usual on-shell renormalization conditions for the
$A^0$-boson, one has
\beq
\delta Z_{H_2}-\delta Z_{H_1}=
-{\tan\beta+\cot\beta\over \,M_Z^2}\,\Sigma_{AZ} (M_{A^0}^2)\,.
\eeq
In addition, there is another mixing term between $H^{\pm}$ and $G^{\pm}$ 
originating from the mass matrix of the Higgs sector~\cite{CPRD}. 
This one loop mixture is contained in:
\begin{equation}
V^b= \left(\begin{array}{cc} {H^+}^b{G^+}^b
\end{array}\right)
\left(\begin{array}{cc} M_{H^\pm}^{b\,2} & \frac{t_{0}^b}{\sqrt{2} v^b} \\
\frac{t_{0}^b}{\sqrt{2} v^b}
& \frac{t_{1}^b}{\sqrt{2} v^b}
\end{array}\right)
\left(\begin{array}{c} {H^-}^b \\ {G^-}^b
\end{array}\right)\,,
\end{equation}
where we have attached a superscript ${\rm b}$ to bare quantities,
and $t_i$ are the tadpole counterterms
\begin{equation}
\begin{array}{lcl}
t_{0}&=&-\sin\left(\beta-\alpha\right)~t_{H^0}
+\cos\left(\beta-\alpha\right)~t_{h^0}\\
t_{1}&=&\sin\left(\beta-\alpha\right)~t_{h^0}
+\cos\left(\beta-\alpha\right)~t_{H^0}\,.
\end{array}
\end{equation}
We are now ready to find an expression for the mixed 2-point Green functions
(Figs.25a and 25b). As the mixing Lagrangian
between $H^{\pm}$ and $W^{\pm}$ on eq.~(\ref{eq:lct}) is the same as on
eq.~(\ref{eq:lclas}), the $2$-point Green function will have the same
expression~(\ref{eq:green}) but with the $W^\pm$ propagator
in the Feynman gauge, namely:
\beqn
& &\frac{i}{k^2-M_{H^\pm}^2} \left[k^\nu\frac{-i\Sigma_{HW} (k^2)}{M_W} 
+ ik^\nu~M_W^2~\frac{\delta Z_{HW}}{M_W}\right] 
\frac{-i g_{\mu\nu} }{k^2-M_W^2}\nonumber\\
& & = \frac{i}{k^2-M_{H^\pm}^2}\,\,\left[k_{\mu}{-i\hat{\Sigma}_{HW}(k^2)
\over M_W}\right]\,\,\frac{-i}{k^2-M_W^2} 
\equiv \Delta^{HW}_{\mu}\,.
\eeqn
Next we impose $\hat\Sigma_{HW} (M_{H^\pm}^2) =0$ as before, 
leading to the same formal expression for $\delta Z_{HW}$
as in eq.(\ref{eq:renorm}). However, as a new ingredient, we now have the 
mixed $H^{\pm}-G^{\pm}$ self-energy:
\begin{equation}
\frac{i}{k^2-M_{H^\pm}^2}\left(-i \Sigma_{HG}(k^2)+i k^2 \delta Z_{HG} - i
\frac{t_0^b}{\sqrt{2} v^b}\right) \frac{i}{k^2-M_W^2}\,.
\end{equation}
This allows us to define renormalized self-energies,~(\ref{eq:sigmahat}) and
\begin{equation}
\hat \Sigma_{HG}(k^2)=\Sigma_{HG}(k^2) - k^2 \delta Z_{HG} 
+ \frac{t_0^b}{\sqrt{2} v^b}\,.
\end{equation}
The mixed self-energies $\hat\Sigma_{HW}(k^2)$ and $\hat\Sigma_{HG}(k^2)$  
obey the following  Slavnov-Taylor identity:
\begin{equation}
 k^2 \hat\Sigma_{HW}(k^2)-M_W^2 \hat\Sigma_{HG}(k^2) = 0\,.
\label{eq:STI}
\end{equation}
This identity is derived from a BRS  
transformation involving the Green function constructed with an anti-FP field and
the charged Higgs field: $<0|\delta_{BRS}\,(\bar{\eta}^+\,H^+)|0>=0$.
Following the standard procedure\,\cite{Collins} one
immediately gets:
\beq
<0|F^{+}\,H^+|0>=<0|\partial^{\mu}W^-_{\mu}+ H^+ - iM_W\,G^+H^+|0>=0\,,
\eeq  
which in momentum space reads
\beq
k^\mu\,\Delta^{HW}_{\mu}+M_W\,\Delta^{HG}=0
\label{eq:MSSTI}
\eeq
with
\beq
\Delta^{HG}\equiv \frac{i}{k^2-M_{H^\pm}^2}\,\,
\left[-i\,\hat{\Sigma}_{HG}(k^2)
\right]\,\,\frac{i}{k^2-M_W^2} 
\,.
\eeq
Clearly, eq.(\ref{eq:MSSTI}) implies eq.(\ref{eq:STI}). The latter identity guarantees
that the contribution to our decay $t\rightarrow H^+\,b$ from the two diagrams in
Figs. 25a and 25b vanishes since no
mixing is generated among the physical boson $H^{\pm}$ and the
renormalized fields  $G^{\pm}$ and  $W^{\pm}$:
 $\hat{\Sigma}_{HG}(M_{H^{\pm}}^2)=\hat{\Sigma}_{HW}(M_{H^{\pm}}^2)=0$.
There is of course another Slavnov-Taylor identity, derived
in a similar manner, which insures that the renormalized
mixing between $G^{\pm}$ and $W^{\pm}$ also vanishes.
Thus we have proven that the expression for $\delta Z_{HW}$ is 
formally the
same in both unitary and Feynman gauges, but that in the latter gauge one
must take into account the additional renormalization of
the mixed self-energy $\Sigma_{HG}$.


\vspace{0.75cm} 
\begin{large}
{\bf Appendix C. Vertex functions}
\end{large}
\vspace{0.5cm}
\setcounter{equation}{0}
\renewcommand{\theequation}{C.\arabic{equation}}

In this appendix we briefly collect, for notational 
convenience, the basic vertex functions frequently referred 
to in the text. Dimensional regularization is used throughout. 
The given 
formulas are exact for arbitrary internal masses and external 
on-shell momenta. Most of them are an adaptation to the 
$g_{\mu\nu}=\{+\,-\,-\,-\}$ metric of the standard formulae 
of ref.\cite{Veltmann}. The basic one, two and 
three-point scalar functions are:
\begin{eqnarray}
    A_{0}(m)&=& \int d^{n}\tilde{q}\, \frac{1}{[q^2-m^2]},\\
    B_{0}(p,m_1,m_2)&=&\int d^{n}\tilde{q}\, \frac{1}{[q^2-m_1^2]\,
    [(q+p)^2-m_2^2]}\,,
\end{eqnarray}
\begin{equation}
    C_{0}(p,k,m_1,m_2,m_3)=\int d^{n}\tilde{q}\, 
    \frac{1}{[q^2-m_1^2]\,[(q+p)^2-m_2^2]\,[(q+p+k)^2-m_3^2]}\,;
\end{equation}
using the integration measure
\begin{equation}
d^n\tilde{q}\equiv {\mu }^{(4-n)}\frac{d^nq}{(2\pi )^n}\,.
\end{equation}
The two and three-point tensor functions needed for our calculation 
are the following
\begin{equation}
    [B_{\mu },B_{\mu \nu }](p,m_1,m_2)=\int d^{n}\tilde{q}\, 
    \frac{[q_{\mu },q_{\mu }q_{\nu}]}{[q^2-m_1^2]\,[(q+p)^2-m_2^2]}\,,
\end{equation}
\begin{eqnarray}
     \lefteqn{[\tilde{C}_0,C_{\mu },C_{\mu \nu }](p,k,m_1,m_2,m_3)=} 
     \nonumber \\
               & &\int d^{n}\tilde{q}\, \frac{[q^2,q_{\mu},q_{\mu} 
               q_{\nu}]}{[q^2-m_1^2]\,[(q+p)^2-m_2^2]\,[(q+p+k)^2-m_3^2]}\,.
\end{eqnarray}
By Lorentz covariance, they can be decomposed in terms of the 
above basic scalar functions and the external momenta:
\begin{eqnarray}
    B_{\mu }(p,m_1,m_2)    &=& p_{\mu } B_1(p,m_1,m_2).\nonumber \\
    B_{\mu \nu }(p,m_1,m_2)&=& p_{\mu } p_{\nu }B_{21}(p,m_1,m_2)+
                            g_{\mu \nu }B_{22}(p,m_1,m_2). \nonumber \\
    \tilde{C}_0(p,k,m_1,m_2,m_3) &=& B_0(k,m_2,m_3)+
                                  m_1^2 C_{0}(p,k,m_1,m_2,m_3). \nonumber\\
    C_{\mu \nu }(p,k,m_1,m_2,m_3)&=& p_{\mu }p_{\nu}C_{21}+
                                  k_{\mu }k_{\nu}C_{22}+
                                  (p_{\mu}k_{\nu}+k_{\mu}p_{\nu})C_{23}+
                                  g_{\mu \nu }C_{24}\,,
\end{eqnarray}
where we have defined the Lorentz invariant functions:
\begin{eqnarray}
B_1(p,m_1,m_2)&=&\frac{1}{2p^2}[A_0(m_1)-A_0(m_2)-
    f_1B_0(p,m_1,m_2)], \\
B_{21}(p,m_1,m_2)&=&\frac{1}{2p^2(n-1)}[(n-2)A_0(m_2)
            -2m_1^2B_0(p,m_1,m_2) \nonumber \\
	    &-& nf_1B_1(p,m_1,m_2)],\\
B_{22}(p,m_1,m_2)&=&\frac{1}{2(n-1)}[A_0(m_2)\!+\!2m_1^2B_0(p,m_1,m_2)
            \!+\!f_1B_1(p,m_1,m_2)],
\end{eqnarray}
\begin{equation}
     \left(\begin{array}{c}
       C_{11}\\C_{12}
     \end{array}\right)=Y
     \left(\begin{array}{c}
       B_0(p+k,m_1,m_3)-B_0(k,m_2,m_3)-f_1C_0 \\ 
       B_0(p,m_1,m_2)-B_0(p+k,m_1,m_3)-f_2C_0
     \end{array}\right)\,,
\end{equation}
\begin{equation}
     \left(\begin{array}{c}
       C_{21}\\C_{23}
     \end{array}\right)=Y 
     \left(\begin{array}{c}
       B_1(p+k,m_1,m_3)+B_0(k,m_2,m_3)-f_1C_{11}-2C_{24} \\ 
       B_1(p,m_1,m_2)-B_1(p+k,m_1,m_3)-f_2C_{11}
     \end{array}\right)\,,
\end{equation}
\begin{equation}
     \begin{array}{l}
       C_{22}=\displaystyle{\frac{1}{2[p^2k^2-(pk)^2]}} 
              \{-pk[B_1(p+k,m_1,m_3)- 
              B_1(k,m_2,m_3)-f_1C_{12}]  \\
              +p^2[-B_1(p+k,m_1,m_3)-f_2C_{12}-2C_{24}]\}\,,
     \end{array}
\end{equation}
\begin{equation}
    C_{24}=\frac{1}{2(n-2)} [B_0(k,m_2,m_3)+
           2m_1^2C_0+f_1C_{11}+f_2C_{12}]\,,\\
\end{equation}
the factors $f_{1,2}$ and the matrix $Y$,
\begin{eqnarray*}
   f_1 &=& p^2+m_1^2-m_2^2, \\
   f_2 &=& k^2+2pk+m_2^2-m_3^2\,,
\end{eqnarray*}
\begin{equation}
   Y=\frac{1}{2[p^2k^2-(pk)^2]} 
   \left(\begin{array}{cc}
        k^2 & -pk \\-pk & p^2
   \end{array}\right)\,.
\end{equation}
The UV divergences for $n\rightarrow 4$ can be parametrized as
\begin{eqnarray}
    \epsilon &=& n-4,\nonumber \\
      \Delta &=& \frac{2}{\epsilon }+
                 {\gamma }_{\scriptscriptstyle E} -\ln (4\pi)\,,
\end{eqnarray}
being ${\gamma }_{\scriptscriptstyle E}$ the Euler constant.
In the end one is left with the evaluation of the scalar one-loop
functions:
\begin{eqnarray}
    A_0(m) &=& \left(\frac{-i}{16{\pi }^2} \right) 
           m^2(\Delta-1+\ln \frac{m^2}{\mu ^2})\,, \\
    B_0(p,m_1,m_2) &=& \left(\frac{-i}{16{\pi }^2} \right) 
           \left[\Delta+\ln \frac{p^2}{{\mu }^2}-2 
           +\ln [(x_1-1)(x_2-1)] \right.\nonumber \\
           & & \left . +x_1\ln \frac{x_1}{x_1-1}+
                        x_2\ln \frac{x_2}{x_2-1}\right], \\
    C_0(p,k,m_1,m_2,m_3) &=& \left(\frac{-i}{16{\pi }^2} \right) 
           \frac{1}{2}\;\frac{1}{pk+p^2\xi }\Sigma\, 
\end{eqnarray}
with
\begin{eqnarray}
    x_{1,2}=x_{1,2}(p,m_1,m_2) &=& \frac{1}{2}+\frac{m_1^2-m_2^2}{2p^2}
           \pm \frac{1}{2p^2}\lambda^{1/2}(p^2,m^2_1,m^2_2), \\
           \lambda (x,y,z) &=& [x-(\sqrt{y}-\sqrt{z})^2]
                               [x-(\sqrt{y}+\sqrt{z})^2]\,, \nonumber
\end{eqnarray}
and where $\Sigma$ is a bookkeeping device for the following alternate 
sum of twelve (complex) Spence functions:
\begin{eqnarray}
    \lefteqn{  
    \Sigma =
      Sp\left(\frac{y_1}{y_1-z_{1}^{i}}\right)-Sp\left(\frac{y_1-1}
      {y_1-z_{1}^{i}}\right)+
      Sp\left(\frac{y_1}{y_1-z_{2}^{i}}\right)-Sp\left(\frac{y_1-1}
      {y_1-z_{2}^{i}}\right) }
\nonumber \\
  & & -Sp\left(\frac{y_2}{y_2-z_{1}^{ii}}\right)+Sp\left(\frac{y_2-1}
      {y_2-z_{1}^{ii}}\right)-
      Sp\left(\frac{y_2}{y_2-z_{2}^{ii}}\right)+Sp\left(\frac{y_2-1}
      {y_2-z_{2}^{ii}}\right)
\nonumber \\
  & & Sp\left(\frac{y_3}{y_3-z_{1}^{iii}}\right)-Sp\left(\frac{y_3-1}
      {y_3-z_{1}^{iii}}\right)+
      Sp\left(\frac{y_3}{y_3-z_{2}^{iii}}\right)-Sp\left(\frac{y_3-1}
      {y_3-z_{2}^{iii}}\right).
\end{eqnarray}
The Spence function is defined as
\begin{equation}
     Sp(z)=-\int_{0}^{1}\frac{\ln (1-zt)}{t}\,dt\,,
\end{equation}
and we have set, on one hand:
\begin{eqnarray}
     z_{1,2}^{i}   & = & x_{1,2}(p,m_2,m_1)\,,   \nonumber \\
     z_{1,2}^{ii}  & = & x_{1,2}(p+k,m_3,m_1)\,, \nonumber \\
     z_{1,2}^{iii} & = & x_{1,2}(k,m_3,m_2)\,;
\end{eqnarray}
and on the other:
\begin{equation}
     y_1=y_0+\xi\,,\;\;\; 
     y_2=\frac{y_0}{1-\xi }\,, \;\;\;\; 
     y_3=-\frac{y_0}{\xi }\,, \;\;\;\; 
     y_0=-\frac{1}{2}\;\frac{g+h\xi }{pk+p^2\xi }\,, \;\;\;\;
\end{equation}
where
\begin{equation}
\begin{array}{ll}
     g=-k^2+m_2^2-m_3^2\,,\;\;\; & h=-p^2-2pk-m_2^2+m_1^2\,, 
\end{array}
\end{equation}
and $\xi $ is a root (always real for external on-shell momenta) of 
\begin{equation}
     p^2\xi^2+2pk\xi +k^2=0\,.
\end{equation}

Derivatives of some 2-point functions are also needed in the 
calculation of selfenergies, and we use the
following notation:
\begin{equation}
     \frac{\partial}{\partial p^2}B_{*}(p,m_1,m_2)\equiv 
                             B^{\prime}_{*}(p,m_1,m_2). 
\end{equation}
We can obtain all the derivatives from the basic $B^{\prime}_0$:
\begin{eqnarray}
  B^{\prime}_0(p,m_1,m_2) 
   &=&      \left( \frac{-i}{16\pi^2} \right)
            \left\{    \frac{1}{p^2} +
            \frac{1}{\lambda^{1/2}(p^2,m^2_1,m^2_2)}\cdot
          \right . \nonumber \\   
   & &    \left . 
            \cdot\left[   x_1(x_1-1)\ln \left( \frac{x_1-1}{x_1} \right)
                         -x_2(x_2-1)\ln \left( \frac{x_2-1}{x_2}
                         \right) 
\right]\right\}\,,
\end{eqnarray}
which has a threshold for $|p|=m_1+m_2$ and a pseudo-threshold
for $|p|=|m_1-m_2|$.





\newpage
\vspace{1cm}
\begin{center}
\begin{Large}
{\bf Figure Captions}
\end{Large}
\end{center}
\begin{itemize}
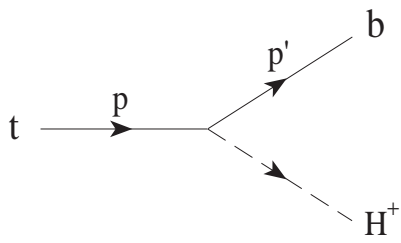
\item{\bf Fig.1} The lowest-order Feynman diagram for the charged Higgs
decay of the top quark. 

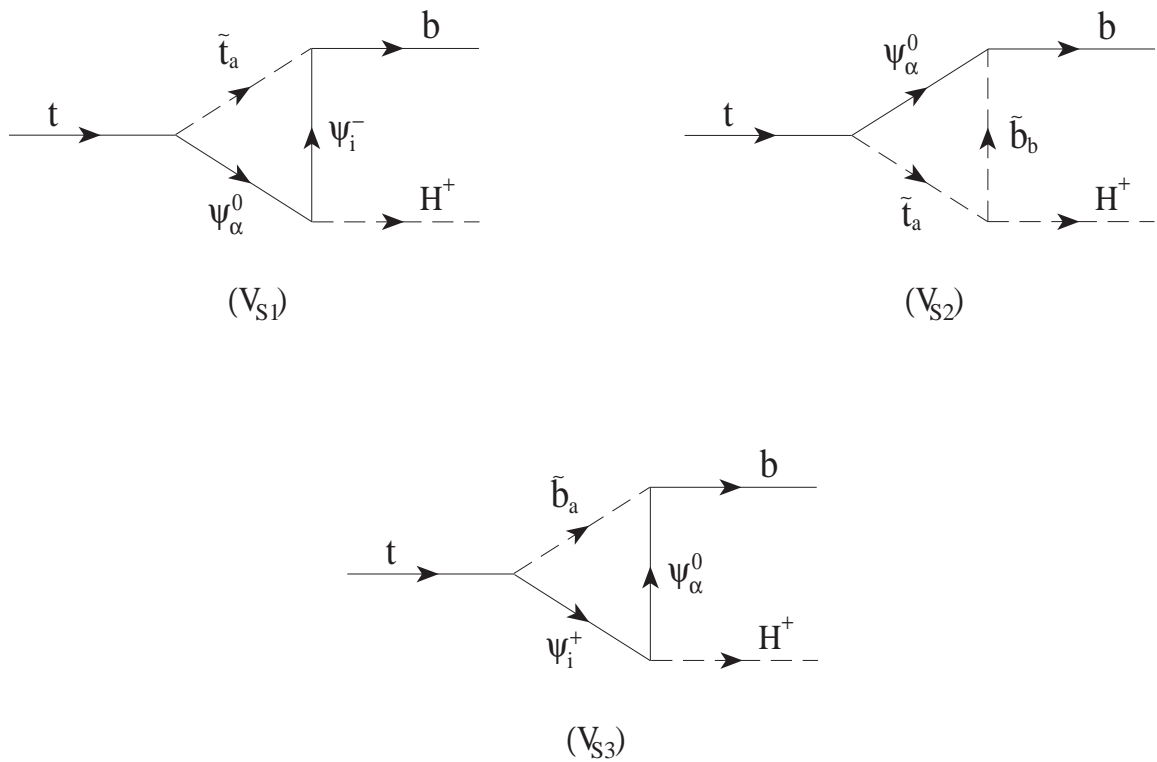
\item{\bf Fig.2} Feynman diagrams, up to one-loop order, for the
electroweak SUSY
vertex corrections to the decay process $t\rightarrow H^{+}\,b$.
Each one-loop diagram is summed over all possible values of the mass-eigenstate
charginos ($\Psi^{\pm}_i\,; i=1,2$), neutralinos
($\Psi^{0}_{\alpha}\,; \alpha=1,...,4$), stop and sbottom squarks
($\tilde{b}_a, \tilde{t}_b\,; a,b=1,2$). 

\item{\bf Fig.3} Feynman diagrams, up to one-loop order, for the 
Higgs and Goldstone boson vertex corrections to the
decay process $t\rightarrow H^{+}\,b$.

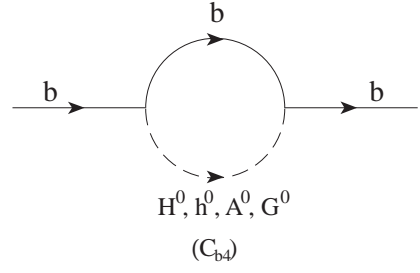
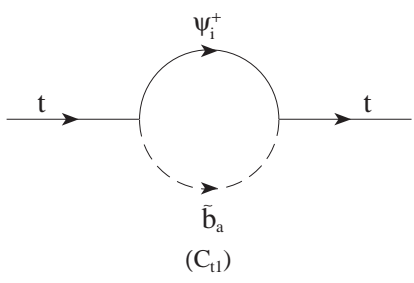
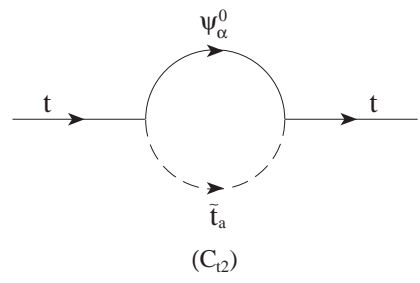
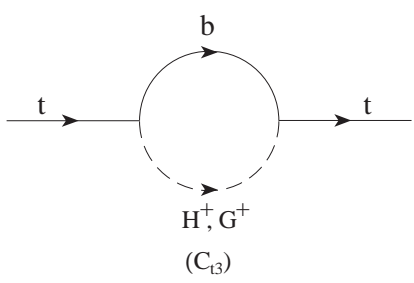
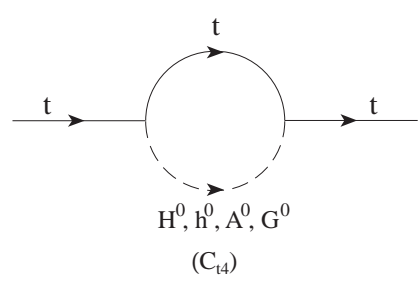
\item{\bf Fig.4} Electroweak self-energy corrections to the top and 
bottom quark external lines from the various
supersymmetric particles, Higgs and Goldstone bosons. 

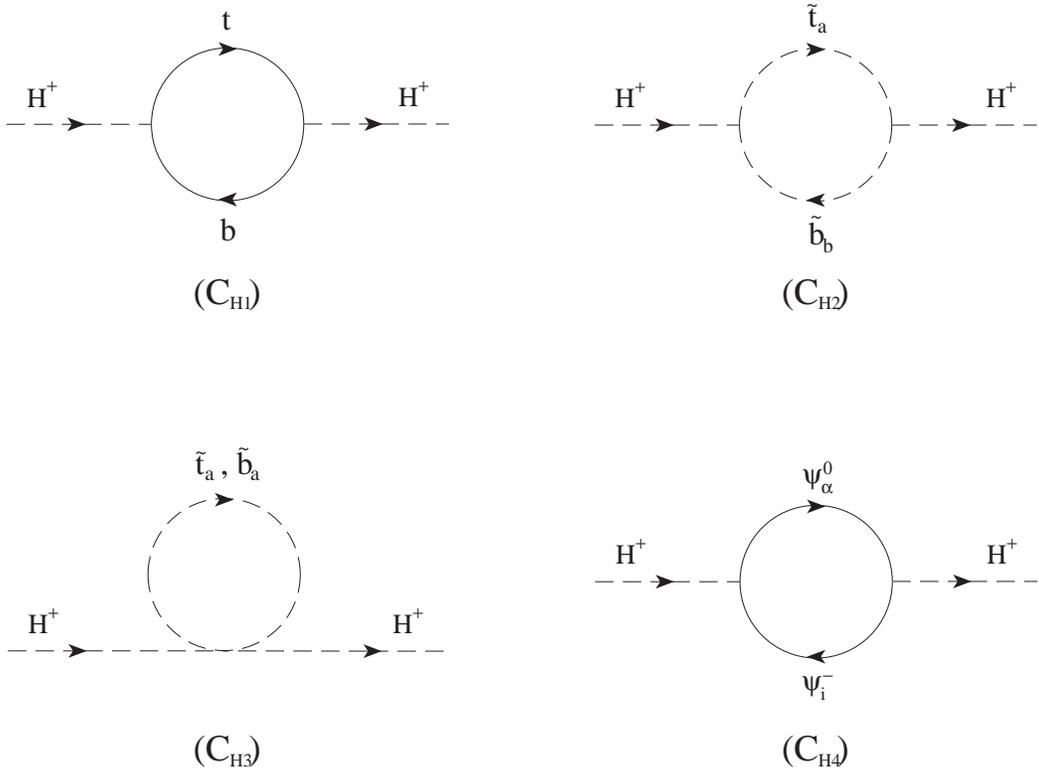
\item{\bf Fig.5} Corrections to the charged
Higgs self-energy from the various
supersymmetric particles and matter fermions. Only the third 
quark-squark generation is illustrated.

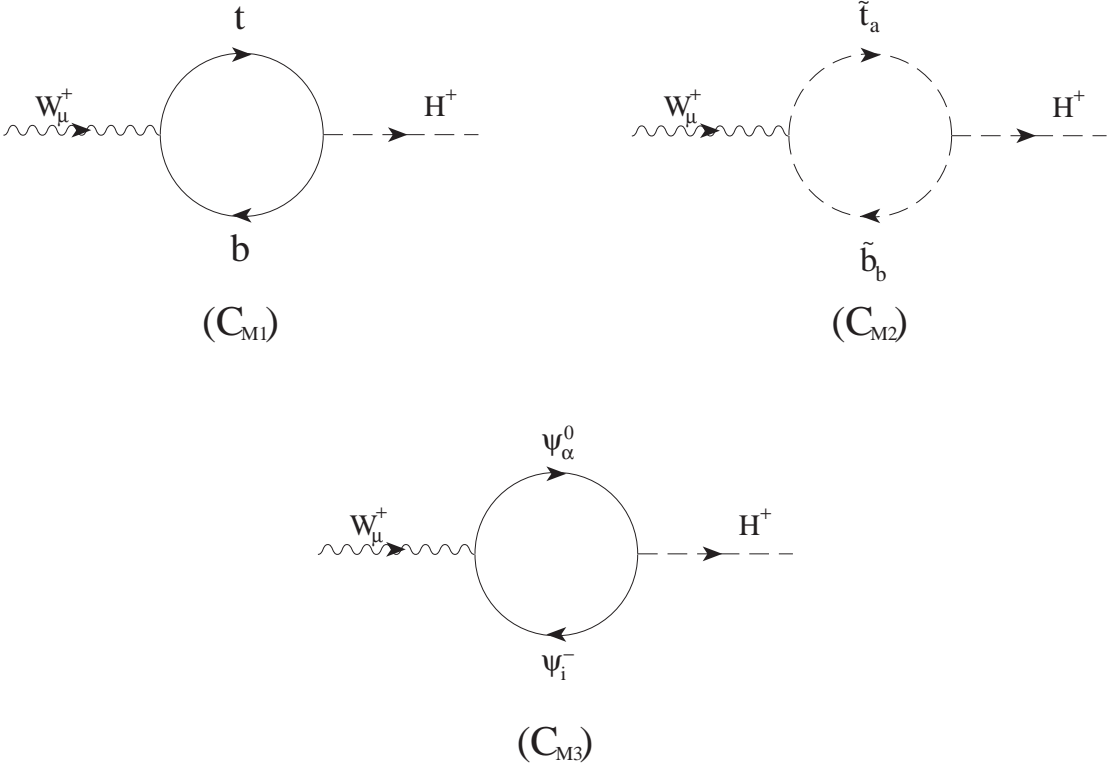
\item{\bf Fig.6} Corrections to the mixed $W^{+}-H^{+}$ self-energy 
from the various supersymmetric particles and matter fermions.
Only the third quark-squark generation is illustrated.

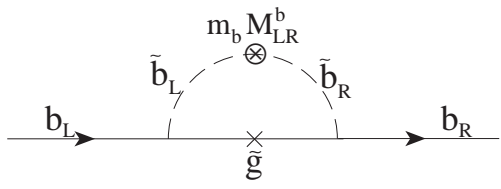
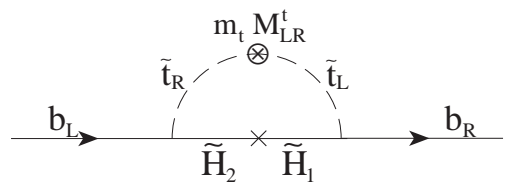
\item{\bf Fig.7} (a) Leading SUSY-QCD  contributions to $\delta m_b/m_b$ in
the electroweak-eigenstate basis; (b)  Leading supersymmetric
Yukawa coupling contributions to $\delta m_b/m_b$ in the 
electroweak-eigenstate basis.

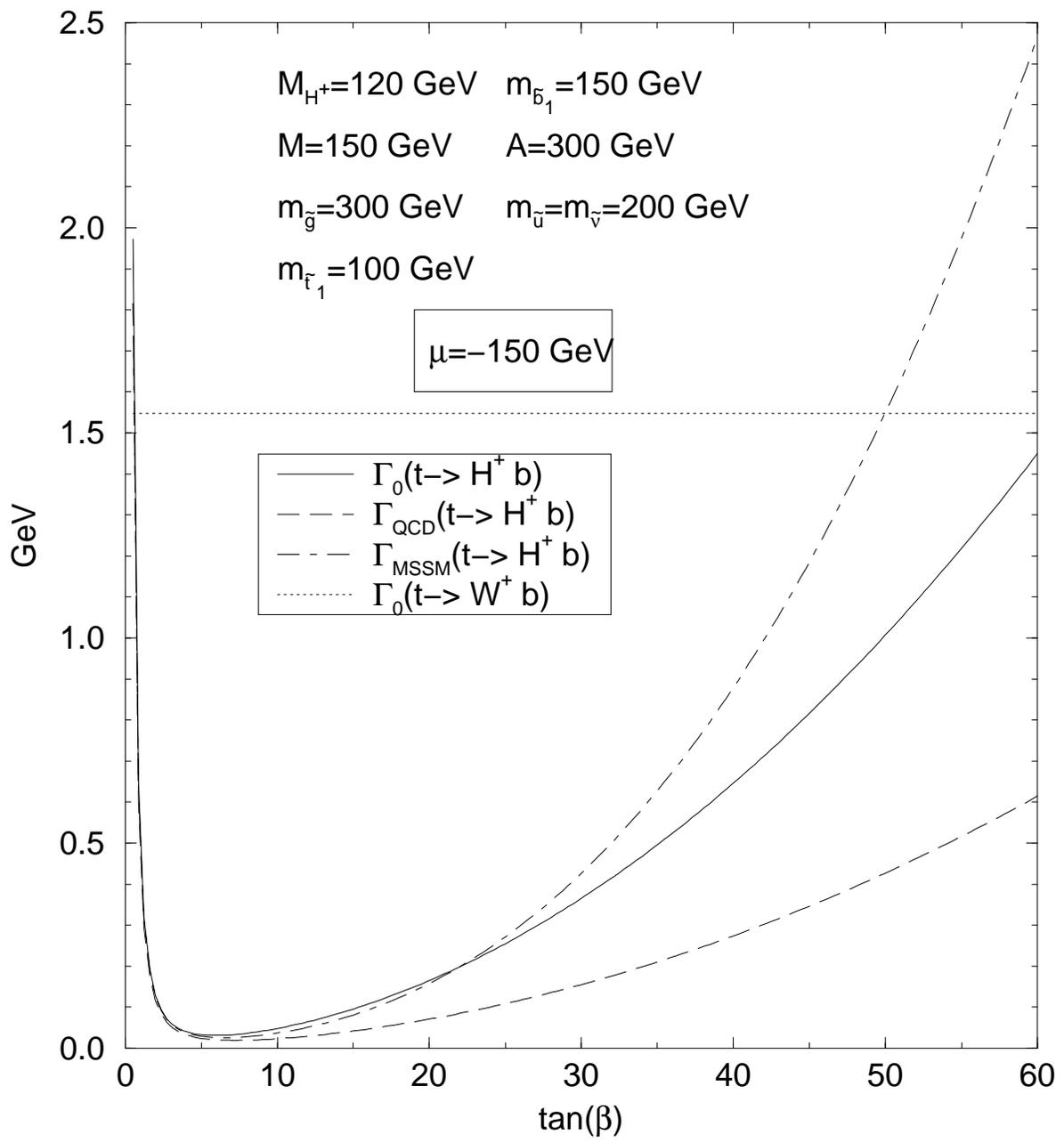
\item{\bf Fig.8} The total partial width, $\Gamma_{MSSM} (t\rightarrow H^+\,b)$,
including all MSSM effects, versus $\tan\beta$, as compared to
the tree-level width and the QCD-corrected width.
Also plotted is the tree-level partial width of the standard
top quark decay, $t\rightarrow W^+\,b$.
The masses of the top and bottom
quarks are  $m_t=175\,GeV$ and $m_b=5\,GeV$, respectively,
and the rest of the inputs are explicitly given. We remark
that $A_t=A_b=...\equiv A$ is a common value of the trilinear coupling for all 
squark and slepton generations.
Unless explicitly stated otherwise, the inputs staying at fixed values 
in the remaining figures are common to the values stated here. 

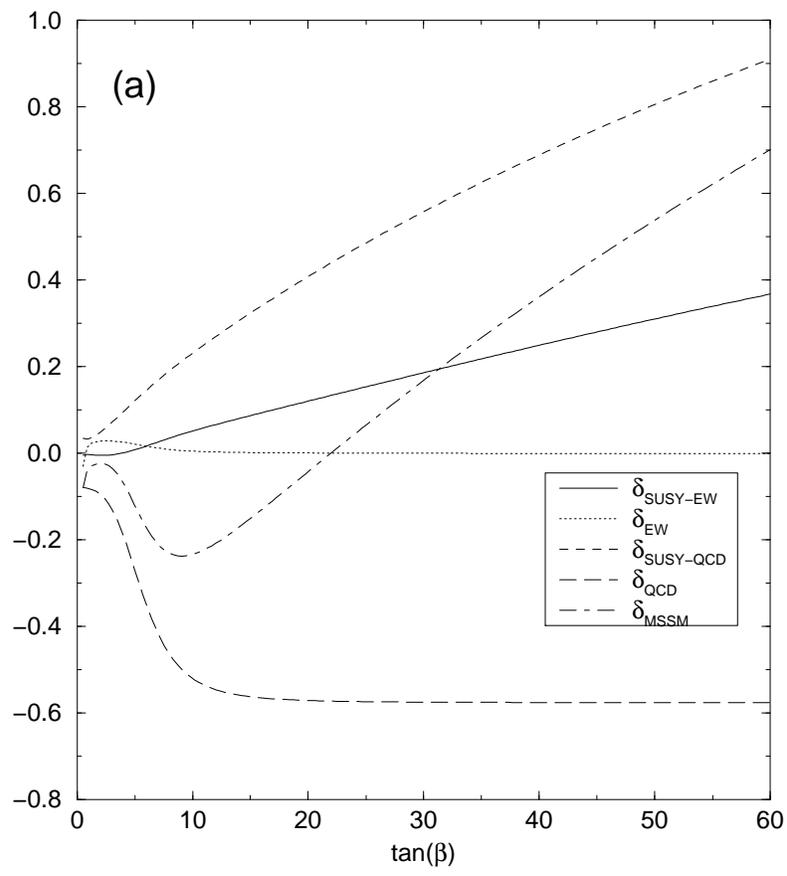
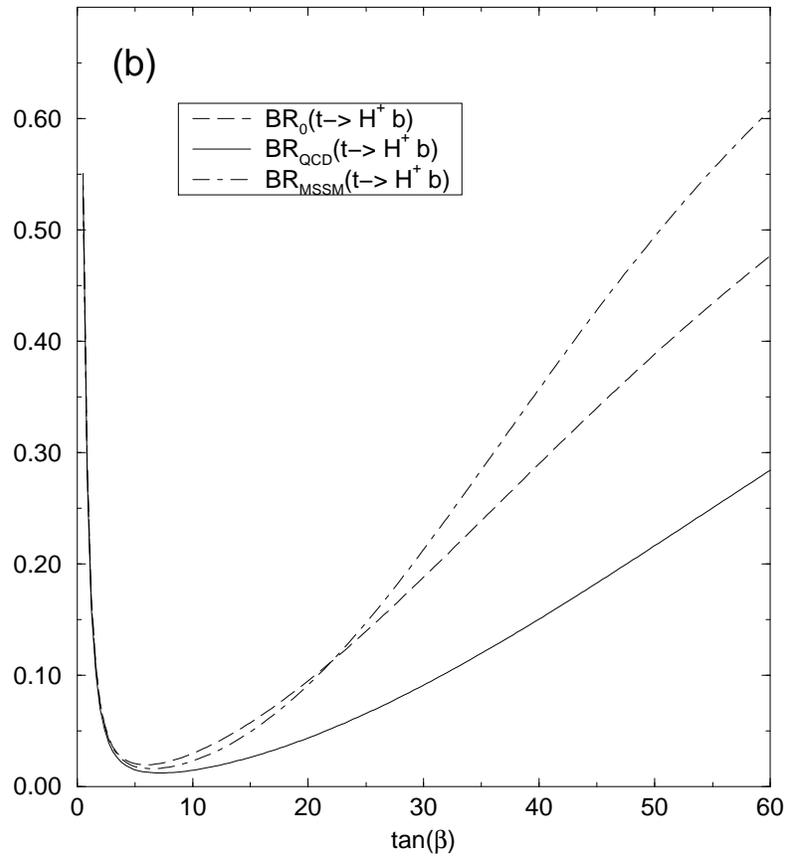
\item{\bf Fig.9} (a) The relative corrections 
$\delta$, eq.(\ref{eq:pito}),
as a function of $\tan\beta$. Shown are the SUSY-EW, standard EW 
(i.e. non-supersymmetric electroweak),
SUSY-QCD, standard QCD, and total MSSM contribution,
eq.(\ref{eq:individ}); 
(b) The branching ratio (\ref{eq:BR}), as a function of $\tan\beta$; separately 
shown are the values of this observable after including standard QCD corrections,
full MSSM corrections, and the tree-level value.

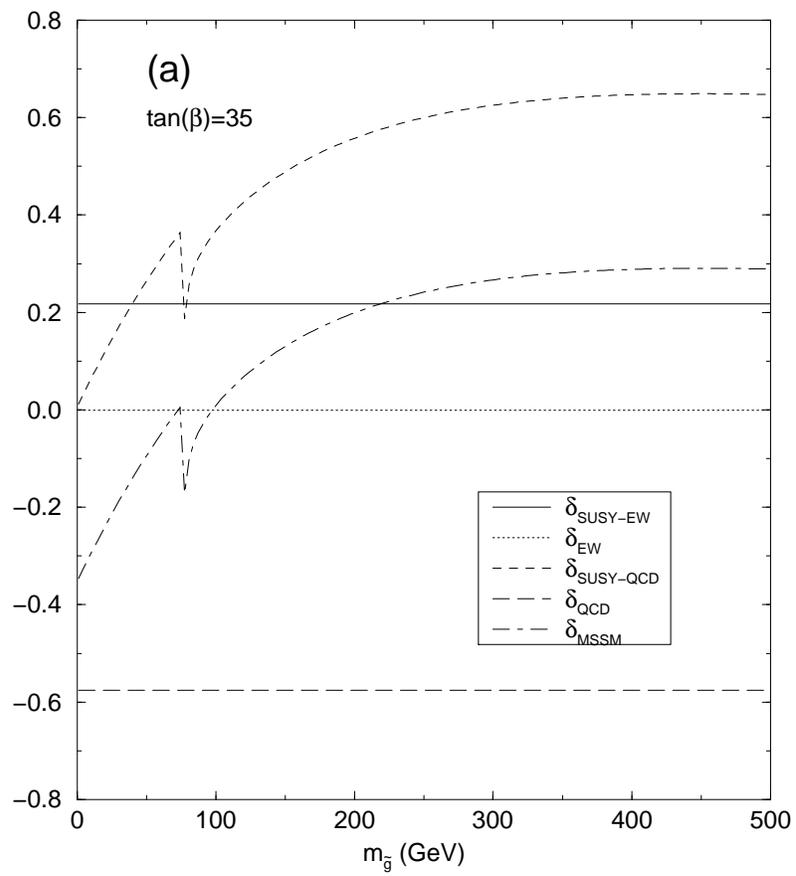
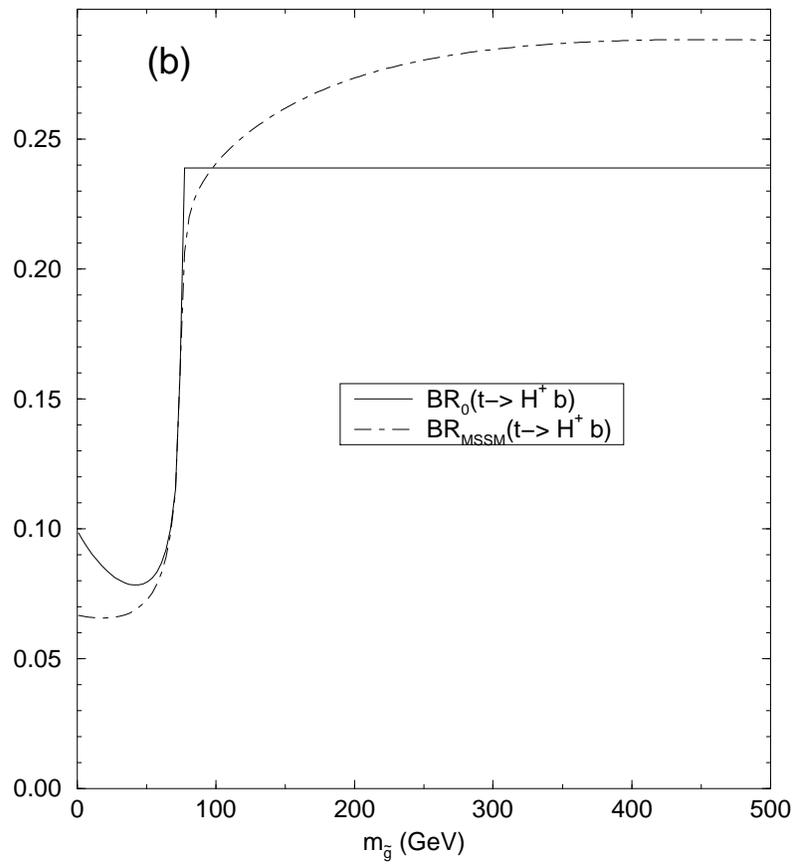
\item{\bf Fig.10} (a) The relative corrections $\delta$, eq.(\ref{eq:pito}),
as a function of the gluino mass, $m_{\tilde{g}}$, for the SUSY-EW, standard EW,
SUSY-QCD, standard QCD contributions, and total MSSM contribution,
(b) As in (a), but for the tree-level and
full MSSM-corrected branching ratios (\ref{eq:BR}).
We have set $\tan\beta=m_t/m_b=35$. This value is maintained in all figures
where $\tan\beta$ is fixed.

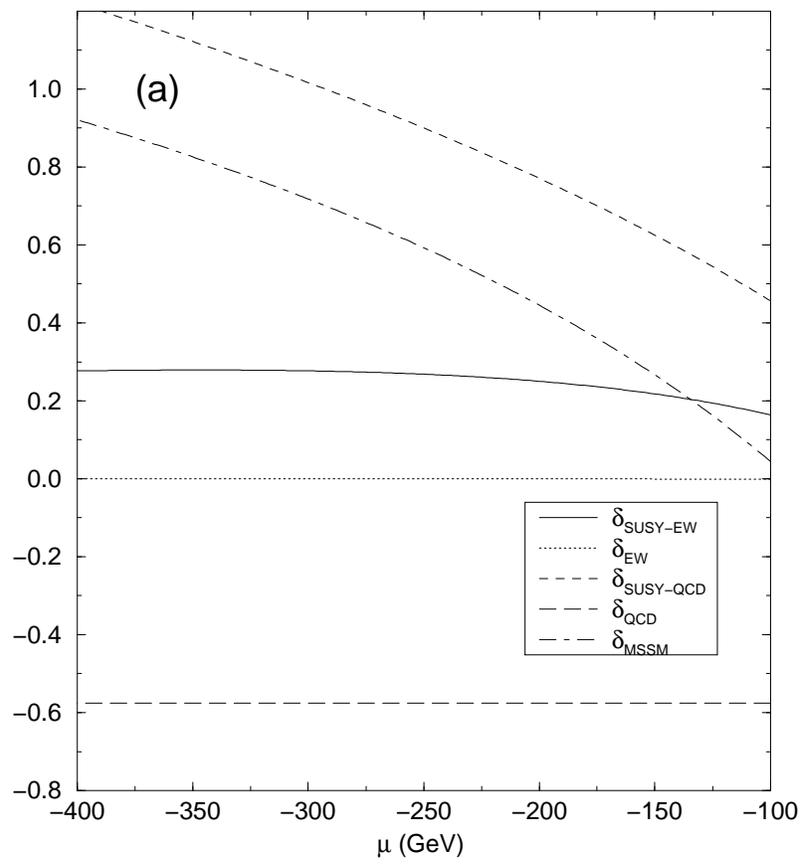
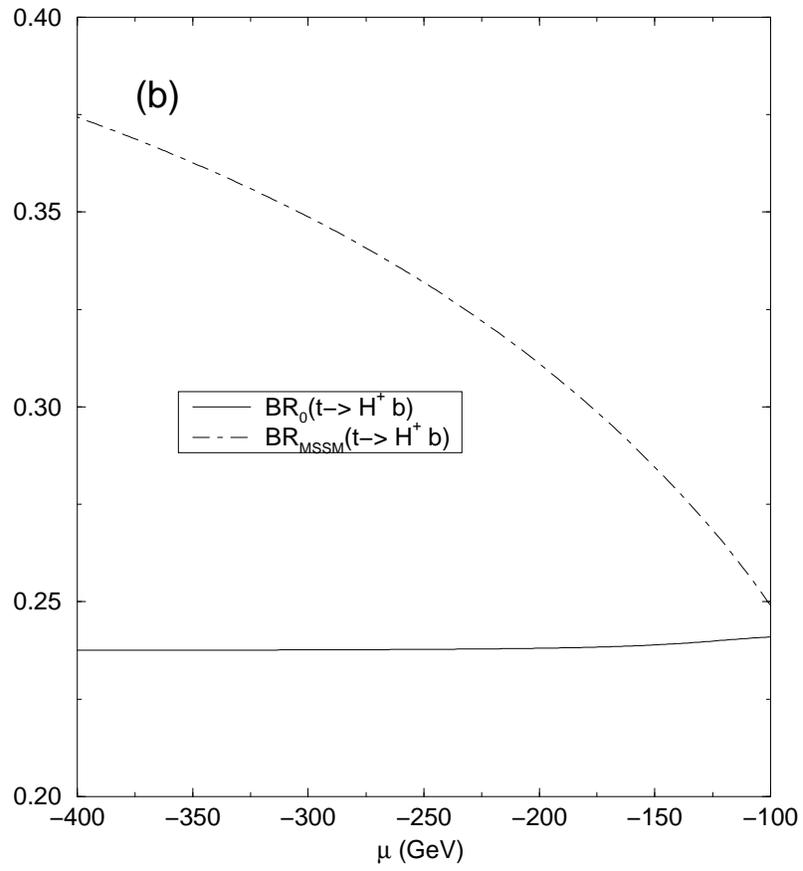
\item{\bf Fig.11} (a) The relative corrections $\delta$, eq.(\ref{eq:pito}),
as a function of the supersymmetric Higgs mixing parameter, $\mu$, for the 
various contributions as in Fig.9a;
(b) As in (a), but for the same branching ratios as in Fig.10b.

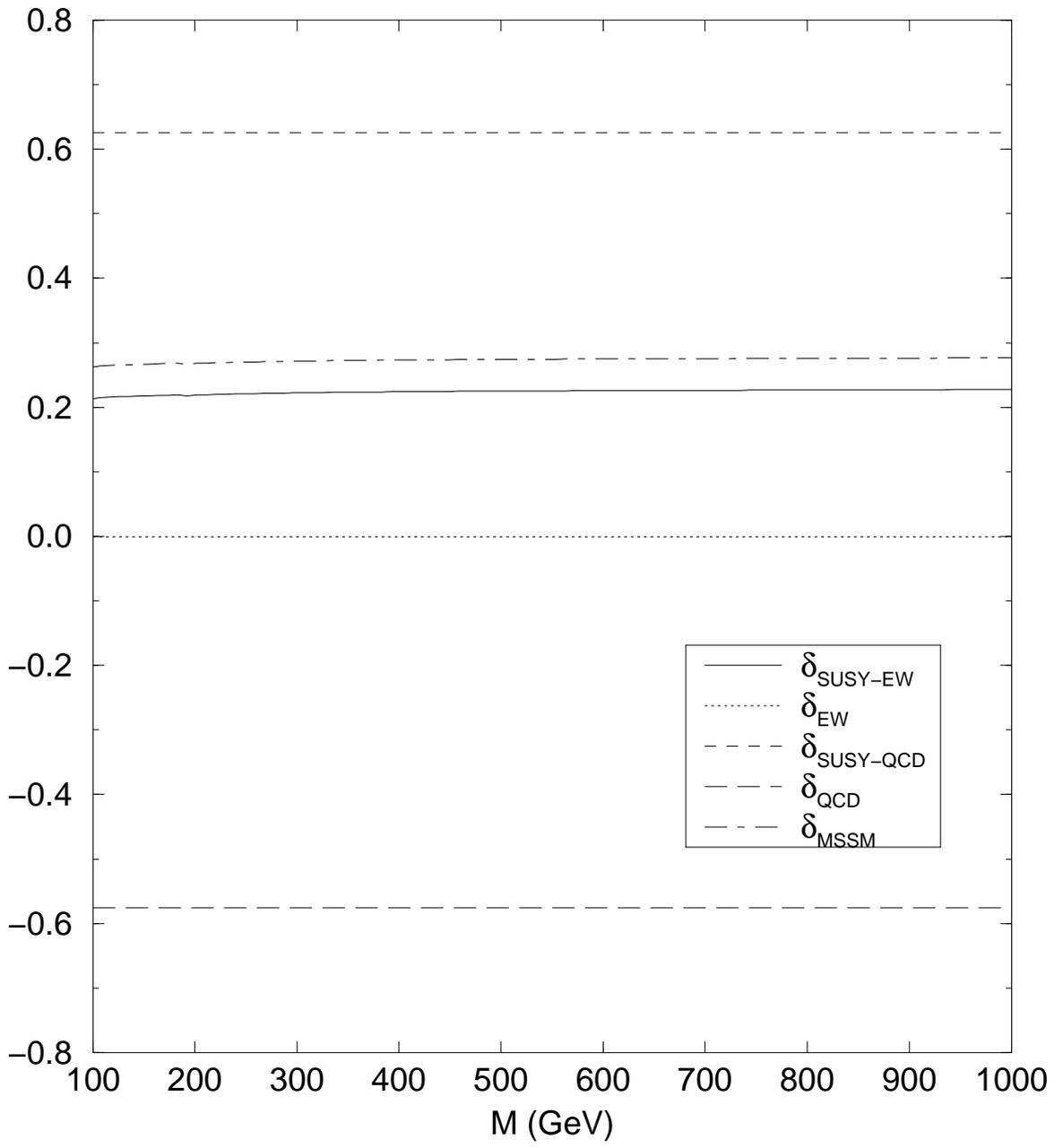
\item{\bf Fig.12} Dependence of $\delta$, eq.(\ref{eq:pito}),
on the  $SU(2)_L$-gaugino 
soft SUSY-breaking mass, $M$, assuming that the $U(1)_Y$ gaugino mass, $M'$, is
related to $M$ through $M'/M={5\over 3}\,\tan^2\theta_W$. The same individual
and total contributions as in Fig.9a are shown.

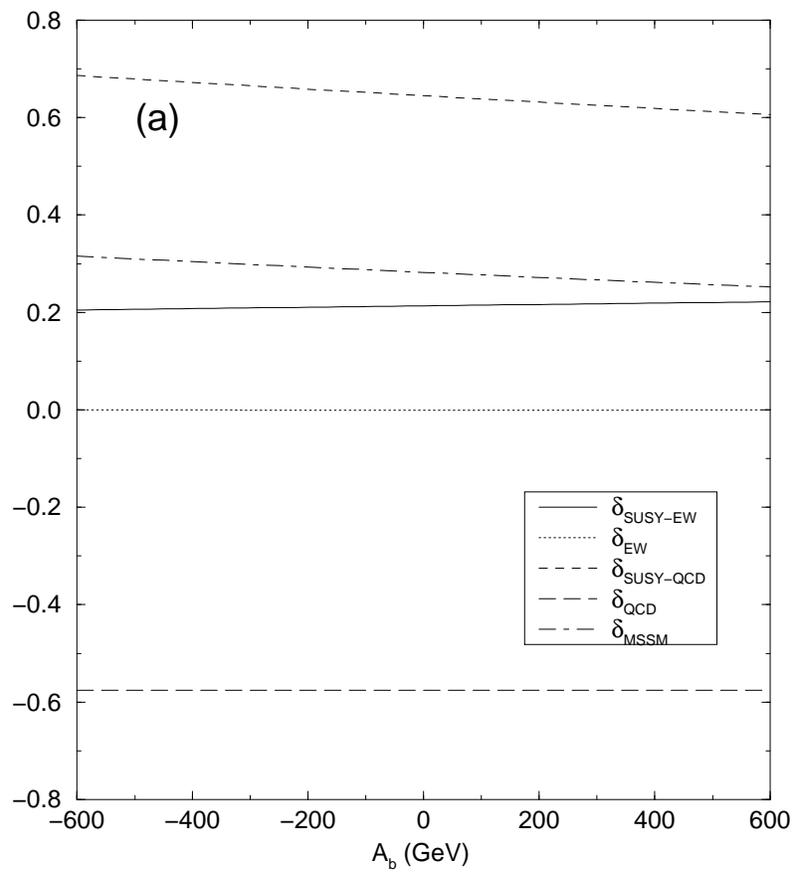
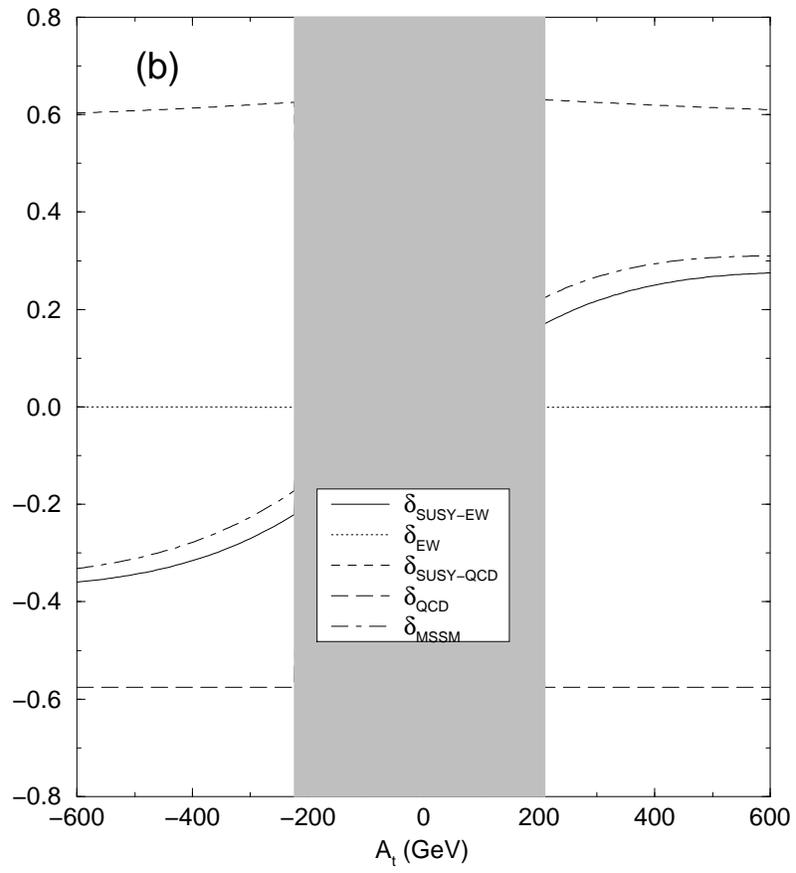
\item{\bf Fig.13} (a) The relative corrections $\delta$, eq.(\ref{eq:pito}),
as a function of the trilinear soft SUSY-breaking parameter $A_b$ in the
bottom sector. The other trilinear couplings are kept as in Fig.8;
(b) As in (a), but for the trilinear soft SUSY-breaking parameter $A_t$ in the
top sector. Shown are the same individual
and total contributions as in Fig.9a.

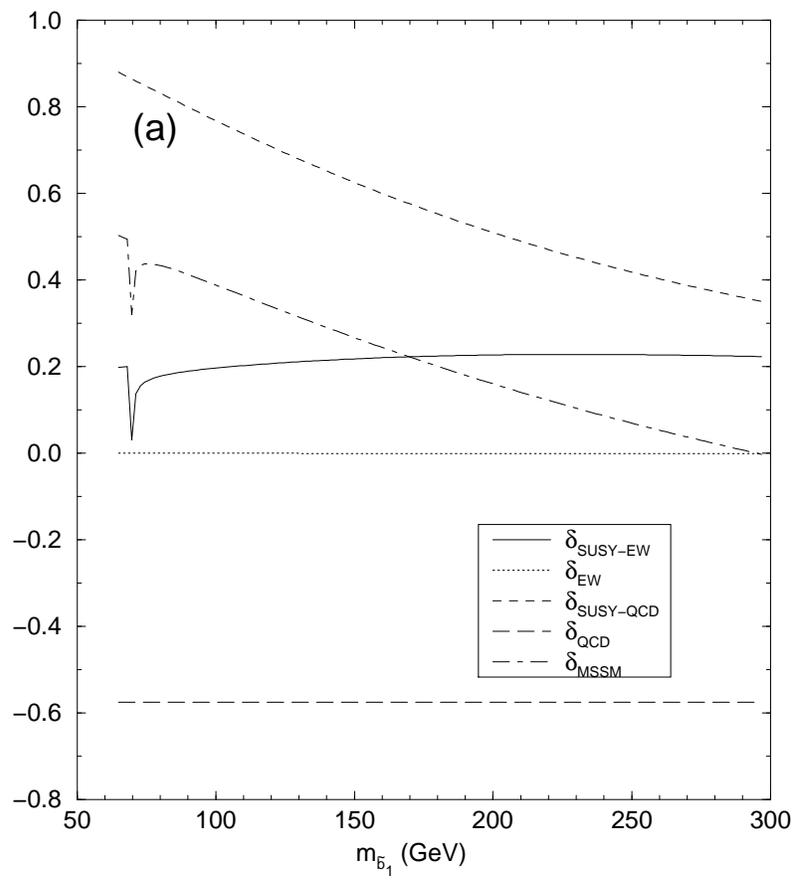
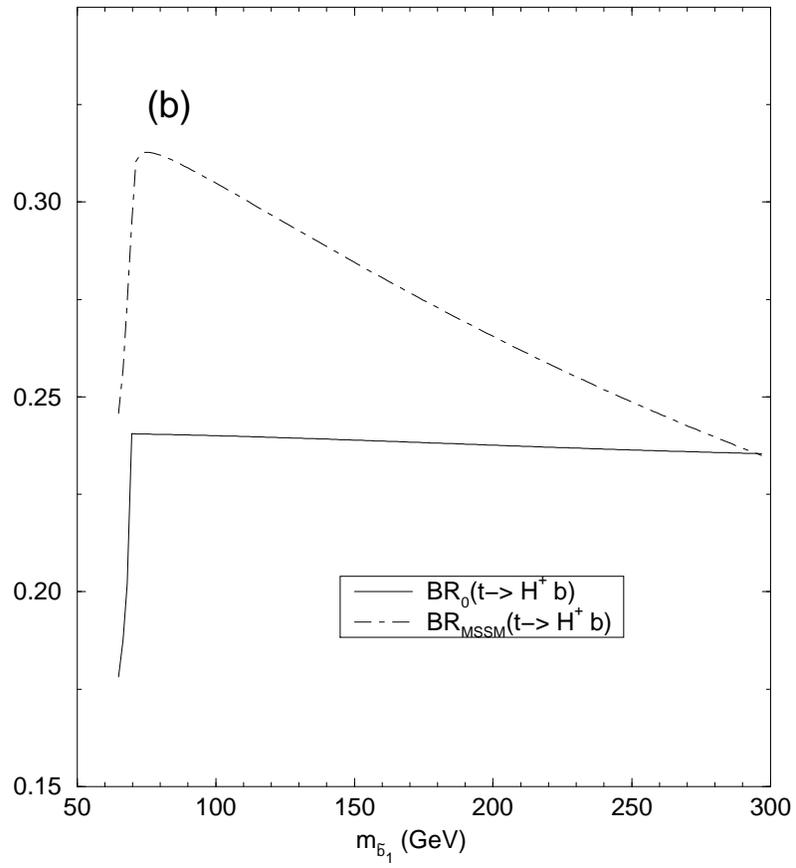
\item{\bf Fig.14}  (a) The relative corrections $\delta$, eq.(\ref{eq:pito}),
as a function of the lightest sbottom mass, $m_{\tilde{b}_1}$,
for the various contributions as in Fig.9a;
(b) As in (a), but for the same branching ratios as in Fig.10b.

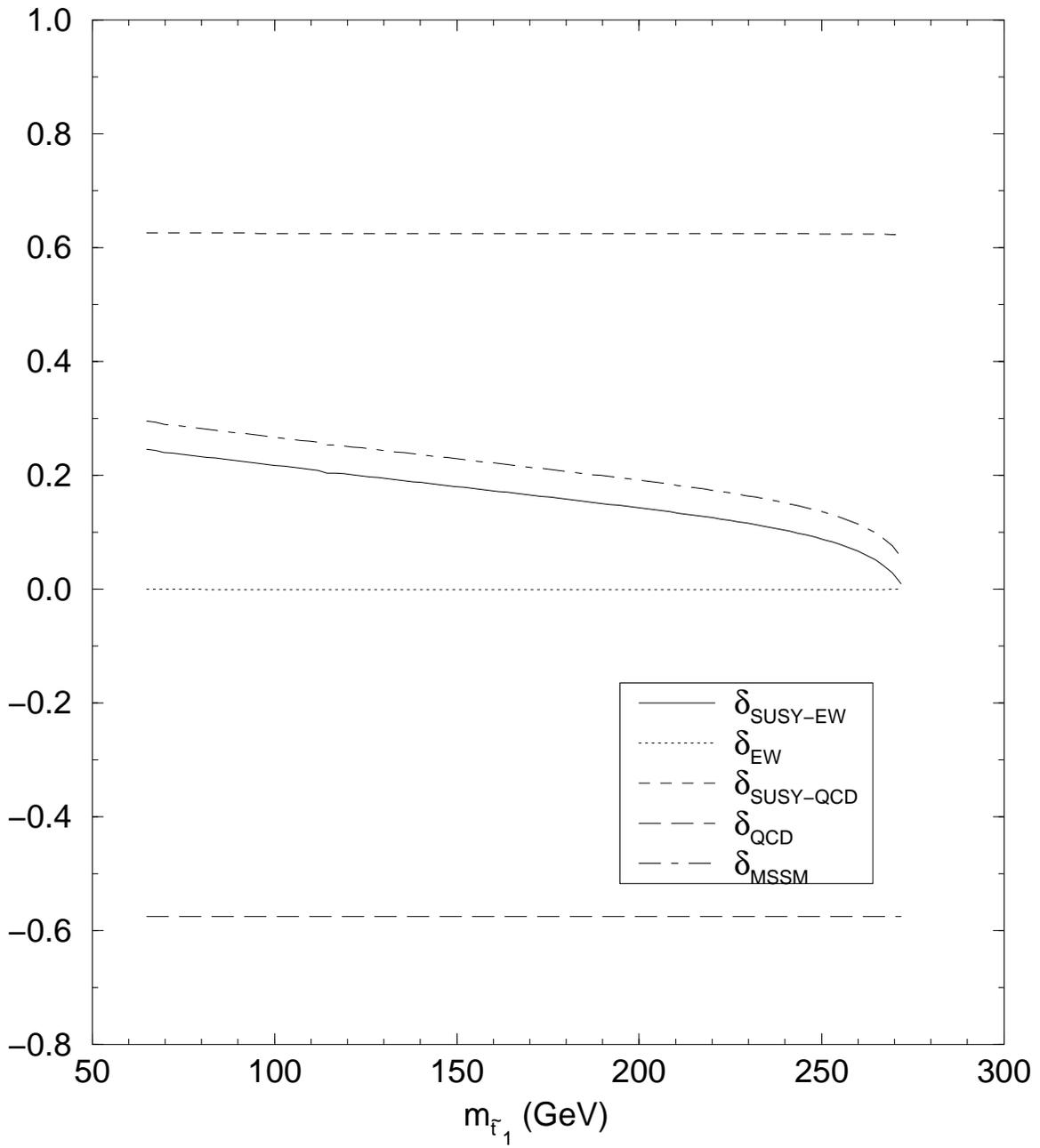
\item{\bf Fig.15} The relative corrections $\delta$, eq.(\ref{eq:pito}),
as a function of the lightest stop mass, $m_{\tilde{t}_1}$, for the 
various contributions as in Fig.9a.

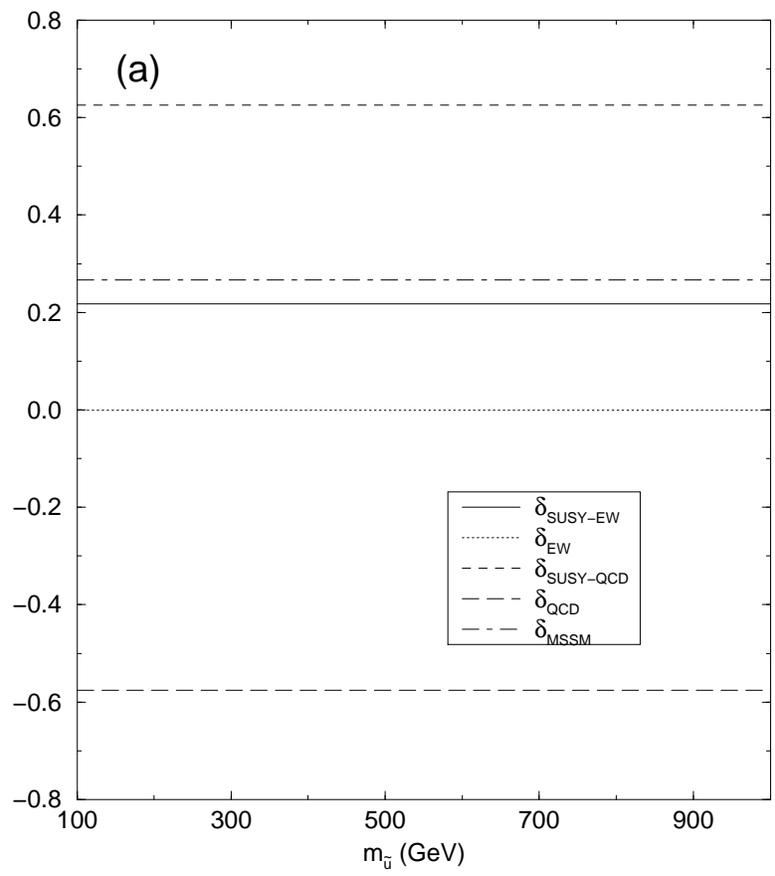
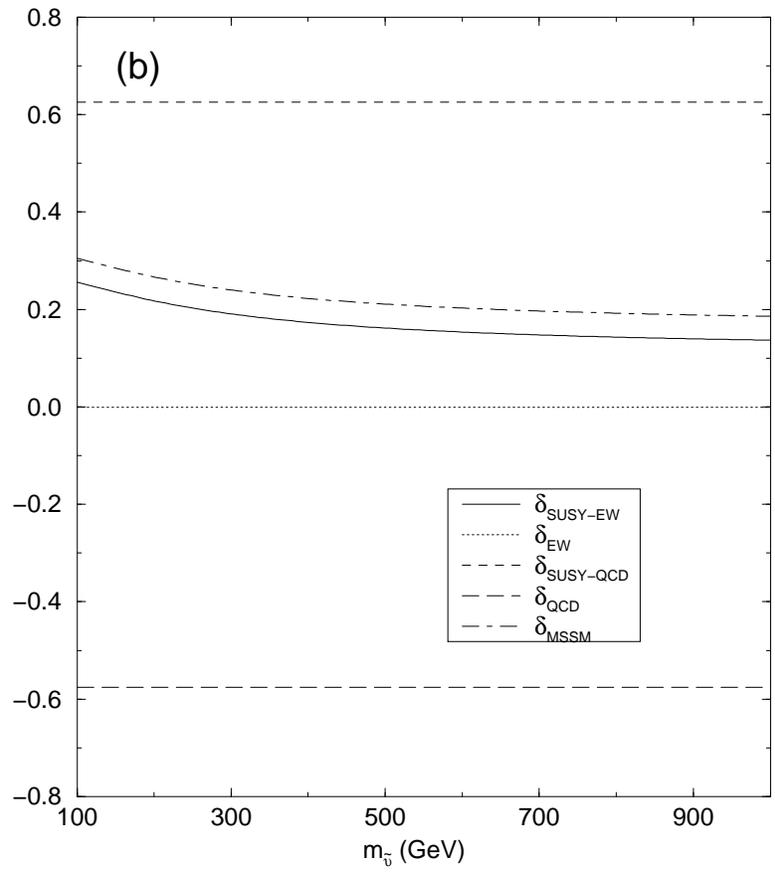
\item{\bf Fig.16}  (a) The relative corrections $\delta$, eq.(\ref{eq:pito}),
as a function of the up-squark masses 
$m_{\tilde{u}}\equiv m_{\tilde{u}_1}=m_{\tilde{u}_2}$. The c-squarks
are assumed to be degenerate with the up-squarks;
(b) $\delta$ as a function of the sneutrino
masses, assumed to be degenerate.

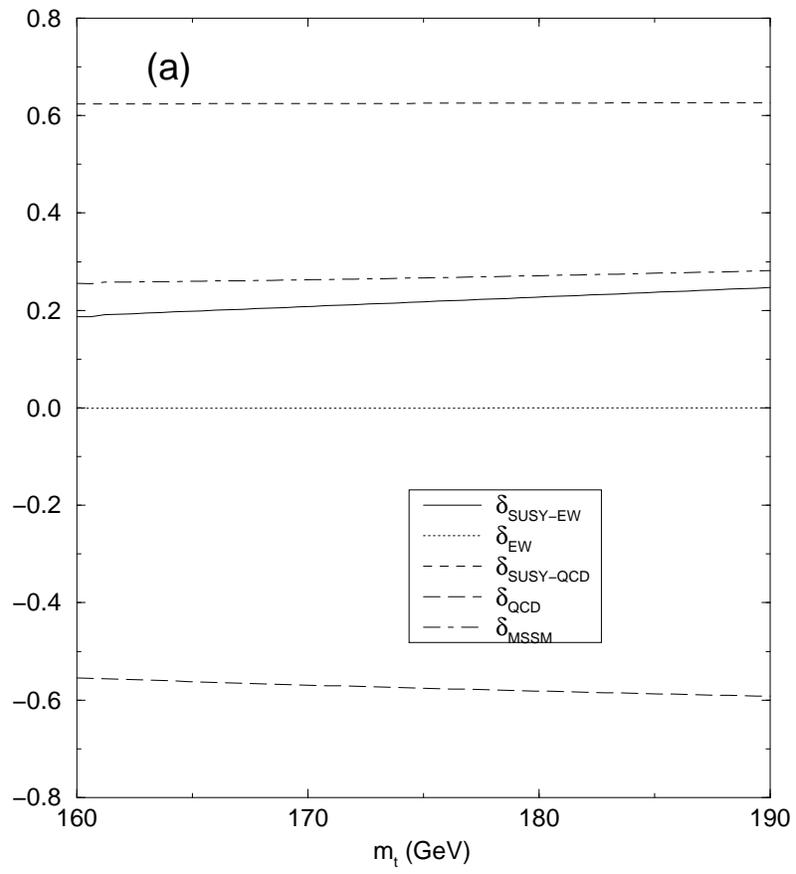
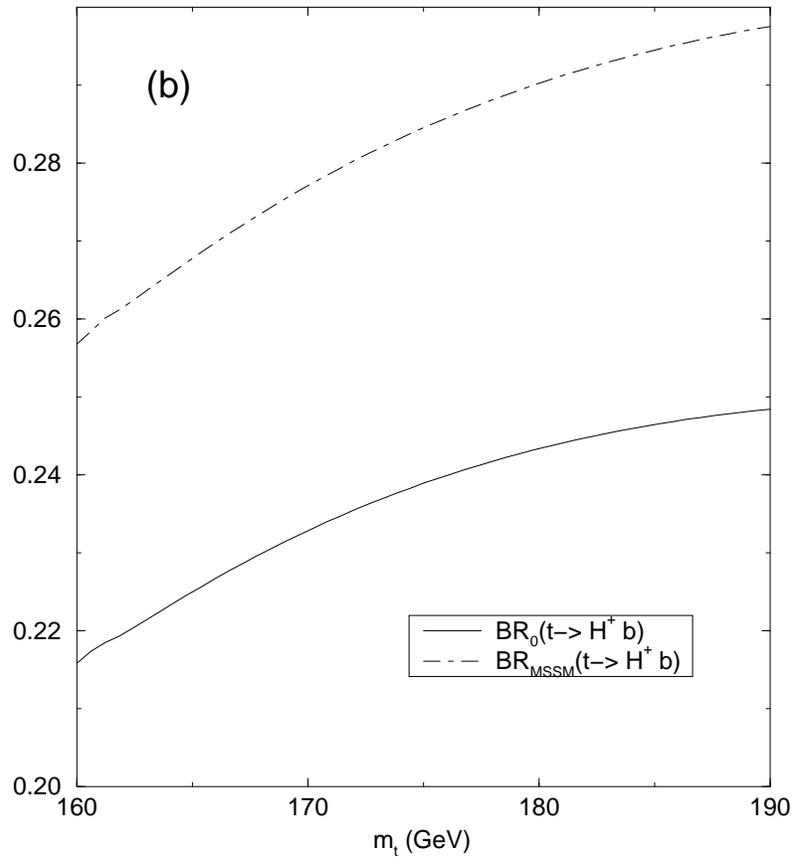
\item{\bf Fig.17} (a) The relative corrections $\delta$, eq.(\ref{eq:pito}),
as a function of the top quark mass within about $2\,\sigma$ of
the present experimental range at the Tevatron. 
(b) As in (a), but for the same branching ratios as in Fig.10b.

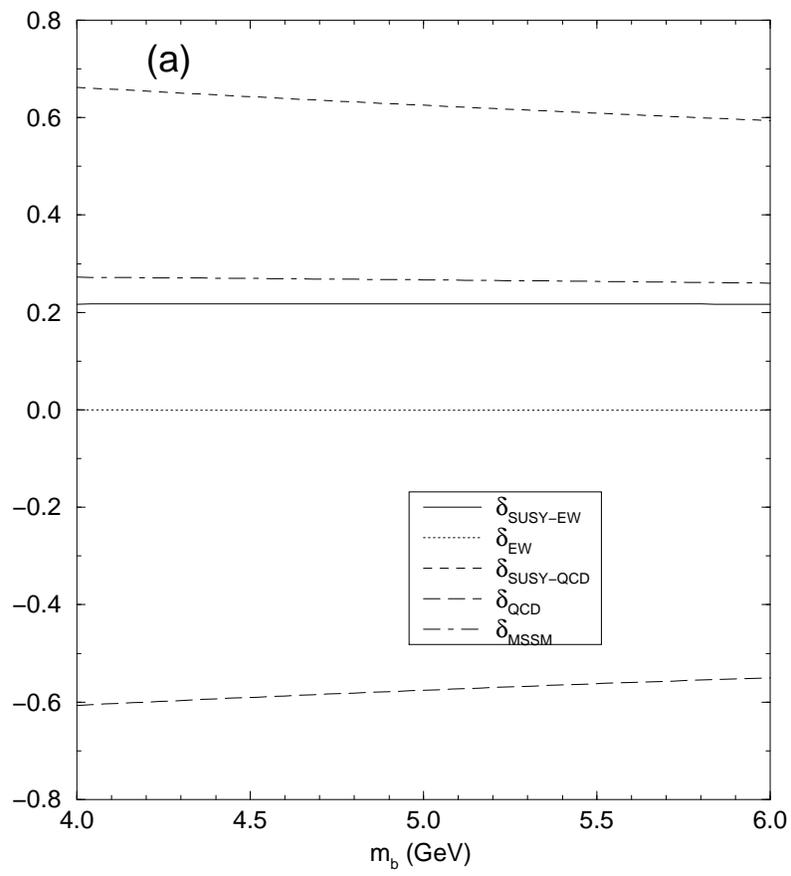
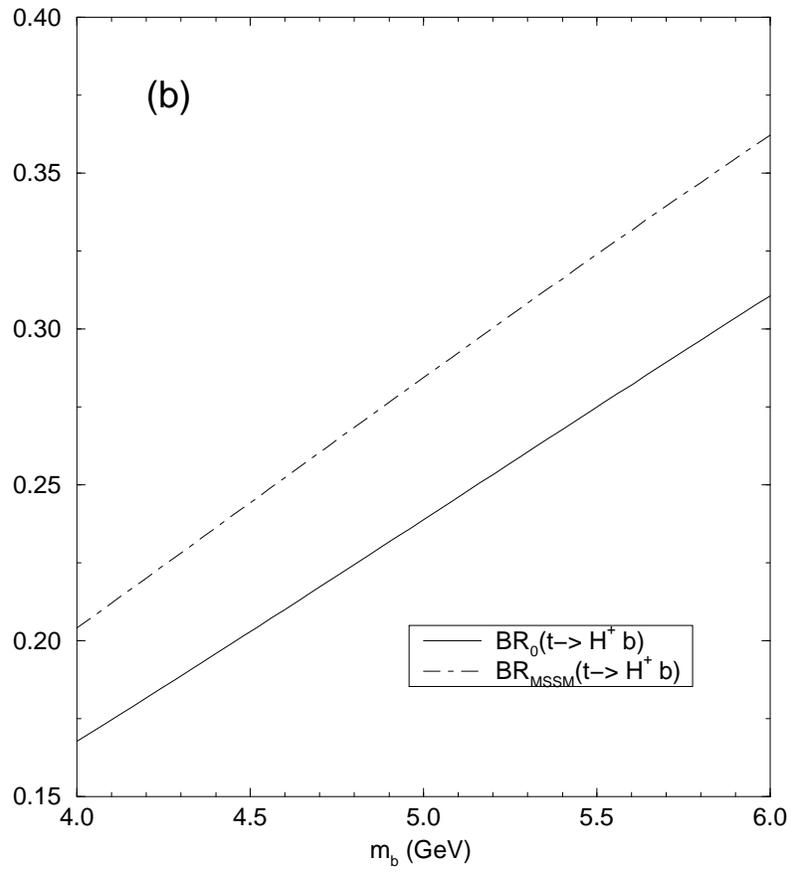
\item{\bf Fig.18}  (a) The relative correction $\delta$, eq.(\ref{eq:pito}),
as a function of the bottom quark mass.
(b) As in (a), but for the same branching ratios as in Fig.10b.

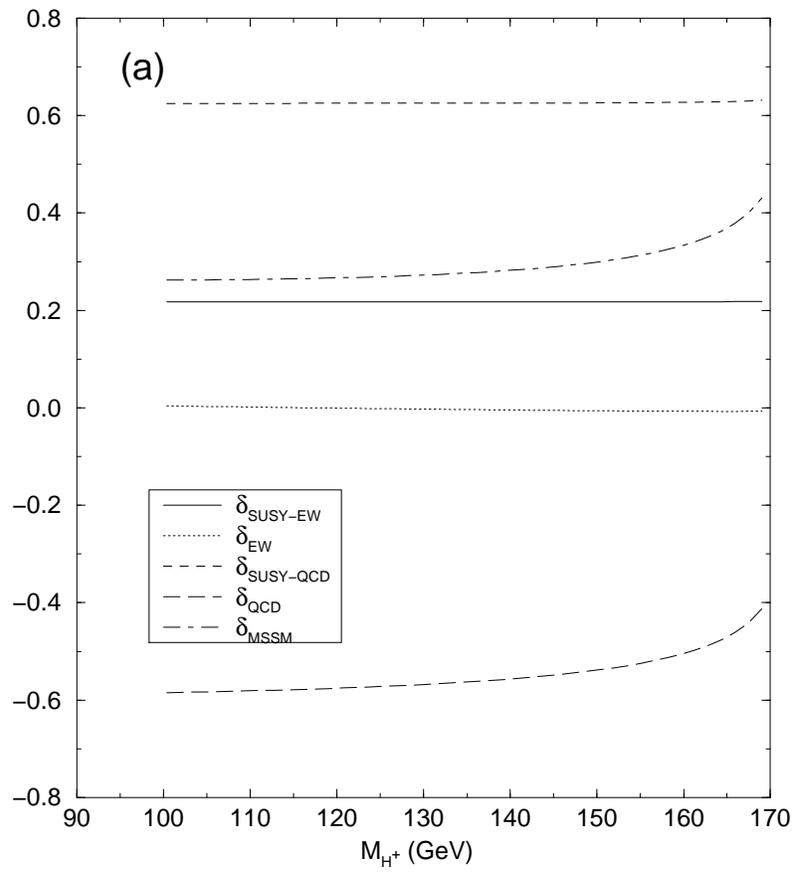
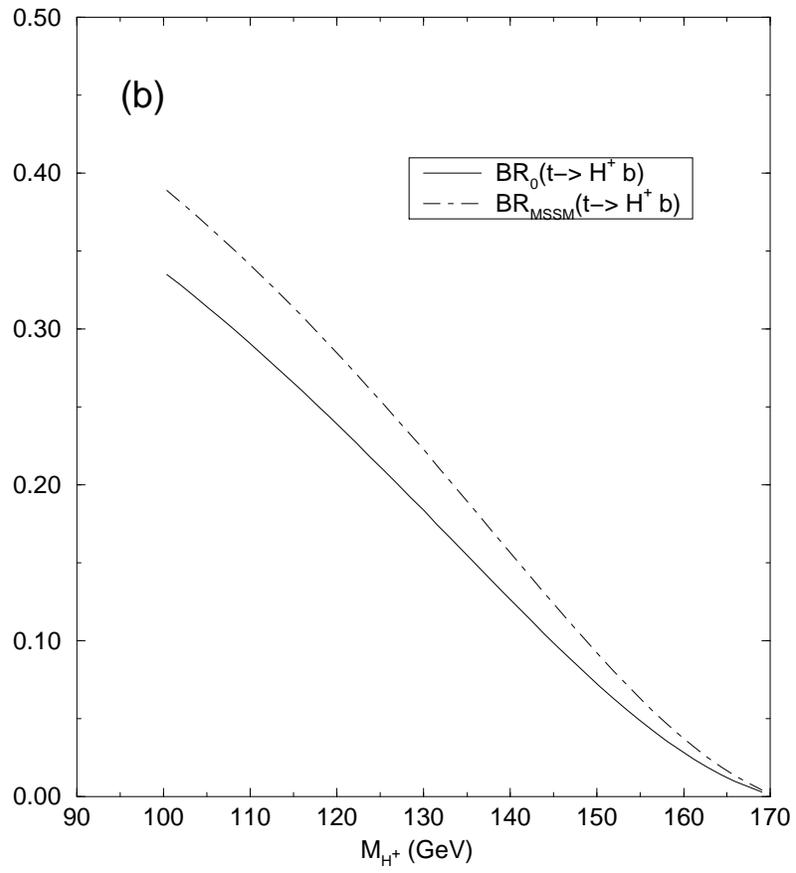
\item{\bf Fig.19}  (a) The relative corrections $\delta$, eq.(\ref{eq:pito}),
as a function of the charged Higgs mass;
(b) As in (a), but for the same branching ratios as in Fig.10b.

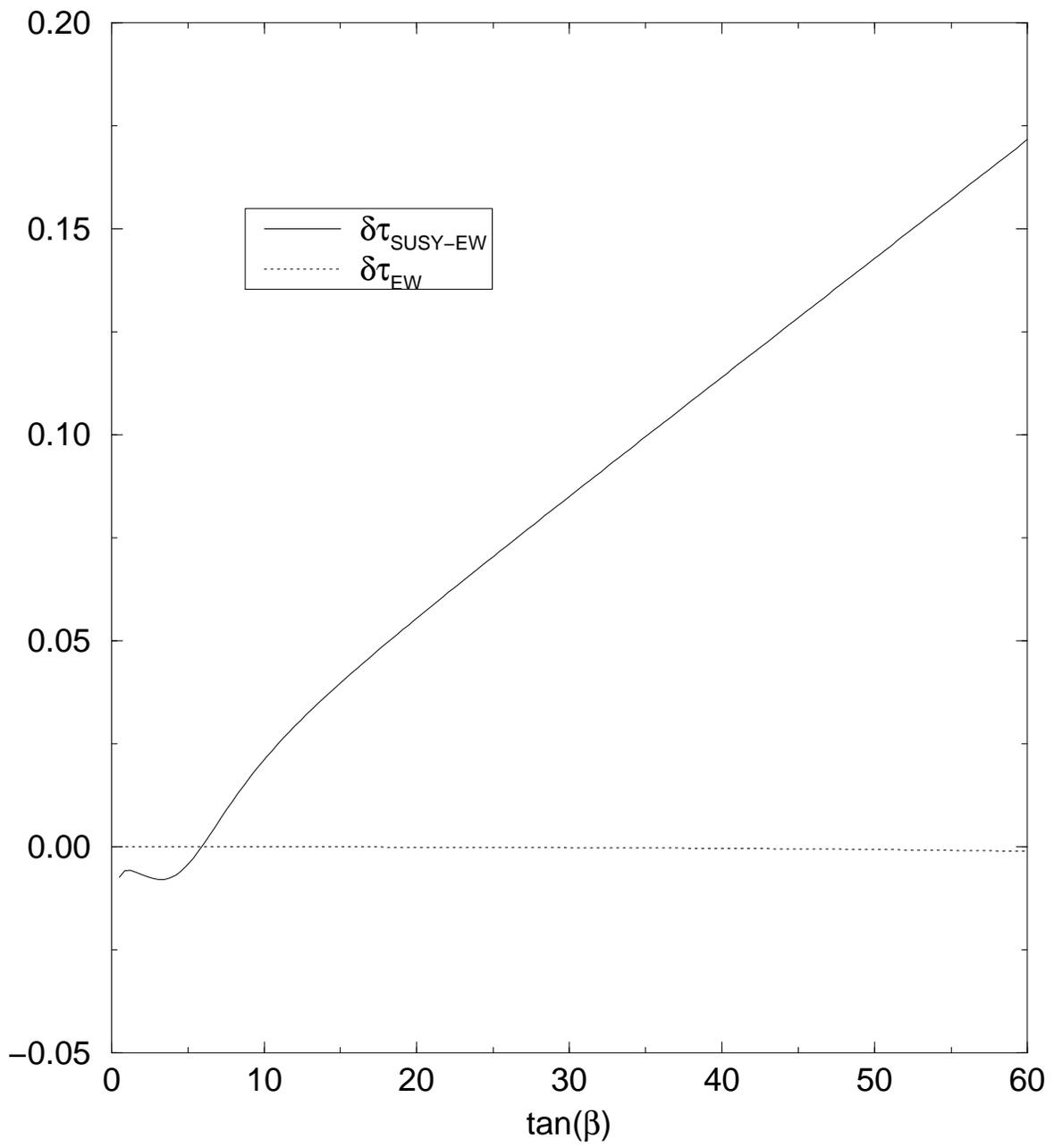
\item{\bf Fig.20} The supersymmetric ($\delta\tau_{\rm SUSY-EW}$)
 and non-supersymmetric ($\delta\tau_{\rm EW}$) 
 electroweak contributions to $\delta$, eq.(\ref{eq:pito}), from
the process-dependent term $\Delta_{\tau}$, eq.(\ref{eq:deltatau}), as
a function of $\tan\beta$.

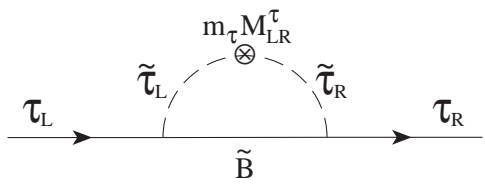
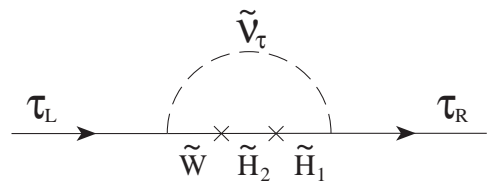
\item{\bf Fig.21} Leading supersymmetric electroweak contributions to
$\delta m_{\tau}/m_{\tau}$ in the electro\-weak-eigenstate basis.

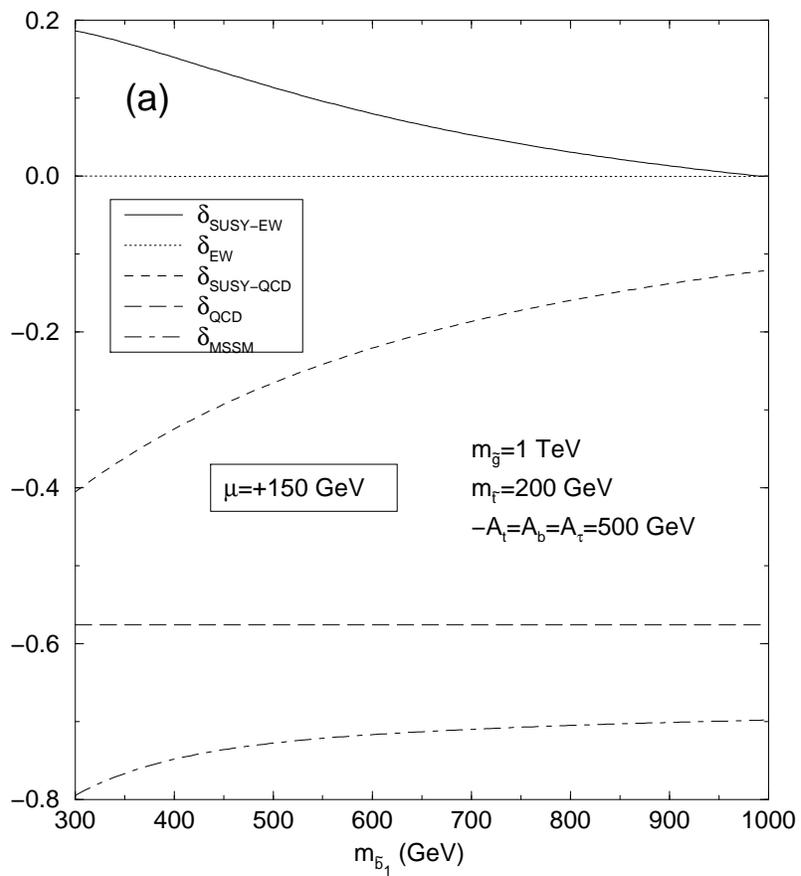
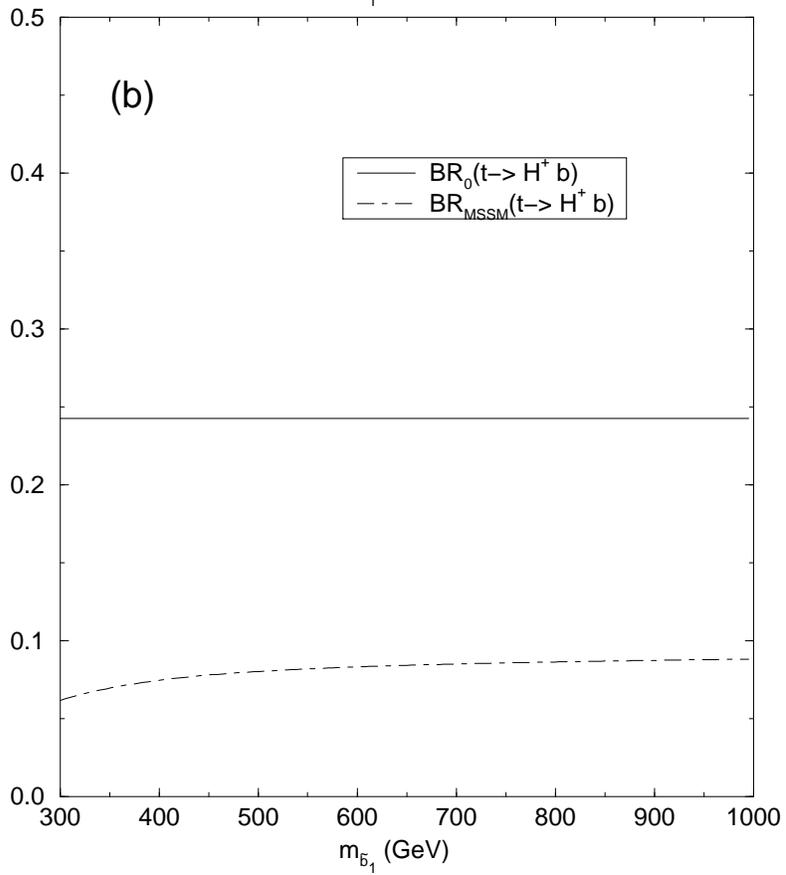
\item{\bf Fig.22} (a) The relative corrections $\delta$, eq.(\ref{eq:pito}),
as a function of the lightest sbottom quark mass, for positive $\mu=+150\,GeV$,
negative $A_t$
and given values of the other parameters. In this case,
huge values of $m_{\tilde{b}_1}$ are needed in order to damp the absolute value
of the total correction down to $100\%$;
(b) As in (a), but for 
 the same branching ratios as in Fig.10b.
 
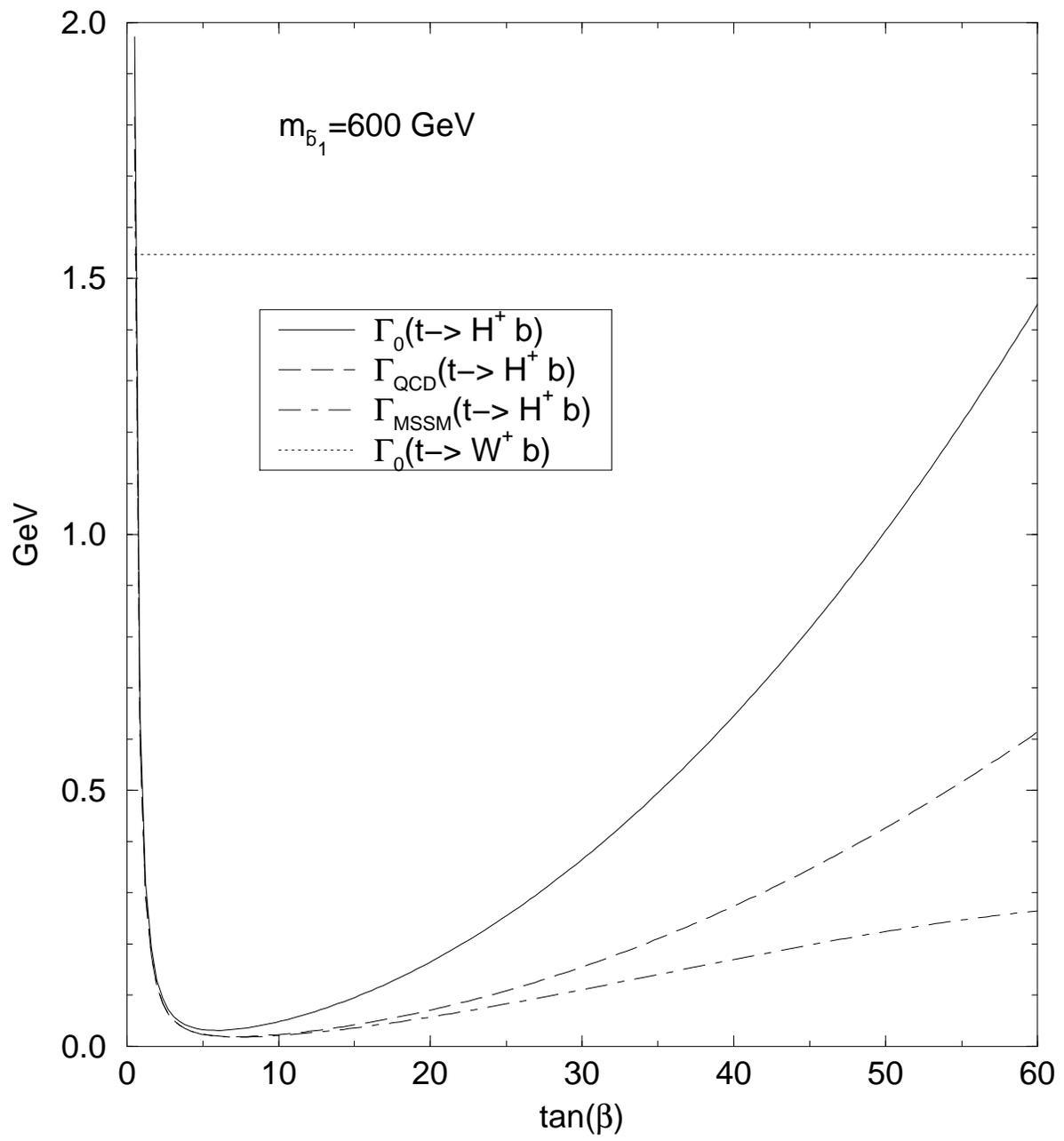
\item{\bf Fig.23 } The total partial width, $\Gamma (t\rightarrow H^+\,b)$,
including all MSSM effects, versus $\tan\beta$,
for the same inputs as in Fig.22a, as compared to
the tree-level width and the QCD-corrected width. A typical value of
$m_{\tilde{b}_1}$ within the range used in Fig.22a is selected.
Also plotted is the tree-level partial width of the standard
top quark decay $t\rightarrow W^+\,b$.

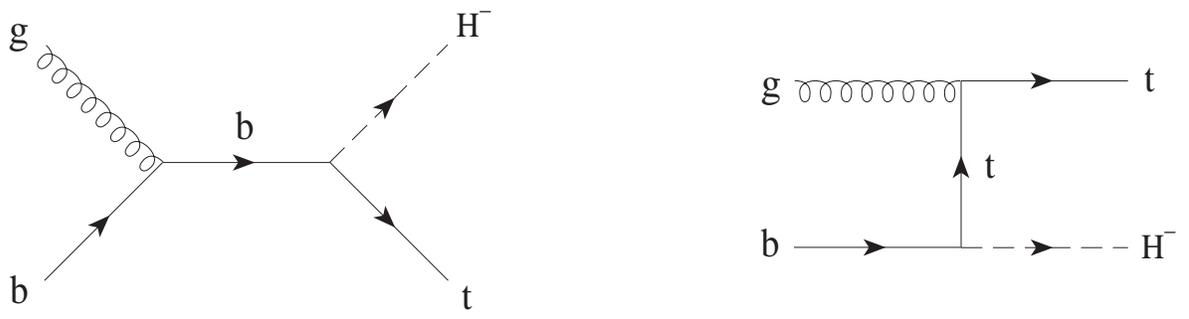
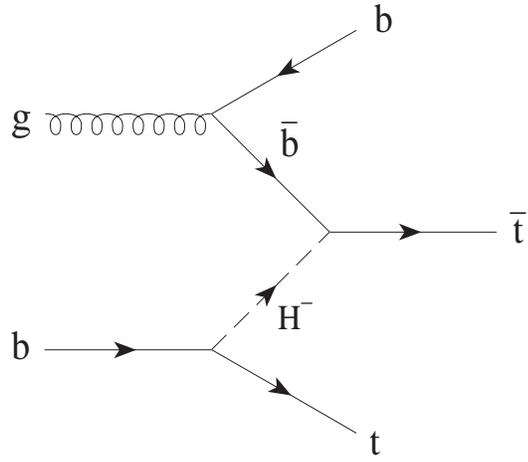
\item{\bf Fig.24}  Typical diagrams for top quark and charged Higgs production
in hadron colliders involving the relevant $t\,b\,H^{\pm}$-vertex.

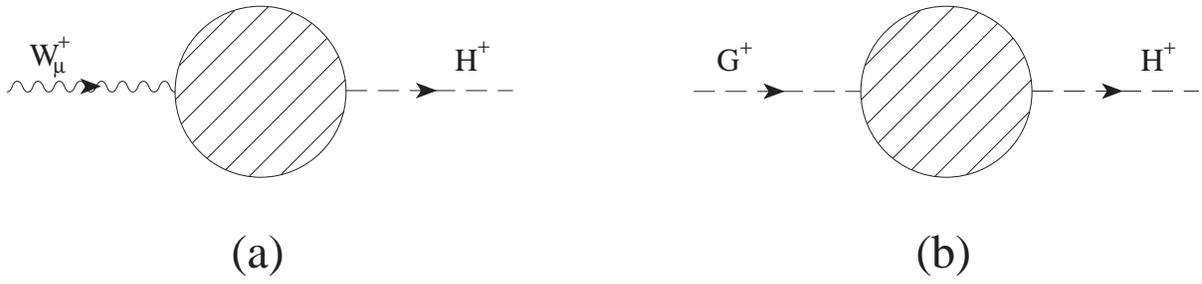
\item{\bf Fig.25}  The mixed blobs $W^{+}-H^{+}$
and $G^{+}-H^{+}$  at any order of
perturbation theory.

\end{itemize}





\vspace{1cm}
\begin{thebibliography}{9999}
\bibitem{Tevatron}
F. Abe {\it et al.} (CDF Collab.), {\it Phys. Rev. Lett.}\,{\bf 74} (1995) 2626;
S. Abachi {\it et al.} (D$0$ Collab.), {\it Phys. Rev. Lett.}\,{\bf 74}
(1995) 2632.
\bibitem{Grannis}
P. Grannis, talk at the
28th International Conference on High Energy Physics, Warsaw, Poland,
July 25-31th 1996.
\bibitem{Zerwas}
D. Shaile, P.M. Zerwas,\, {\it Phys. Rev.}\,{\bf D 45} (1992) 3262.
\bibitem{Bardeen}
W.A. Bardeen, C.T. Hill, M. Lindner,\, {\it Phys. Rev.} {\bf D 41}
(1990) 1647;\, C.T. Hill, {\it Phys. Lett.}\,{\bf B 266} (1991) 419;
\, {\it ibid.}\,{\bf B 345} (1995) 483.
\bibitem{Chikuvula}
R.S. Chikuvula, E.H. Simmons, J. Terning,\, {\it Phys. Lett.} {\bf B 331} (1994) 383.
\bibitem{Georgi}
H. Georgi,\, {\it Weak Interactions and Modern Particle Theory}
(The Benjamin/Cum\-mings Publishing Company, 1984).
\bibitem{Peccei}
R.D. Peccei, X. Zhang,\, {\it Nucl. Phys.} {\bf B 337} (1990) 269;
C. Hill, X. Zhang\, {\it Phys. Rev.} {\bf D 51} (1995) 3563. 
\bibitem{MSSM}
H. Nilles,\, {\it Phys. Rep.} {\bf 110} (1984) 1;
 H. Haber, G. Kane, {\it Phys. Rep.}
{\bf 117} (1985) 75;
 A. Lahanas, D. Nanopoulos, {\it Phys. Rep.} {\bf 145} (1987) 1;
 See also the exhaustive reprint collection {\bf Supersymmetry}
(2 vols.), ed. S. Ferrara (North Holland/World Scientific, Singapore, 1987).
\bibitem{LangMaryland}
P. Langacker, in: {\it Tests and status of the Standard Model}\,, talk
given at the 4th International Workshop on Supersymmetry (SUSY 96), 
Univ. of Maryland, College Park, USA, May 29th-June 1st 1996;
J. Ellis, G.L. Fogli, E. Lisi,\, {\it Phys. Lett.} {\bf B 333} (1994) 118;
P. Chankowski, S. Pokorski,\, 
 preprint IFT-95/6 [hep-ph/9505308];
\bibitem{WdeBoer}
W.de Boer, A. Dabelstein, W. Hollik, W. M\"osle, U. Schwickkerath,
{\it Updated global fits of the SM and the MSSM to electroweak precision data},
preprint IEK-KA.96-07 [hep-ph/9609209].
\bibitem{SO10}
M. Carena, S. Pokorski, 
C.E.M. Wagner,\,{\it Nucl. Phys.} {\bf B 426} (1994) 269;
L.J. Hall, R. Rattazzi, U. Sarid,\, {\it Phys. Rev.} {\bf D 50}
 (1994) 7048; V. Barger, M.S. Berger, P. Ohmann\,
{\it Phys. Rev.} {\bf D 47} (1993) 1093.
\bibitem{Bigi}
I. Bigi, Y. Dokshitzer, V. Khoze, J. K\"uhn, P. Zerwas,\,
{\it Phys. Lett.}\, {\bf B 181} (1986) 157;
V. Barger, R.J.N. Phillips,\, {\it Phys. Rev.}\, {\bf D 41} (1990) 884;
A.C. Bawa, C.S. Kim, A.D. Martin\, {\it Z. Phys.}\,{\bf C 47} (1990) 95;
M. Drees, D.P. Roy,\, {\it Phys. Lett.} {\bf B 269} (1991) 155;
B.K. Bullock, K. Hagiwara, A.D. Martin,\, {\it Phys. Rev. Lett.}\, 
{\bf 67} (1991) 3055; W. Bernreuther et al., in 
Proc. of the Workshop on $e^+e^-$ Collisions at 500 GeV: The Physics
Potential, Hamburg, 1991, ed. P.M. Zerwas; preprint DESY 92--123A (1992).
\bibitem{Kunszt}
Z. Kunszt, F. Zwirner,\,{\it Nucl. Phys.} {\bf B 385} (1992) 3;
D.P. Roy,\, {\it Phys. Lett.} {\bf B 283} (1992) 403. 
\bibitem{CD}
A. Czarnecki, S. Davidson,\, {\it Phys. Rev.} {\bf D 48} (1993) 4183; 
{\bf D 47} (1993) 3063, and references therein.
\bibitem{GJS}
J. Guasch, R.A. Jim\'enez, J. Sol\`a,\,{\it Phys. Lett.}\, {\bf B 360} (1995) 47.
\bibitem{JSP}
J. Sol\`a,\, {\it  Supersymmetric effects on top quark
and Higgs boson decay widths in the MSSM}, talk given at the
4th International Workshop on Supersymmetry and Unification
(SUSY 96), Univ. of Maryland,
College Park, USA, May 29th-June 1st 1996; and talk given at the Cracow
International Symposium on Radiative 
Corrections (CRAD 96), Cracow, Poland, 1-5 August 1996; and talk given
at the CERN Theory Division, September 20th, 1996. 
\bibitem{GJSH}
D. Garcia, W. Hollik, R.A. Jim\'enez, J. Sol\`a,\, {\it Nucl. Phys.} {\bf B 427}
(1994) 53.
\bibitem{DHJJS}
A. Dabelstein, W. Hollik, R.A. Jim\'enez, C. J\"unger,
J. Sol\`a,\, {\it Nucl. Phys.}\, {\bf B 456} (1995) 75.
\bibitem{GH}
B. Grzadkowski, W. Hollik,\, {\it Nucl. Phys.} {\bf B 384} (1992) 101;
A. Denner, A.H. Hoang,\, {\it Nucl. Phys.} {\bf B 397} (1993) 483.
\bibitem{GJS2}
D. Garcia, R.A. Jim\'enez, J. Sol\`a,\, {\it Phys. Lett.} 
{\bf B 347} (1995) 321;
D. Garcia, J. Sol\`a,\, {\it Phys. Lett.}\, {\bf B 354} (1995) 335 and
{\bf B 357} (1995) 349.
\bibitem{Rb2} 
J.D. Wells, C. Kolda, G.L. Kane, {\it Phys. Lett.}
 {\bf B 338} (1994) 219;
 X. Wang, J.L. Lopez,
D.V. Nanopoulos,\, {\it Phys. Rev.}\, {\bf D 52} (1995) 4116 and
preprint CERN-TH.7553/95 [hep-ph/9501258];
A. Dabelstein, W. Hollik, W. M\"osle,\, 
preprint KA-THEP-5-1995 [hep-ph/9506251];
P. Chankowski, S. Pokorski, preprint IFT-96/6 [hep-ph/9603310].
\bibitem{Warsaw}
A. Blondel, Plenary talk at the
28th International Conference on High Energy Physics, Warsaw, Poland, 
July 25-31th 1996; 
G. Altarelli, summary talk at the Cracow
International Symposium on Radiative Corrections (CRAD 96), 
Cracow, Poland, 1-5 August 1996. 
\bibitem{GuaschS}
J. Guasch, J. Sol\`a,\, preprint UAB-FT-389 [hep-ph/9603441] 
({\it Z. Phys.}\, {\bf C}, in press).
\bibitem{Hunter}
J.F. Gunion, H.E. Haber, G.L. Kane, S. Dawson,\, {\it The Higgs Hunters'
Guide} (Addison-Wesley, Menlo-Park, 1990).
\bibitem{Higgsloop} 
J. Ellis, G. Ridolfi, F. Zwirner,\, {\it Phys. Lett.} {\bf B 262} (1991) 477; 
A. Brignole, J. Ellis, G. Ridolfi, F. Zwirner,\,
{\it Phys. Lett.} {\bf B 271} (1991) 123; 
H.E. Haber, R. Hempfling,\, {\it Phys. Rev. Lett.} {\bf 66} (1991) 1815;
{\it Phys. Rev. }{\bf D 48} (1993) 4280;
H.E. Haber, in: {\it Perspectives on
Higgs Physics},\, Advanced Series on Directions in High Energy Physics,
Vol.13, ed. by G.L. Kane (World Scientific, Singapore, 1993).
\bibitem{Moriond} J. Nachtman and P. Rebecchi, talks given at the
  XXXIst Rencontres de Moriond on Electroweak Interactions and
  Grand Unified Theories, Les Arcs, March 1996, to appear in the
  Proceedings (eds. Fronti\`eres).
\bibitem{Tartarelli}
G.F. Tartarelli, talk given at the
  XXXIst Rencontres de Moriond on Electroweak Interactions and
  Grand Unified Theories, Les Arcs, March 1996, to appear in the
  Proceedings (eds. Fronti\`eres).
\bibitem{CPYuan}
C.-P. Yuan,\, {\it Top Quark Physics},\, preprint-MSUHEP-50228 [hep-ph/9503216].
\bibitem{Kim}
J. Kim, J.L. Lopez, D.V. Nanopoulos, R. Rangarajan,\, preprint ACT-05/96
[hep-ph/9605419]; J.M. Yang, C.S. Li, [hep-ph/9603442];
W. Beenakker, R.H. H\"opker,\, [hep-ph/9606290 and 9607345].
\bibitem{Unpub}
J. Guasch and J. Sol\`a, unpublished; A. Djouadi, W. Hollik, C. J\"unger,
preprint KA-TP-20-96 [hep-ph/9609419].
\bibitem{KaneM}
G.L. Kane, S. Mrenna,\,preprint ANL-HEP-PR-96-43 [hep-ph/9605351];
T. Kon, T. Nonaka,\, {\it Phys. Rev.} {\bf D 50} (1994) 6005.
\bibitem{tauTev}
F. Abe {\it et al.} (CDF Collab.),\, {\it Phys. Rev. Lett.} {\bf 72} (1994) 1977.
\bibitem{Matorras}
F. Matorras, private communication.
\bibitem{Atlas}
Atlas Collab.,\, {\it Atlas technical proposal for a general-purpose pp experiment
at the Large Hadron Collider at CERN}, preprint CERN/LHCC/94-43, December 1994. 
\bibitem{Conway}
J. Conway, talk
given at the 4th International Workshop on Supersymmetry, Univ. of Maryland,
College Park, USA, May 29th-June 1st 1996.
\bibitem{GuraRoy}
D.P. Roy, private communication. 
\bibitem{Raychau}
S. Raychaudhuri, D.P. Roy,\, {\it Phys. Rev.}\, {\bf D 52} (1995) 1556, 
and {\it Phys. Rev.}\, {\bf D 53} (1996) 4902. 
\bibitem{RossTaylor}
D.A. Ross, J.C. Taylor, {\it Nucl. Phys.} {\bf B 51} (1973) 25;
 A. Sirlin,\, {\it Phys. Rev.} {\bf D 22} (1980) 971;
 W. Marciano, A. Sirlin,\, {\it Phys. Rev.} {\bf D 22} (1980) 2695.
\bibitem{Yellow}
See e.g. M. Consoli, W. Hollik, G. Burgers, F. Jegerlehner,
in: {\it Z physics at LEP 1}, eds. G. Altarelli, R. Kleiss,
 C. Verzegnassi, Yellow Report CERN 89-08 (1989).
\bibitem{BSH}
M. B\"ohm, H. Spiesberger, W. Hollik,\, {\it Fortschr. Phys.}
{\bf 34} (1986) 687;
W. Hollik,\,  {\it Fortschr. Phys.} {\bf 38} (1990) 165;  
W. Hollik, in: {\it Precision
Tests of the Standard Electroweak Model}, Advanced Series in Directions in 
High Energy Physics, ed. by P. Langacker (World Scientific, Singapore, 1995).
\bibitem{Jegerlehner}
 F. Jegerlehner,\,
{\it Renormalizing the Standard Model}, lectures given at the
Theoretical Advanced Study Institute in Elementary Particle Physics, 
Boulder, Colorado, June 1990 (TASI 90) p. 476;
A. Denner,\,  {\it Fortschr. Phys.} {\bf 41} (1993) 307.
\bibitem{GS}
D. Garcia, J. Sol\`a,\, {\it Mod. Phys. Lett.} {\bf A 9} (1994) 211;
P.H. Chankowski, A. Dabelstein, W. Hollik, W. M\"osle, S. Pokorski, J. Rosiek,\,
{\it Nucl. Phys. } {\bf B 417} (1994) 101.
\bibitem{CPRD}
A.~Yamada {\it Z. Phys.}\,{\bf C 61} (1994) 247;
P.H. Chankowski, S. Pokorski, J. Rosiek,\, {\it Nucl. Phys.}
{\bf B 423} (1994) 437;
A. Dabelstein,\, {\it Z. Phys.} {\bf C 67} (1995) 495; 
{\it Nucl. Phys.} {\bf B 456} (1995) 25.
\bibitem{Godbole}
R.M. Barnett, J.F. Gunion, H.E. Haber, I. Hinchliffe, B. Hubbard,
H.J. Trost, preprint SDC-90-00141; R.M. Godbole, D.P. Roy,\,
{\it Phys. Rev.} {\bf D 43} (1991) 3640; J.F. Gunion, L.H. Orr,\, 
{\it Phys. Rev.}\, {\bf D 46} (1992) 2052.     
\bibitem{Hollikprivate}
W. Hollik, private conversation; W. Hollik, talk given at the IFAE, 
Univ. Aut\`onoma de Barcelona, February 1996.
\bibitem{GrifSol}
J. Grifols, J. Sol\`a,\, {\it Nucl. Phys.}\, {\bf B 253} (1985) 47;
{\it Phys. Lett.} {\bf B 137} (1984) 257; For a detailed review, see:
 J. Sol\`a, in: {\it Phenomenological
Aspects of Supersymmetry}, ed. W. Hollik, R. R\"uckl and J. Wess (Springer
-Verlag, Lecture Notes in Physics {\bf 405}, 1992) p. 187.
\bibitem{Bertolini}
S. Bertolini, {\it Nucl. Phys.}\,{\bf B 272} (1986) 77.
\bibitem{Koenig}
H. K\"onig, {\it Phys. Rev.} {\bf D 50} (1994) 3310.
\bibitem{LiYangHu}
C.S. Li, J.M. Yang, B.Q. Hu, {\it Phys. Rev.} {\bf D 48} (1993) 5425. 
\bibitem{Mendez}
A. M\'endez, A. Pomarol,\,{\it Phys. Lett.} {\bf B 265} (1991) 177. 
\bibitem{LiHuYang}
C.S. Li, B.Q. Hu, J.M. Yang,\, {\it Phys. Rev.} {\bf D 47} (1993) 2865.
\bibitem{Yamada2}
A. Yamada,\,{\it Phys. Lett.} {\bf B 263} (1991) 233; A. Brignole,\,
{\it Phys. Lett.} {\bf B 281} (1992) 284;
H.E. Haber, R. Hempfling,\,
{\it Phys. Rev. }{\bf D 48} (1993) 4280; 
M.A. Diaz, H.E. Haber,\, {\it Phys. Rev.}\, {\bf D 45} (1992) 4246;
M.A. Diaz,\, {\it Phys. Rev.}\, {\bf D 48} (1993) 2152.
\bibitem{Yamada3}
Y. Yamada, preprint MADPH-96-956 [hep-ph/9608382].
\bibitem{CPW}
M. Carena, S. Pokorski, C.E.M. Wagner\,{\it Nucl. Phys.} {\bf B 406} (1993) 59.
\bibitem{MC}
M. Carena and C. Wagner, private conversation.
\bibitem{LEPEWWG}
LEP Electroweak Working Group, CERN preprint LEPEWWG/96-02, July 1996.
\bibitem{TopSM}
M. Jezabek, J.H. K\"uhn,\, {\it Nucl. Phys.}\, {\bf B 314} (1989) 1;
{\sl ibid} {\bf B 320} (1989) 20; 
C.S. Li, R.J. Oakes, T.C. Yuan,\, {\it Phys. Rev.}\,{\bf D 43} (1991) 3759;
G. Eilam, R.R. Mendel, R. Migneron, A. Soni,\, {\it Phys. Rev. Lett.}
{\bf 66} (1991) 3105; 
A. Denner, T. Sack,\, {\it Z. Phys.}\, {\bf C 46} (1990) 653;
{\it Nucl. Phys.}\, {\bf B 358} (1991) 46;
B.A. Irwin, B. Magnolis, H.D. Trottier, {\it Phys. Lett.}\, {\bf B 256} (1991) 533;
T.C. Yuan, C.-P. Yuan,\, {\it Phys. Rev.}\, {\bf D 44} (1991) 3603.
\bibitem{Djuadi}
A. Djouadi, W. Hollik, C. J\"unger,\, preprint KA-TP-14-96 [hep-ph/9605340];
C.S. Li, R.J. Oakes, J.M. Yang,\, preprint NUHEP-TH-96-3 [hep-ph/9606385].
\bibitem{YangLi}
J.M. Yang, C.S. Li,\, {\it Phys. Lett.}\, {\bf B 320} (1994) 117.
\bibitem{LiYuan}
C.S. Li, T.C. Yuan,\, {\it Phys. Rev.}\, {\bf D 42} (1990) 3088.
\bibitem{Mehen}
T. Mehen, preprint JHU-TIPAC-96004 [hep-ph/9602410].
\bibitem{Ng}
R. Garisto, J.N. Ng,\, {\it Phys. Lett.}\, {\bf B 315} (1993) 372;
R. Barbieri, G.F. Giudice,\, {\it Phys. Lett.}\, {\bf B 309} (1993) 86;
M. Carena, C.E.M. Wagner, {\it Nucl. Phys.}\, {\bf B 452} (1995) 45.
\bibitem{RicardToni}
R.A. Jim\'enez, J. Sol\`a, preprint UAB-FT-375 [hep-ph/9511292]
({\it Phys. Lett.}\,{\bf B},
in press); J. A. Coarasa, R.A. Jim\'enez, J. Sol\`a, preprint UAB-FT-376
[hep-ph/9511402] ({\it Phys. Lett.}\,{\bf B}, in press).
\bibitem{Willenbrock}
S. Willenbrock, {\it Top Quark Physics at Fermilab}, talk 
presented at the International Symposium on Particle Theory and
Phenomenology, Iowa State University, May 1995 [hep-ph/9508212],
and talk presented in the XXXIst Rencontres de Moriond 
on Electroweak
Interactions and Grand Unified Theories, Les Arcs, March 1996, to appear in the
Proceedings (eds. Fronti\`eres);  
C. Quigg,\, preprint Fermilab-CONF-95/139-T [hep-ph/9507257].
\bibitem{Heinson}
A.P. Heinson,\, {\it Electroweak Top Quark Production at the Fermilab Tevatron},
\,Talk given at the Workshop on Physics of the Top Quark, 
Ames, Iowa, May 1995; A.P. Heinson, A.S. Belyaev,
E.E. Boos, preprint UCR/95-17 [hep-ph/9509274].
\bibitem{Fujii}
K. Fujii, T. Matsui,\, {\it Phys. Rev.}\,{\bf 50} (1994) 4341.
\bibitem{Frey}
R. Frey, preprint OREXP 96-04 [hep-ph/9606201].
\bibitem{Mathematica}
S. Wolfram, {\it Mathematica-
A System for doing Mathematics by Computer}, Addi\-son-Wesley, Redwood City,
CA, 1991).
\bibitem{Gunion}
J.F. Gunion, H. Haber,\, {\it Nucl. Phys.}\,{\bf B 272} (1986) 1.
\bibitem{Frere}
J.M. Fr\`ere, D.R.T. Jones, S. Raby,\, {\it Nucl. Phys.}\, 
{\bf B 222} (1983) 11; M. Claudson,
L. Hall, I. Hinchliffe,\, {\it Nucl. Phys.}\, {\bf B 228} (1983) 501;
C. Kounnas, A.B. Lahanas, D.V. Nanopoulos, M. Quir\'os,\,
{\it Nucl. Phys.}\,{\bf B 236} (1984) 438; J.F. Gunion, H.E. Haber,
M. Sher,\, {\it Nucl. Phys.}\, {\bf B 306} (1988) 1.
\bibitem{Collins}
See e.g. J. Collins, {\it Renormalization} (Cambridge University Press, 1984).
\bibitem{Veltmann}
G. `t Hooft, M. Veltman, {\it Nucl. Phys.}\, {\bf B 153} (1979) 365;
G. Passarino, M. Veltman,\, {\it Nucl.Phys.}\, {\bf B 160} (1979) 151;
A. Axelrod,\, {\it Nucl. Phys.}\, {\bf B 209} (1982) 349.



\end{thebibliography}
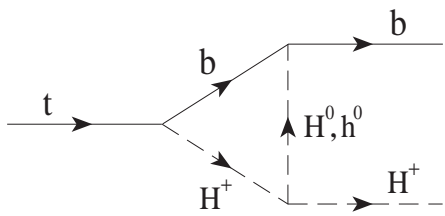
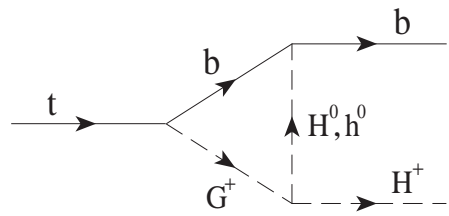
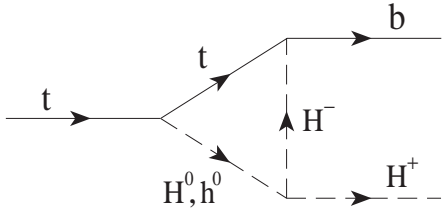
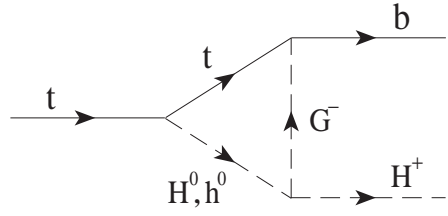
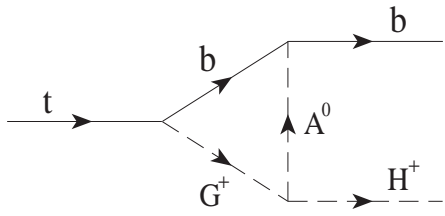
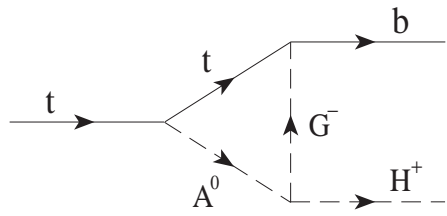
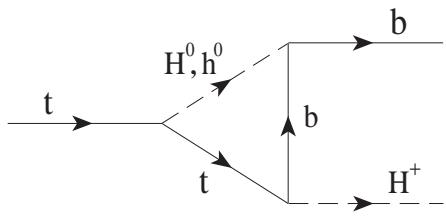
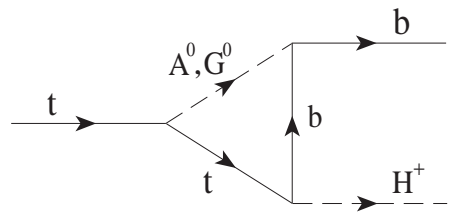
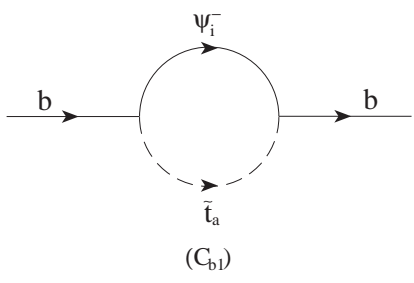
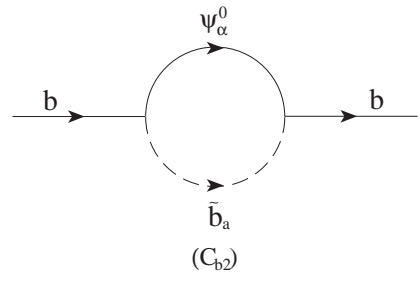
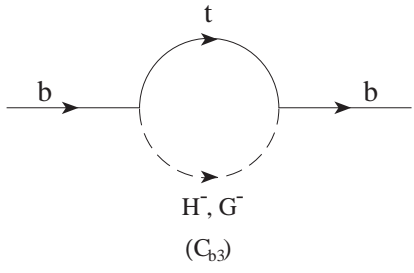
\end{document}